# Twistronics and moiré superlattice physics in 2D transition metal dichalcogenides

Dawei Zhai[1,2*], Hongyi Yu[3,4*], Wang Yao[1,2]

[1] New Cornerstone Science Lab, Department of Physics, The University of Hong Kong, Hong Kong, China
[2] HK Institute of Quantum Science & Technology, The University of Hong Kong, Hong Kong, China
[3] Guangdong Provincial Key Laboratory of Quantum Metrology and Sensing & School of Physics and Astronomy, Sun Yat-Sen University (Zhuhai Campus), Zhuhai 519082, China
[4] State Key Laboratory of Optoelectronic Materials and Technologies, Sun Yat-Sen University (Guangzhou Campus), Guangzhou 510275, China

[*] Correspondence to: dzhai@hku.hk, yuhy33@mail.sysu.edu.cn

**Abstract:** The moiré superlattices formed by stacking 2D semiconducting transition metal dichalcogenides (TMDs) with twisting angle or lattice mismatch have provided a versatile platform with unprecedented tunability for exploring many frontier topics in condensed matter physics, including optical, topological and correlation phenomena. This field of study advances rapidly and a plethora of exciting experimental and theoretical progresses have been achieved recently. This review aims to provide an overview of the fundamental properties of TMDs moiré superlattices, as well as highlight some of the major breakthroughs in this captivating field.

## Table of Contents

## 1. Introduction

Van der Waals (vdW) assembly of 2D atomic crystals has revolutionized the way we understand and manipulate materials. The resultant layered structures can exhibit customizable and intriguing properties, beyond those of the individual building blocks. Particularly, a small rotational misalignment or lattice constant mismatch can lead to the formation of long-wavelength moiré patterns at the interfaces with a periodicity typically one or two orders of magnitude larger than the monolayer lattice constant. Interlayer couplings in the moiré define a new landscape for low-energy carriers and elementary excitations, thereby modulating the electronic, optical and topological properties.

Moiré superlattice physics and twist control of electronic structures first caught attention in the context of graphene, where early experimental signatures include the energy band hybridization and flattening in twisted bilayer graphene [1], emergence of secondary Dirac points in graphene deposited on hexagonal Boron Nitride (hBN) [2, 3] and fractal electric structures in the moiré of monolayer/bilayer graphene aligned with hBN in an out-of-plane magnetic field [3-5]. The big motivation in advancing the research of moiré materials and twistronics came a few years later with the groundbreaking discovery of superconductivity and correlated insulating behavior in magic-angle twisted bilayer graphene hosting moiré flat bands [6, 7]. Importantly, the nontrivial band topology in moiré superlattices make it a viable platform to explore the interplay of band topology and electron correlation, with opportunities that have not been possible in conventional correlated electron systems, thanks to the unprecedented controllability in moiré materials via twist angle control, electrostatic gating, in-plane and out-of-plane fields, dielectric environment, strain and pressure. Many novel phenomena have been discovered in graphene moiré superlattices and the exploration have also been actively expanded to systems consisting of multilayer graphene and multiple twisted interfaces [8-14].

Another versatile moiré platform is those consisting of monolayer transition metal dichalcogenides (TMDs), especially the semiconducting compounds with a direct band gap where strong spin-orbit coupling (SOC) leads to spin-valley locked band edges. In contrast to graphene, the TMDs family has a number of compounds with disparity in band gap, electron affinity, spin splitting, and modestly different lattice constant, offering more flexibility in designing moiré superlattices for the various frontier topics of interest. For instance, in the long-wavelength moiré of TMDs heterobilayer, the band offset between the two compounds in general confines the carriers on an individual layer, effectively realizing triangular superlattices, which can be employed for simulating strongly correlated physics with geometrical frustration. In several heterobilayer TMDs moiré systems, experimental endeavors have discovered varieties of charge ordered states and magnetic ordered states at integer and fractional filling of the moiré superlattices. On the other hand, in small-angle twisted homobilayer TMDs, a superlattice Kane-Mele model can be realized, which yields topologically nontrivial moiré bands with valley-contrasted Chern numbers. Such topologically nontrivial moiré platforms have also led to many exciting experimental breakthroughs, including the observation of fractional quantum anomalous Hall effects at zero external magnetic field in twisted bilayer $MoTe_2$. In addition to electronic properties, semiconducting TMDs also exhibit interesting optical properties. For example, the Coulomb binding of interlayer electron-hole pairs in a moiré landscape can give rise to moiré excitons exhibiting unique spatially modulated optical selection rules. In small-angle twisted homobilayer TMDs, intralayer and interlayer excitons can hybridize due to interlayer carrier hopping, leading to the formation of hybridized moiré excitons that can exhibit distinct optical and even topological properties. In the presence of exciton-exciton interactions, correlated excitonic phenomena, e.g. optically entangled light emitting and excitonic crystals, can also occur.

In this review, we introduce the fundamental properties of TMDs moiré superlattices, with an overview of the experimental and theoretical progresses in the exploration of optical, topological and correlation phenomena.

## 1.1. Monolayer building block

Monolayer TMDs with the chemical formula $MX_2$ ($M$ = Mo/W, $X$ = S/Se/Te) exhibit great potentials for applications in novel optoelectronic, photonic and valleytronic devices [15-18]. They have a 2D hexagonal crystal structure consisting of covalently bonded $X$-$M$-$X$ tri-atomic planes [19]. The monolayer crystal structure has the $D_{3h}$ point group symmetry, containing the in-plane $2\pi/3$-rotational ($\hat{C}_3$) symmetry about the rotation center $M$, $X$ or $h$, in-plane mirror ($\hat{\sigma}_v$) symmetry across the vertical planes connecting $M$ and $X$ atoms, and the mirror reflection symmetry ($\hat{\sigma}_\mathrm{h}$) about the horizontal plane crossing all $M$ atoms (see Fig. 1(a, b)). Each monolayer is a direct band gap semiconductor with the gap value ranging from infrared to visible frequencies depending on material compounds [20, 21]. Both the conduction and valence band edges are located at the two degenerate but inequivalent corners of the hexagonal Brillouin zone (denoted as $\pm\mathbf{K}$, see Fig. 1(c)). Low-energy carriers in monolayer TMDs thus have a binary quantum degree-of-freedom (termed as the valley pseudospin) describing their positions in momentum space, which can be exploited as an information carrier in future electronic and optoelectronic applications [17, 22]. Near $\pm\mathbf{K}$, the valence (conduction) bands mainly consist of $d_{x^2-y^2} \pm i d_{xy}$ ($d_{z^2}$) orbitals of the transition metal atom $M$, where the SOC effect splits each band into two sub-bands with opposite spins [19, 23, 24]. Due to the $\hat{\sigma}_\mathrm{h}$ symmetry, each Bloch state has a spin expectation value pointing along the out-of-plane ($z$) direction and the SOC is dominated by the Ising term ($\propto \hat{S}_z \hat{L}_z$, where $\hat{L}_z$ denotes the $z$-component of the atomic orbital angular momentum). Near $\pm\mathbf{K}$, the strong SOC of transition metal $d_{x^2-y^2} \pm i d_{xy}$ orbitals with $L_z = \pm 2$ results in a giant spin splitting $\lambda_\mathrm{v} \sim 0.15$ to 0.45 eV for the valence band [23-27], whereas the $d_{z^2}$ orbital with $L_z = 0$ leads to a much weaker SOC spin splitting $\lambda_\mathrm{c}$ for the conduction band (several to several tens meV) [24, 28-37]. The time-reversal symmetry between $\pm\mathbf{K}$ valleys results in opposite signs for their SOC splitting values, which then leads to the spin-valley locking of band-edge carriers, see Fig. 1(d).

The simplest $\mathbf{k}\cdot\mathbf{p}$ Hamiltonian for spin-$S_z$ conduction and valence states at $\tau\mathbf{K} + \mathbf{k}$ with $\tau$

$= \pm$ corresponds to a two-band massive Dirac model [23], with the Hamiltonian given by

$$\hat{H}_{\mathbf{k}\cdot\mathbf{p}}^{(1\mathrm{L})} = \begin{pmatrix} \Delta_{\tau s_z}/2 & \hbar v_{\mathrm{F}}(\tau k_x - ik_y) \\ \hbar v_{\mathrm{F}}(\tau k_x + ik_y) & -\Delta_{\tau s_z}/2 \end{pmatrix}. \tag{1}$$

Here $\Delta_{\tau s_z}$ is the spin- and valley-dependent band gap and $v_{\mathrm{F}}$ is the Fermi velocity. Despite omitting the asymmetry in effective masses of electrons and holes (except possibly for $MoS_2$ [27, 38]), $\hat{H}_{\mathbf{k}\cdot\mathbf{p}}^{(1\mathrm{L})}$ can already give a good description to band-edge dispersions and quantum geometries characterized by momentum-space Berry curvatures. With a three-band tight-binding model consisting of $d_{z^2}$ and $d_{x^2-y^2} \pm id_{xy}$ orbitals, nearest-neighbor hopping can give a rather accurate account of the electron-hole asymmetry, and the trigonal warping effects near $\tau\mathbf{K}$, while including up to third-nearest-neighbor hopping can fit the three bands dominated by $d_{z^2}$, $d_{xy}$ and $d_{x^2-y^2}$ orbitals from first-principles well in the entire Brillouin zone [28].

The Bloch states at the high-symmetry points $\tau\mathbf{K}$ with spin $S_z$, denoted as $|\psi_{n,\tau\mathbf{K},S_z}\rangle$ with $n$ = c (v) labeling the conduction (valence) band, are constrained by the $\hat{C}_3$ and $\hat{\sigma}_{\mathrm{h}}$ symmetries. Specifically, they need to satisfy $\hat{C}_3|\psi_{n,\tau\mathbf{K},S_z}\rangle = e^{-i\frac{2\pi}{3}(\tau C_{3,n}+S_z)}|\psi_{n,\tau\mathbf{K},S_z}\rangle$ and $\hat{\sigma}_{\mathrm{h}}|\psi_{n,\tau\mathbf{K},S_z}\rangle = e^{-i\pi S_z}|\psi_{n,\tau\mathbf{K},S_z}\rangle$. Here $\tau C_{3,n}$ corresponds to the orbital $\hat{C}_3$ quantum number of $|\psi_{n,\tau\mathbf{K},S_z}\rangle$, which has opposite signs at $\pm\mathbf{K}$. Although both $C_{3,\mathrm{c}}$ and $C_{3,\mathrm{v}}$ varies with the choice of the rotation center $M$, $X$ or $h$, their difference is always given by $C_{3,\mathrm{c}} - C_{3,\mathrm{v}} = +1$ (mod 3), see Table 1. These symmetry constraints give rise to the valley- and spin-dependent optical selection rules in monolayer TMDs [23, 39, 40] and determine the stacking-dependent selection rules of the interlayer tunneling and optical transition in a bilayer, which will be discussed in detail in later sections.

**Table 1.** The orbital $\hat{C}_3$ quantum numbers $C_{3,\mathrm{c}}$ and $C_{3,\mathrm{v}}$ for the conduction and valence band Bloch states at $\mathbf{K}$, respectively, under three rotation centers ($M$, $X$ or $h$) of monolayer TMDs.

| | $M$ | $X$ | $h$ |
|---|---|---|---|
| $C_{3,\mathrm{c}}$ | 0 | −1 | +1 |
| $C_{3,\mathrm{v}}$ | −1 | +1 | 0 |

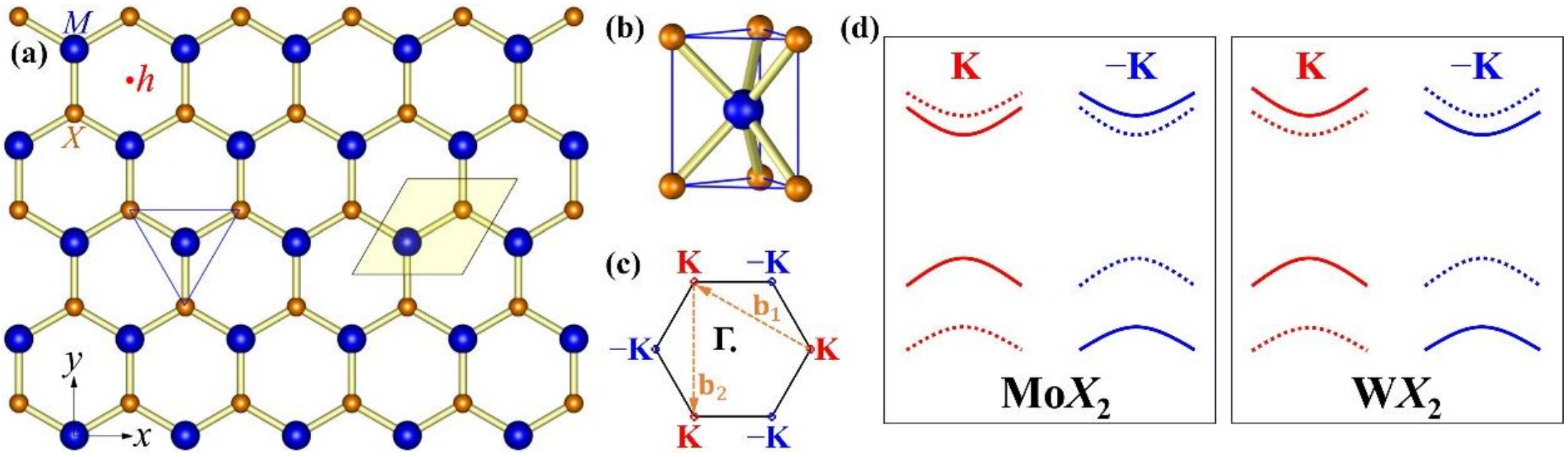

**Figure 1**. The top view of the 2D hexagonal crystal structure of a TMDs monolayer. The blue and orange dots denote $M$ and $X$ atoms, respectively. The yellow diamond encloses a unit cell. (b) The side view showing the arrangement of $M$ and $X$ atoms in the blue triangle in (a). (c) The hexagonal Brillouin zone of monolayer TMDs with $\pm\mathbf{K}$ being inequivalent corners. (d) Schematics of $\pm\mathbf{K}$ valley band dispersions and SOC splittings in monolayer $MoX_2$ (left panel) and $WX_2$ (right panel), where solid (dashed) curves correspond to spin-up (spin-down) state. (a) Adapted from Ref. [19] with permission from the Royal Society of Chemistry.

## 1.2. Spatially varying local registries and tunable superlattice geometry of moiré pattern

Two monolayer TMDs of the same or different compound can be vertically stacked to form a homo- or hetero- bilayer. As the two layers are bound together by the weak vdW force, such interface is not subject to the requirement of lattice matching. In the presence of different lattice constants and/or a small interlayer twist angle, a long-wavelength moiré pattern can emerge, which corresponds to a spatially periodic modulation of the local stacking registry. The moiré pattern can be described in terms of the twist angle $\theta$ (Fig. 2(a)) or mismatch $\delta = |1 - a'/a|$ between their lattice constants $a$ and $a'$ (Fig. 2(b)), or the combination of the two. When both quantities are small, the moiré pattern forms a $\hat{C}_3$-symmetric triangular superlattice with a wavelength $\lambda \approx a/\sqrt{\theta^2 + \delta^2}$. Moiré patterns can be further tuned by applying a heterostrain (differential strain between the two layers), which can give rise to various lattice geometries with or without $\hat{C}_3$ symmetry [41-46]. In experiments, one-dimensional moiré patterns have been realized in twisted $WSe_2/MoSe_2$ heterobilayers by applying a small uniaxial strain [47], and incidentally, similar anisotropic moiré patterns have also been obtained in twisted bilayer $WTe_2$ composed of monolayers with a 1T distorted structure [48, 49]. Note that under an area-conserving heterostrain with a rather small magnitude, the moiré superlattice is still in the triangular type when $\theta = 0$ but without $\hat{C}_3$ symmetry, see Fig. 2(c). The moiré periodicity sets up a new length scale for the low-energy carriers, excitons and other collective excitations, which in the momentum space defines the moiré mini-Brillouin zone (mBZ) (see the lower panel of Fig. 2(a) and insets in Figs. 2(b, c)). Combined with the registry-dependent interlayer coupling effects, the moiré pattern can have profound impacts on the electronic, optical and mechanical properties of the system [18, 50-53]. It can serve as a novel platform for simulating various correlated and topological phenomena as well as having potential applications in next-generation optoelectronic devices.

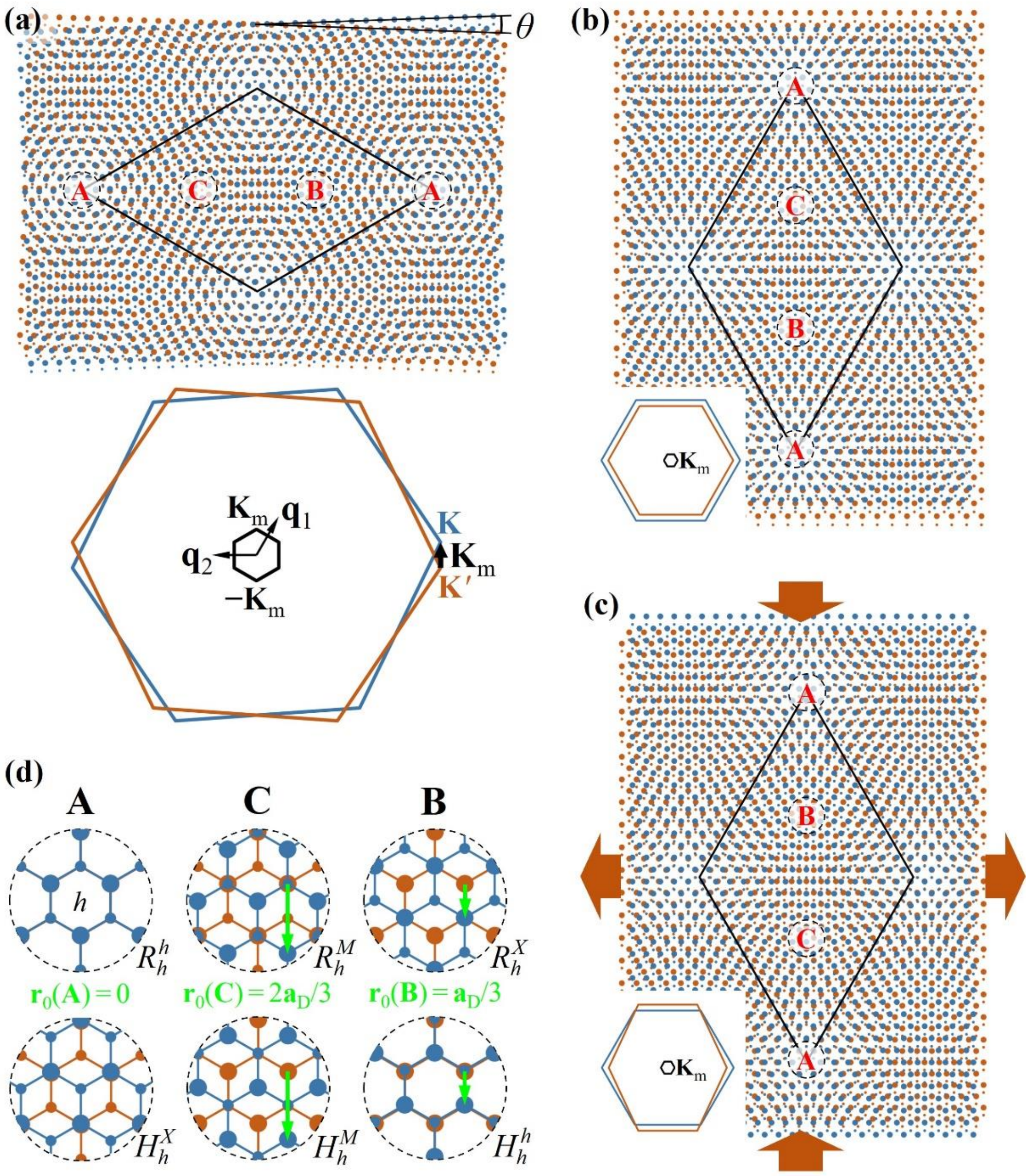

**Figure 2.** (a) The upper panel illustrates a moiré pattern formed by an interlayer twist angle $\theta$ with the diamond enclosing a supercell, the lower panel schematically shows monolayer Brillouin zones (blue and orange hexagons) and the moiré mini-Brillouin zone (small black hexagon) with the corners at $\pm\mathbf{K}_m$. $\mathbf{q}_{1,2}$ correspond to the primitive reciprocal lattice vectors of the moiré pattern. (b) A moiré pattern formed by a lattice constant mismatch between the two layers, whose crystalline directions are aligned. (c) A moiré pattern formed by applying an area-conserving uniaxial strain on the orange layer of the commensurate homobilayer. Thick arrows indicate the strain direction. (d) The enlarged views of atomic stacking registries in the three local regions **A**, **B** and **C**, which resemble lattice-matched bilayer stackings $R_h^h$, $R_h^X$ and $R_h^M$ ($H_h^X$, $H_h^h$ and $H_h^M$) with interlayer translations $\mathbf{r}_0 = 0$, $\mathbf{a}_D/3$ and $2\mathbf{a}_D/3$, respectively, for close to R-type (H-type) bilayer stackings. $\mathbf{a}_D$ is schematically defined in Fig. 3(c).

For a local region in the moiré pattern with a spatial extension much smaller than the moiré period, the stacking registry is nearly indistinguishable from that of R- or H-type (i.e., parallel or anti-parallel) commensurate bilayer. The local-to-local variation of the stacking registry over the moiré length scale can be characterized by a spatially varying interlayer translation $\mathbf{r}_0(\mathbf{r})$ with $\mathbf{r}$ the center position of the local region. This leads to the local approximation widely adopted for describing long-wavelength moiré, in which band structures of commensurate bilayers with different $\mathbf{r}_0$ are used to approximate local electronic structures of the corresponding stacking regions in moiré. Compared to brute-force first-principles calculations, with a moiré supercell having hundreds to thousands of atoms, the local approximation greatly reduces the computational cost. Among the R-type (H-type) commensurate stacking registries, the three $\hat{C}_3$-symmetric ones will be denoted as $R_h^h$, $R_h^X$, $R_h^M$ ($H_h^X$, $H_h^h$ and $H_h^M$): in $R_h^\mu$ ($H_h^\mu$) the upper-layer rotation center $\mu = M/X/h$ is vertically aligned with the lower-layer rotation center $h$ (Fig. 2(d)). The three $\hat{C}_3$-symmetric stacking registries appear once each in a moiré supercell, which qualitatively dictate the spatially varying properties of the moiré superlattices. When the two layers are treated as rigid lattices, the mapping between spatial location $\mathbf{r} \equiv (x, y)$ and local registry $\mathbf{r}_0(\mathbf{r}) \equiv (x_0(\mathbf{r}), y_0(\mathbf{r}))$ corresponds to a linear function. For moiré patterns introduced by the combination of a twist angle $\theta$, a lattice constant mismatch $\delta$ and/or heterostrain, the function is $\mathbf{r}_0(\mathbf{r}) = \mathbf{r}_0(0) + (1 - \hat{R}^{-1}\hat{S}^{-1})\mathbf{r}$, where $\hat{R}$ is the rotation matrix by angle $\theta$, and $\hat{S} = 1 + \delta + \hat{\epsilon}$ with $\hat{\epsilon}$ the strain tensor [43]. Figure 2(d) illustrates $\mathbf{r}_0(\mathbf{r})$ at the $\hat{C}_3$-symmetric stacking registries, where the coordinate origin is chosen as the location where two metal atoms from the two layers overlap and $\mathbf{r}_0(0) = 0$ is adopted. As the geometry of the moiré pattern is highly sensitive to the twist angle and the heterostrain, it has been demonstrated that distorted moiré superlattices with spatially inhomogeneous wavelengths can emerge by introducing continuous varying twist angles and strains [54].

### 1.3. Spatially modulated interlayer distance and interlayer hopping in the moiré pattern

In an R-/H-type commensurate bilayer structure, the interlayer vdW interaction varies sensitively with the interlayer translation $\mathbf{r}_0$, which determines the equilibrium interlayer distance $D$ [55-58]. Here $D$ is defined as the vertical separation between the two metal-atom planes, which is typically in the order of 6 Å in bilayer TMDs. Stacking registries with $X$ atoms from the two layers vertically aligned, i.e., $R_h^h$ and $H_h^M$, have the largest $D$ values [55-59]. Minimum $D$ values occur at $R_h^X$ and $R_h^M$ for the R-type stacking and $H_h^h$ for the H-type stacking (see Table 2 and Fig. 3(a)). The difference between the minimum and maximum $D$ values is approximately 0.6 Å. The spatially varying $\mathbf{r}_0$ values in a moiré pattern then result in a modulation of $D$ across different positions in a single moiré supercell, as confirmed by scanning tunneling microscopy/spectroscopy (STM/S) measurements [55, 59]. In bilayers of type-II band alignment with conduction and valence band edges located in opposite layers [60, 61], interlayer distance $D$ and its spatial variation allows tuning of local bandgaps, as well as their differences at different locations, through a vertical electric field [56]. Meanwhile, the interlayer vdW binding energy for an R-/H-type commensurate bilayer also varies similarly with $\mathbf{r}_0$, as summarized in Table 2, which affects the moiré landscape when lattice relaxation is considered (Sec. 1.5).

**Table 2.** Relative size of the interlayer distance or binding energy of $\hat{C}_3$-symmetric local stacking configurations. Max/min/medium indicates the corresponding value is the largest/smallest/intermediate among various stacking configurations with different $\mathbf{r}_0$. Values for $R_h^M$ and $R_h^X$ are slightly different in heterobilayers with distinct chalcogen atoms.

| | | | |
|---|---|---|---|
| R-type | $R_h^h$, max | $R_h^M$, min | $R_h^X$, min |
| H-type | $H_h^X$, medium | $H_h^M$, max | $H_h^h$, min |

Now we consider the interlayer hopping of carriers in commensurate and moiré-patterned bilayers. The finite interlayer hopping can give rise to layer-hybridized Bloch states. Low-energy carriers thus exhibit a layer degree-of-freedom in addition to spin and valley. Most heterobilayers have the type-II band alignment with band offset typically in the order of several hundred meV [62, 63], much larger than the interlayer hopping strength in ±**K** valleys. Electronic states around these band edges thus have negligible layer hybridizations. In homobilayers, interlayer hopping becomes important, and the states can become strongly layer-hybridized.

Starting from two decoupled monolayers, with $\psi_{n,\mathbf{k},s_z}$ ($\psi_{n',\mathbf{k}',s_z'}$) denoting the $n$-th ($n'$-th) band Bloch state in the lower (upper) layer, their interlayer hopping can occur provided $\mathbf{k}+\mathbf{G}=\mathbf{k}'+\mathbf{G}'$ is satisfied, where $\mathbf{G}$ and $\mathbf{G}'$ denote the reciprocal lattice vector of the lower and upper layer. Adopting the two-center approximation $\langle n,\mathbf{R}|\hat{H}|n',\mathbf{R}'\rangle=\tilde{t}_{nn'}(\mathbf{R}'-\mathbf{R})$ [64-66], where $|n,\mathbf{R}\rangle$ denotes atomic orbitals of band $n$ localized at position $\mathbf{R}$ and $\tilde{t}_{nn'}(\mathbf{R}'-\mathbf{R})$ is the hopping integral which is assumed to only depend on the relative position $\mathbf{R}'-\mathbf{R}$, the interlayer hopping matrix element between $\psi_{n,\mathbf{k},s_z}$ and $\psi_{n',\mathbf{k}',s_z'}$ can be expressed as the sum of a series of harmonic terms, with the form $\langle\psi_{n,\mathbf{k},s_z}|\hat{H}|\psi_{n',\mathbf{k}',s_z'}\rangle=\sum_{\mathbf{G}\mathbf{G}'}\delta_{\mathbf{k}+\mathbf{G},\mathbf{k}'+\mathbf{G}'}t_{nn'}(\mathbf{k}+\mathbf{G})e^{-i\mathbf{G}\cdot\mathbf{r}_0}$ [67]. Here $t_{nn'}(\mathbf{k}+\mathbf{G})$ is the Fourier transform of the hopping integral $\tilde{t}_{nn'}(\mathbf{R}'-\mathbf{R})$ between localized orbital compositions of the states. Note that $|t_{nn'}(\mathbf{k}+\mathbf{G})|$ decreases quickly with the increase of $|\mathbf{k}+\mathbf{G}|$, thus usually only harmonic terms with the smallest and second-smallest $|\mathbf{k}+\mathbf{G}|$ need to be kept. In a commensurate TMDs bilayer, at the band edge **K**, the smallest (second-smallest) $|\mathbf{K}+\mathbf{G}|$ corresponds to $\mathbf{K}_0\equiv\mathbf{K}$, $\mathbf{K}_1\equiv\hat{C}_3\mathbf{K}$ and $\mathbf{K}_2\equiv\hat{C}_3^2\mathbf{K}$ ($-2\mathbf{K}_0$, $-2\mathbf{K}_1$ and $-2\mathbf{K}_2$). One then gets

$$T_{nn',\mathbf{K}}\approx e^{i\mathbf{K}\cdot\mathbf{r}_0}\left[\sum_{j=0}^{2}t_{nn'}(\mathbf{K}_j)e^{-i\mathbf{K}_j\cdot\mathbf{r}_0}+\sum_{j=0}^{2}t_{nn'}(-2\mathbf{K}_j)e^{i2\mathbf{K}_j\cdot\mathbf{r}_0}\right], \tag{2}$$

and $T_{nn',-\mathbf{K}}$ is simply its time reversal. For spin-conserved interlayer hopping processes with $S_z=S_z'$, the $\hat{C}_3$ symmetry requires the momentum-space hopping integrals to satisfy $t_{nn'}(\hat{C}_3\mathbf{q})=e^{i\frac{2\pi}{3}(C_{3,n'}-C_{3,n})}t_{nn'}(\mathbf{q})$, with $C_{3,n/n'}$ the orbital $\hat{C}_3$ quantum number of $\psi_{n/n',\mathbf{K},S_z}$ (see Table 1) [67]. This underlies the stacking selection rules for the $\hat{C}_3$-symmetric interlayer registries as summarized in Table 3. This can be straightforwardly generalized to spin-flip processes by including the spin contribution to the $\hat{C}_3$ quantum number, but the hopping magnitude is generally expected to be small $O(1)$ meV [68].

**Table 3.** A list of whether the spin-conserved interlayer hopping at **K** vanishes or not, for $\hat{C}_3$-symmetric stacking registries $R_h^\mu$ and $H_h^\mu$ ($\mu=M,X,h$). Only those involving the lowest conduction and top valence bands are shown.

| | $T_{cc',\mathbf{K}}$ | $T_{vv',\mathbf{K}}$ | $T_{cv',\mathbf{K}}$ | $T_{vc',\mathbf{K}}$ |
|---|---|---|---|---|
| $R_h^h$ | finite | finite | 0 | 0 |
| $R_h^X$ | 0 | 0 | 0 | finite |
| $R_h^M$ | 0 | 0 | finite | 0 |
| $H_h^X$ | finite | 0 | 0 | 0 |
| $H_h^h$ | 0 | finite | 0 | 0 |
| $H_h^M$ | 0 | 0 | finite | finite |

For the interlayer hopping slightly away from **K**, one can start from the monolayer $\mathbf{k}\cdot\mathbf{p}$ model (Eq. (1)) and write the commensurate bilayer Hamiltonian as [69]

$$\widehat{H}^{(2L)}_{\mathbf{k}\cdot\mathbf{p}} = \begin{pmatrix} V_c(\mathbf{r}_0) + \Delta_{s_z}/2 & \hbar v_F(k_x - ik_y) & T_{cc',\mathbf{K}} & T_{cv',\mathbf{K}} \\ \hbar v_F(k_x + ik_y) & V_v(\mathbf{r}_0) - \Delta_{s_z}/2 & T_{vc',\mathbf{K}} & T_{vv',\mathbf{K}} \\ T^*_{cc',\mathbf{K}} & T^*_{vc',\mathbf{K}} & V_{c'}(\mathbf{r}_0) + \Delta_{\tau' s_z}/2 & \hbar v'_F(\tau' k_x - ik_y) \\ T^*_{cv',\mathbf{K}} & T^*_{vv',\mathbf{K}} & \hbar v'_F(\tau' k_x + ik_y) & V_{v'}(\mathbf{r}_0) - \Delta_{\tau' s_z}/2 \end{pmatrix}. \quad (3)$$

Here $V_n(\mathbf{r}_0)$ ($V_{n'}(\mathbf{r}_0)$) is a diagonal energy shift at **K** for the $n$-th ($n'$-th) band in the bottom (top) layer, which can account for the effects from band gap modulations by the interlayer coupling and externally applied interlayer bias. $\tau' = +$ and $-$ for R- and H-type stacking registries, respectively. The interlayer hopping at **K** + **k** can be obtained from $\widehat{H}^{(2L)}_{\mathbf{k}\cdot\mathbf{p}}$ by diagonalizing its $2 \times 2$ monolayer blocks. Taking $n = \mathrm{v}$ and $n' = \mathrm{c}'$ as an example:

$$T_{vc',\mathbf{K}+\mathbf{k}} = T_{vc',\mathbf{K}} + \frac{T_{vv',\mathbf{K}} v'_F}{\Delta_{\tau' s_z}} \hbar(\tau' k_x + ik_y) - \frac{T_{cc',\mathbf{K}} v_F}{\Delta_{s_z}} \hbar(k_x + ik_y) + O(k^2). \quad (4)$$

In the $R_h^X$ and $H_h^M$ stacking, $T_{vc'}$ has a finite **k**-independent leading term, meaning that it has an $s$-wave form as a function of **k**. For other stackings, the **k**-independent term vanishes, and the leading term is $T_{vc'} \propto k_x + ik_y$, of a $p$-wave character. The forms of $T_{vc'}$ can make a crucial difference upon the band inversion when an interlayer bias drives the bilayer from type-II to type-III (or inverted type-II) band alignment. Interlayer hybridization between the conduction and valence band can either be a trivial avoided band crossing when $T_{vc'}$ takes an $s$-wave form, or be a topological band inversion when $T_{vc'}$ has $p$-wave form [69]. It also makes a crucial difference upon the formation of interlayer excitonic insulators. An $s$-wave interlayer tunneling tends to fix the phase of the interlayer electron-hole coherence, thereby quenching the excitonic superfluidity. In contrast, tunneling of the $p$-wave form does not affect excitons condensed in the Coulomb favored $s$-wave channel, instead such chiral tunneling induces a background coherence in the $p$-wave channel, whose interference with the condensate in the $s$-wave channel can map the condensate phase to a measurable in-plane electrical polarization [70].

For the valence band local maximum **Γ** and conduction band local minimum **Q**, one gets

$$\langle \psi_{n,\boldsymbol{\Gamma},s_z} | \widehat{H} | \psi_{n',\boldsymbol{\Gamma},s_z} \rangle \approx T_{nn',\boldsymbol{\Gamma}},$$
$$\langle \psi_{n,\mathbf{Q},s_z} | \widehat{H} | \psi_{n',\mathbf{Q},s_z} \rangle \approx T_{nn',\mathbf{Q}}. \quad (5)$$

Under the leading harmonic approximation, the interlayer hopping strength at **Γ**/**Q** shows no explicit dependence on $\mathbf{r}_0$. However, $T_{nn'}$ decays exponentially with the increase of the interlayer distance, whose equilibrium value $D$ varies with $\mathbf{r}_0$ (see Fig. 3(a)). This gives rise to position-dependent modulations of $T_{nn',\boldsymbol{\Gamma}}$ and $T_{nn',\mathbf{Q}}$ within a moiré supercell. Due to the significant fractions of chalcogen atom $p_z$ orbital at **Γ** and **Q**, the magnitude of $T_{nn',\boldsymbol{\Gamma}}$ and $T_{nn',\mathbf{Q}}$ is found to reach several hundred meV [19], much larger than the interlayer hopping strength at **K** (several to several tens meV). Thus, **Γ**- and **Q**-valley in TMDs bilayers generally have strong layer hybridizations, and their energies are also shifted significantly compared to the monolayer case [19] resulting in direct to indirect band gap transition (for $MoTe_2$ and $WSe_2$ bilayers the valence band edges locate at the **K** valley). It has been proposed that these strongly layer-hybridized states can mediate the interlayer charge transfer process in heterobilayer TMDs: an upper-layer $\tau$**K**-valley electron (hole) can be transferred to the strongly layer-hybridized **Q**-valley (**Γ**-valley) by scattering with a phonon, and the subsequent scattering with another phonon can further bring the electron (hole) to the lower-layer $\tau$**K**′-valley [67]. In contrast, the direct interlayer hopping near $\tau$**K** is expected to be inefficient in heterobilayer TMDs as the hopping strength $T_{nn',\tau\mathbf{K}}$ is much weaker than the corresponding band offset between the two layers. The above **Γ**/**Q**-valley mediated mechanism has been confirmed by a series of experiments [71-73], which is found to dominate the ultra-fast interlayer charge transfer process [74-76].

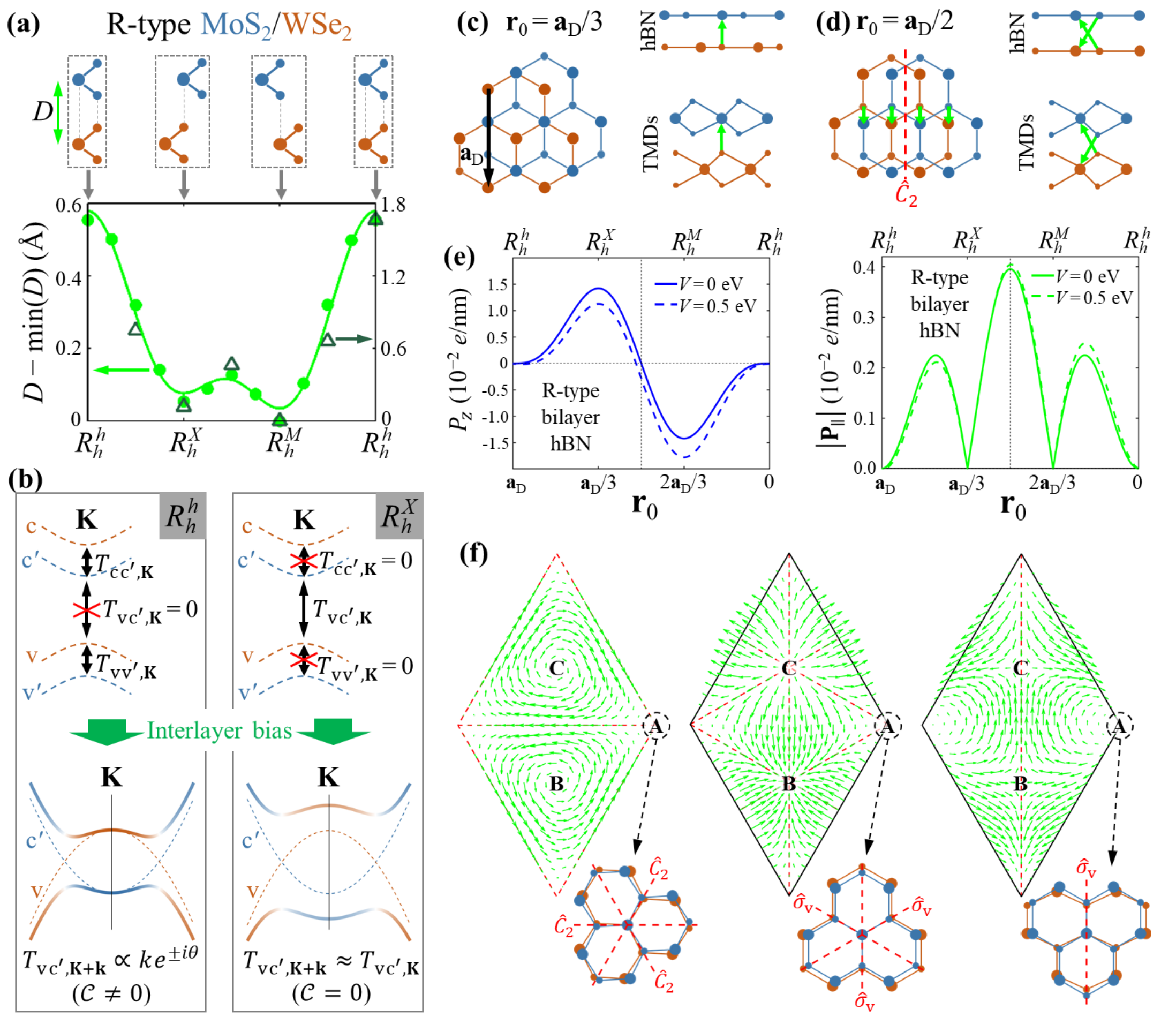

**Figure 3.** (a) Dependence of the equilibrium interlayer distance $D$ on the atomic registries in R-type $MoS_2/WSe_2$. Solid circles are first-principles calculations, and triangles are the STM measured local $D$ values in a $MoS_2/WSe_2$ moiré pattern. (b) Thin dashed (thick solid) curves show band dispersions near **K** without (with) the interlayer hopping. Double arrows correspond to the interlayer hopping at **K** under $R_h^h$ (left panel) and $R_h^X$ (right panel) stacking-registries. When a large interlayer bias $V$ is applied which leads to band inversion between c′ and v bands, the interlayer hopping $T_{\mathrm{vc'},\mathbf{K+k}} \propto ke^{i\theta}$ ($T_{\mathrm{vc'},\mathbf{K+k}} \approx T_{\mathrm{vc'},\mathbf{K}}$) under $R_h^h$ ($R_h^X$) gives rise to topologically nontrivial (trivial) layer-hybridized bands with Chern numbers $\mathcal{C} \neq 0$ ($\mathcal{C} = 0$). (c) An intuitive picture for the interfacial out-of-plane electric polarization in $R_h^X$ stacking-registry with $\mathbf{r}_0 = \mathbf{a}_D/3$. The green arrows indicate the local electric dipole pointing from a B atom in one hBN layer to its nearest N atom in the other (or from an $X$ atom in one TMDs layer to the nearest $M$ atom in the other), induced by the interlayer charge redistribution. (d) Emergence of the interfacial in-plane electric polarization in the IM stacking-registry with $\mathbf{r}_0 = \mathbf{a}_D/2$. The out-of-plane component is forbidden by the in-plane $\hat{C}_2$-symmetry (red dashed line). (e) The calculated out-of-plane electric polarization $P_z$ (left panel) and in-plane electric polarization $|\mathbf{P}_\parallel|$ (right panel) in bilayer hBN as functions of $\mathbf{r}_0$. $V = 0$ or 0.5 eV corresponds to the interlayer bias. (f) Illustrations of $\mathbf{P}_\parallel$ vector fields in bilayer hBN moiré patterns originating from an interlayer twist (left panel), a biaxial-heterostrain (middle panel) and an area-conserving uniaxial-heterostrain (right panel), which correspond to Fig. 2(a), (b) and (c), respectively. The atomic registries at **A** locations are enlarged, with red dashed lines indicating the in-plane $\hat{C}_2$ axes or vertical mirror planes ($\hat{\sigma}_v$) of the three moiré patterns. (a) From [56], adapted with permission from AAAS. (e, f) From [77], reprinted with permission from Science China Press.

## 1.4. Interfacial ferroelectricity from interlayer charge redistribution

The interlayer coupling effect can also lead to modest charge redistribution between the two layers, giving rise to an electric polarization **P** across the vdW interface at charge neutrality [78, 79]. From symmetry considerations, the $\hat{\sigma}_{\mathrm{h}}$ and $\hat{C}_3$ symmetries forbid the presence of a finite **P** in monolayer TMDs. However, both symmetries can be broken by the interlayer coupling under certain stacking registries, leading to finite electric polarization. For $\hat{C}_3$-symmetric stackings, **P** must point along the out-of-plane ($\pm z$) direction, see Fig. 3(c). Additional symmetries in homobilayers can enforce **P** to take specific forms: H-type homobilayers with any $\mathbf{r}_0$ values are inversion-symmetric thus $\mathbf{P} = 0$; the $\hat{\sigma}_{\mathrm{h}}$ and $\hat{C}_3$ symmetries in $R_h^h$ homobilayers lead to $\mathbf{P} = 0$; the $\hat{C}_3$-symmetric $R_h^X$ and $R_h^M$ homobilayers are related through the $\hat{\sigma}_{\mathrm{h}}$ operation, resulting in their opposite out-of-plane **P** values; the intermediate stacking registry between $R_h^X$ and $R_h^M$ (denoted as IM, see Fig. 3(d)) has the $\hat{C}_2$ symmetry about an axis within the 2D plane, allowing an in-plane **P**. For heterobilayers, an out-of-plane component in **P** is always allowed as both $\hat{\sigma}_h$ and inversion symmetries are absent [80]. These conclusions imply that such electric polarization can be controlled by the interlayer translation, which is thus termed as the sliding ferroelectricity which has caught remarkable interest both in R-stacking TMDs and hBN [81-87].

A simple and intuitive picture to understand the interfacial electric polarization is the electron redistribution between different kinds of atoms located in opposite layers [78, 79]. Electron can transfer from the upper-layer (lower-layer) $M$ atom to a nearby lower-layer (upper-layer) $X$ atom due to their different electron affinities, with the transferred amount decaying exponentially with their separation. Such an interlayer charge redistribution gives rise to local electric dipoles pointing from $X$ to $M$, whose summation determines the overall **P** in the bilayer. For the stacking registries $R_h^X$ ($R_h^M$), the upper-layer $M$ ($X$) and lower-layer $X$ ($M$) atoms vertically align thus have the smallest separation (Fig. 3(c)), the magnitudes of the charge redistribution and the resultant $P_z$ is therefore the largest. For the IM stacking registry, the out-of-plane components of local electric dipoles sum to zero and the overall **P** lies in-plane (Fig. 3(d)), consistent with the symmetry consideration.

The quantitative value of **P** can be obtained through a first-principles calculation [88-90], or from an analytical treatment to the interlayer hopping [77]. Below we use the latter to give a more tractable physical model, which indicates that **P** originates from the cross-bandgap interlayer hopping between Bloch states below and above the Fermi energy. In a commensurate homobilayer, the empty conduction bands $\psi_{\mathrm{c/c'},\mathbf{k}}$ and filled valence bands $\psi_{\mathrm{v/v'},\mathbf{k}}$ are separated by a large band gap $\Delta_{\mathbf{k}}$. $T_{\mathrm{cc'},\mathbf{k}}$ and $T_{\mathrm{vv'},\mathbf{k}}$ correspond to spin-conserved interlayer hopping terms between closely aligned bands, whereas $T_{\mathrm{cv'},\mathbf{k}}$ and $T_{\mathrm{vc'},\mathbf{k}}$ correspond to cross-bandgap terms. In TMDs and hBN, $|T_{nn',\mathbf{k}}| \ll \Delta_{\mathbf{k}}$, the effects of cross-bandgap terms $T_{\mathrm{cv'},\mathbf{k}}$ and $T_{\mathrm{vc'},\mathbf{k}}$ can then be well accounted for in the perturbation expansion of valence states $|\tilde{\psi}_{\mathrm{v},\mathbf{k}}\rangle \approx \left(1 - \frac{1}{2}\left|\frac{T_{\mathrm{vc'},\mathbf{k}}}{\Delta_{\mathbf{k}}}\right|^2\right)|\psi_{\mathrm{v},\mathbf{k}}\rangle - \frac{T^*_{\mathrm{vc'},\mathbf{k}}}{\Delta_{\mathbf{k}}}|\psi_{\mathrm{c'},\mathbf{k}}\rangle$ and $|\tilde{\psi}_{\mathrm{v'},\mathbf{k}}\rangle \approx \left(1 - \frac{1}{2}\left|\frac{T_{\mathrm{cv'},\mathbf{k}}}{\Delta_{\mathbf{k}}}\right|^2\right)|\psi_{\mathrm{v'},\mathbf{k}}\rangle - \frac{T_{\mathrm{cv'},\mathbf{k}}}{\Delta_{\mathbf{k}}}|\psi_{\mathrm{c},\mathbf{k}}\rangle$. Although $T_{\mathrm{vv'},\mathbf{k}}$ can strongly hybridize $\tilde{\psi}_{\mathrm{v},\mathbf{k}}$ and $\tilde{\psi}_{\mathrm{v'},\mathbf{k}}$, it doesn't change the overall interlayer charge redistribution up to the second order as both $\tilde{\psi}_{\mathrm{v},\mathbf{k}}$ and $\tilde{\psi}_{\mathrm{v'},\mathbf{k}}$ are fully occupied. The out-of-plane electric polarization is then given by

$$P_z = \frac{2D}{(2\pi)^2}\int d\mathbf{k}\left(\left|\frac{T_{\mathrm{cv'},\mathbf{k}}}{\Delta_{\mathbf{k}}}\right|^2 - \left|\frac{T_{\mathrm{vc'},\mathbf{k}}}{\Delta_{\mathbf{k}}}\right|^2\right). \tag{6}$$

The in-plane component $\mathbf{P}_{\parallel}$ corresponds to the summation of the Berry connections for the occupied valence states [77, 91]

$$\mathbf{P}_{\parallel} = \frac{-2}{(2\pi)^2}\int d\mathbf{k}\left(i\left\langle \tilde{u}_{\mathrm{v},\mathbf{k}}\middle|\frac{\partial \tilde{u}_{\mathrm{v},\mathbf{k}}}{\partial \mathbf{k}}\right\rangle + i\left\langle \tilde{u}_{\mathrm{v}',\mathbf{k}}\middle|\frac{\partial \tilde{u}_{\mathrm{v}',\mathbf{k}}}{\partial \mathbf{k}}\right\rangle\right), \tag{7}$$

with $\tilde{u}_{\mathrm{v}/\mathrm{v}',\mathbf{k}}$ the periodic part of $\tilde{\psi}_{\mathrm{v}/\mathrm{v}',\mathbf{k}}$. $P_z$ and $\mathbf{P}_{\parallel}$ as functions of $\mathbf{r}_0$ can then be calculated with $\psi_{n,\mathbf{k}}$ and $\psi_{n',\mathbf{k}}$ from the tight-binding models of monolayer TMDs or hBN and the interlayer tunneling $T_{nn',\mathbf{k}}$ from the two-center approximation. The analytical result indicates that the maximal $|\mathbf{P}_{\parallel}|$ is achieved under the IM stacking, which is comparable to max($P_z$), see Fig. 3(e).

In R-type homobilayer TMDs or hBN moiré, the local stacking registry dependent charge redistribution gives rise to spatially varying $\mathbf{P}$ with both out-of-plane and in-plane components [77, 88-90]. Note that the homobilayer moiré pattern can be introduced by a twist angle or a biaxial or area-conserving uniaxial heterostrain, all of which show the same spatial patterns of $P_z$ and $|\mathbf{P}_{\parallel}|$. However, the three directional patterns of $\mathbf{P}_{\parallel}$ are sharply different [77]. The winding number, defined as the number of counterclockwise rotations of $\mathbf{P}_{\parallel}$ when the path circles around the $R_h^X$ or $R_h^M$ region once, is +1 for the twist and biaxial heterostrain cases, and $-1$ for the area-conserving uniaxial heterostrain case (Fig. 3(f)). The topological charge density can be defined as $\rho_{\mathrm{Q}} \equiv \frac{1}{4\pi}\hat{\mathbf{P}}\cdot\left(\frac{\partial \hat{\mathbf{P}}}{\partial x}\times\frac{\partial \hat{\mathbf{P}}}{\partial y}\right)$, with $\hat{\mathbf{P}} \equiv \mathbf{P}/|\mathbf{P}|$ the unit vector of the electric polarization. In a half moiré supercell which corresponds to the equilateral triangle centered at $R_h^X$ (or $R_h^M$), the total topological charge is +1/2 (−1/2) for twist and biaxial-heterostrain cases, and −1/2 (+1/2) for the uniaxial-heterostrain case. The spatial pattern of the electric polarization in an equilateral triangle thus corresponds to a topological meron or anti-meron structure.

Experimental observations of the out-of-plane electric polarization $P_z$ have been reported both in commensurate and marginally twisted homobilayer hBN and TMDs systems [81-87]. Furthermore, controllability on the bilayer stacking registry has been demonstrated by utilizing the coupling between $P_z$ and an out-of-plane electric field. For R-type commensurate homobilayers whose lowest-energy registry corresponds to $R_h^X$ or $R_h^M$ with opposite $P_z$ values, applying an out-of-plane electric field raises the energy of one stacking but lower the other. A large enough electric field can then introduce an interlayer sliding and switch the stacking registry from one to the other, thereby flipping the sign of $P_z$ [83, 86]. As we will discuss next in Sec. 1.5, in marginally twisted homobilayers with long-period moiré patterns, triangular $R_h^X$ and $R_h^M$ domains appear alternatively separated by narrow domain walls. Due to the opposite $P_z$ values in the two domains, applying an out-of-plane electric field leads to the expansion of one and contraction of the other through the movement of domain walls [81, 83-85]. Very recently, experiments have also reported the observation of the in-plane electric polarization $\mathbf{P}_{\parallel}$ and the corresponding topological structure in twisted homobilayer hBN and TMDs [92-97].

### 1.5. Spontaneous lattice reconstruction in the moiré and its effects on electronic properties

The discussions so far have been primarily focused on rigid moiré patterns. However, the vdW binding energy varies for different stacking configurations (Table 2), thus a long-wavelength moiré tends to spontaneously adjust the local area and interlayer spacing to minimize the total energy. For instance, in R-type homobilayer moiré, the energetically favorable $R_h^X$ and $R_h^M$ regions tends to expand, while the unfavorable $R_h^h$ regions tends to shrink (Fig. 4(a) upper panel). The energy gain from the binding energy competes with the cost from the strain elastic energy due to atomic displacements, and the lattice structure reaches equilibrium when they are balanced. Such spontaneous lattice reconstruction is pronounced in long-wavelength moiré patterns, i.e., in small-angle twisted homobilayers, and heterobilayers with identical chalcogen atoms $MX_2/M'X_2$. The relaxed structural landscapes are qualitatively similar in homobilayers and $MX_2/M'X_2$ (Figs. 4(b, c)). In R-type twisted bilayers, large triangular domains of $R_h^X$ and $R_h^M$ stacking with small interlayer distance emerge, separated by narrow domain walls intersecting at the small nodes of $R_h^h$ stacking with large interlayer distance; while in twisted H-type bilayers, large hexagonal domains of $H_h^h$ stacking appear. Early theoretical studies predicted that lattice relaxation occurs for $\theta$ below $\sim 2.5^\circ$ ($\sim 1^\circ$) in R-type (H-type) moiré [98], while it was suggested that even for $\theta \approx 3-4^\circ$ lattice reconstruction can be important for the experimental observation of the FQAH effect in R-type twisted $MoTe_2$ [99-102].

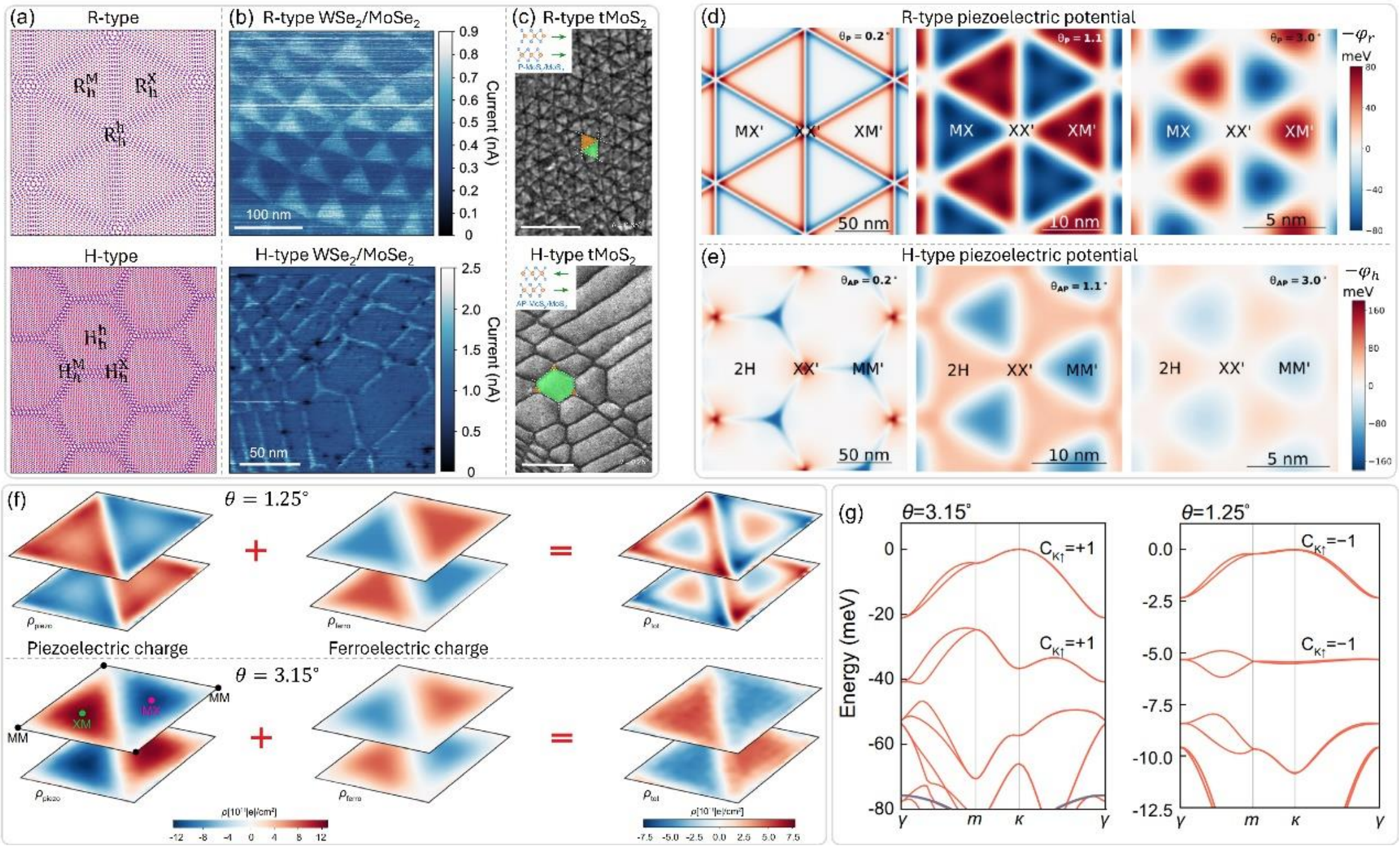

**Figure 4.** (a) Schematics of relaxed R-type and H-type twisted TMDs bilayers. (b) Conductive AFM image of R-type and H-type twisted bilayer $WSe_2/MoSe_2$. (c) Low-magnification annular dark-field scanning transmission electron microscopy image of R-type and H-type twisted bilayer $MoS_2$. (d) Piezoelectric potential $-\varphi_r$ in R-type $MoS_2$ at different twist angles. Here MX′, XX′ and XM′ correspond to $R_h^M$, $R_h^h$, and $R_h^X$, respectively. (e) Piezoelectric potential $-\varphi_h$ in H-type $MoS_2$ at different twist angles. Here 2H, XX′, and MM′ correspond to $H_h^h$, $H_h^M$, and $H_h^X$, respectively. (f) The piezoelectric charge density, the ferroelectric charge density, and the total charge density in R-type twisted bilayer $WSe_2$. Here MM, XM and MX correspond to $R_h^h$, $R_h^M$, and $R_h^X$, respectively. (g) Energy bands of R-type twisted bilayer $WSe_2$ corresponding to (f). (a) adapted with permission from Ref. [103]. (b) adapted with permission from Ref. [104]. Copyright 2020 American Chemical Society. (c) adapted with permission from Ref. [105]. (d, e) adapted with permission from Ref. [106]. (f, g) adapted from Ref. [107].

Several approaches have been developed to quantitatively characterize the relaxed moiré pattern. Of paramount importance is the accurate description of the intralayer and interlayer energy terms, the minimization of which determines the equilibrium interlayer distance and atomic positions. The classical force field method [108, 109] employs simple inter-atomic potentials to model the intralayer and interlayer interactions. The continuum approach [98, 103] simulates the interlayer binding energy and atomic displacements (thus the intralayer elastic energy) by using harmonic expansions with the moiré period. The model parameters in both approaches can be obtained by fitting to DFT calculations of the relevant energies in commensurate bilayer configurations. The relaxed lattice geometry from the continuum method was shown to well reproduce the experimentally observed structure [105]. Some recent studies employ machine learning-assisted classical force fields to capture the moiré lattice reconstruction [107, 110, 111], which can be incorporated into DFT calculations for obtaining the moiré energy bands at twist angles down to $\sim 1.1°$.

As the interlayer hopping $T(\mathbf{r}_0)$ and intralayer potential $V_l(\mathbf{r}_0)$ in a moiré depend on the local stacking registry $\mathbf{r}_0(\mathbf{r})$ between the two layers, one immediate consequence of the lattice reconstruction is that $\mathbf{r}_0(\mathbf{r})$ should be replaced by $\mathbf{r}_0(\mathbf{r}) + \mathbf{u}^t(\mathbf{r}) - \mathbf{u}^b(\mathbf{r})$ in the moiré potential, where $\mathbf{u}^l(\mathbf{r})$ denotes the in-plane atomic displacement of layer $l$ upon the reconstruction. Such a replacement captures the redistribution of local stacking areas, while the potential's strength in each specific local stacking configuration is treated as unaffected. This results in rather uniform potential in the large domains bounded by narrow domain walls with sharp potential variations when lattice reconstruction is significant. This change in the potential's landscape ($\mathbf{r}_0 \to \mathbf{r}_0 + \mathbf{u}^t - \mathbf{u}^b$) is not explicitly manifested in the continuum models cited in Table 4, although the DFT calculations used to fit the model parameters have taken lattice reconstruction into account. The spatial variation of the atomic displacements results in a local strain tensor $\epsilon^l$ with the components $\epsilon^l_{ij} \equiv \left(\partial_i u^l_j + \partial_j u^l_i\right)/2$, and a local twist angle $\delta\theta^l \equiv \partial_x u^l_y - \partial_y u^l_x$. Both $\epsilon^l$ and $\delta\theta^l$ vary periodically in the moiré, with dependence on layer index $l$, whereas $\epsilon^l$ can also substantially change the electronic structures of the band edge carriers in the moiré.

The strain tensor $\epsilon^l$ generates piezoelectric charges with layer dependence. The piezoelectric charge in layer $l$ reads $\rho_l = e_l\left[2\,\partial_x\epsilon^l_{xy} + \partial_y\left(\epsilon^l_{xx} - \epsilon^l_{yy}\right)\right]$, where $e_l$ is the piezoelectric coefficient. Strains in the two layers are opposite $\epsilon^t = -\epsilon^b$, while the sign of $e_l$ depends on the stacking configuration: $e_t = e_b$ ($e_t = -e_b$) for R-type (H-type) bilayers. These features indicate that the piezoelectric potential is layer anti-symmetric (symmetric) in R-type (H-type) bilayers, which modifies the intralayer potentials $V_t$ ($V_b$) by $\varphi_r$ and $\varphi_h$ ($-\varphi_r$ and $\varphi_h$) in R-type and H-type bilayers, respectively. Figures 4(d, e) show the distribution of $\varphi_r$ and $\varphi_h$ in twisted bilayer $MoS_2$ at different twist angles, which becomes more localized towards the domain boundaries as $\theta$ is reduced. When $\theta$ is large, $\varphi_r \approx \varphi_0 \sum_{i=1}^{3} \sin(\mathbf{q}_i \cdot \mathbf{r})$ and $\varphi_h \approx \varphi_0 \sum_{i=1}^{3} \sin(\mathbf{q}_i \cdot \mathbf{r} + \phi)$ with $\varphi_0$ and $\phi$ parameters characterizing the magnitude and profile, $\boldsymbol{q}_i$ are the primitive moiré reciprocal lattice vectors related by $\hat{C}_3$, while higher harmonics are necessary for small $\theta$ [112]. Figure 4(f) shows the sum of ferroelectric charge density from interlayer charge redistributions (Sec. 1.4) and piezoelectric charge density from lattice relaxation in R-type twisted bilayer $WSe_2$ for two different $\theta$. Both charge contributions to the charge density will contribute electrostatically to the moiré potential experienced by the doped carriers. As the ferroelectric and piezoelectric charge densities compete with opposite signs, the calculations find the charge distributions around the $R^X_h$ and $R^M_h$ regions to be opposite in the small vs large $\theta$ chosen. The resultant overall intralayer potentials also show significant distinctions for different $\theta$ [112].

The evolution of the piezoelectric effect with $\theta$ has an important impact on the moiré potential. This further suggests that moiré superlattices with different $\theta$ could exhibit distinct electronic and topological properties, and the continuum model and its parameters (see Sec. 1.6) should be adjusted accordingly when $\theta$ is varied [107, 113]. Figure 4(g) exemplifies this by showing distinct energy bands with opposite Chern numbers for $\theta = 1.25°$ vs $3.15°$ from DFT, where the latter cannot be explained with the continuum model by varying $\theta$ only. Recent studies have also shown that, with the lowest harmonics in the continuum model, only the first two moiré bands from DFT can be well fitted [111, 114], while including second-lowest harmonics can help to capture the third moiré band from DFT [111]. This improvement might be attributed to a better account of the lattice reconstruction effects by the higher harmonics.

In addition to the above effects, strain can also modify the electronic properties of each layer through a scalar potential from the modulation of the band edge position, and a pseudo-vector potential $\mathbf{A}_l \propto \pm(\epsilon_{xx}^l - \epsilon_{yy}^l, -2\epsilon_{xy}^l)$ in the $\pm\mathbf{K}$ valley [115-117]. In H-type twisted heterobilayer $MoSe_2/WSe_2$, the experimentally measured deep moiré potentials (0.3 eV for the valence band and 0.15 eV for the conduction band) are attributed to the inhomogeneous strain from lattice reconstruction [118]. In R-type $WSe_2/WS_2$, the ~ 4% lattice constant mismatch between the two layers yields a moiré pattern with wavelength ~ 8 nm, the resultant inhomogeneous strain introduces a band gap variation as large as ~ 90 meV to the $WSe_2$ layer, with the minimum gap located at the $R_h^h$ region [119]. An extremely narrow moiré flat band (~ 10 meV) from the **K**-point valence band in $WSe_2$ revealed by STS measurement is also attributed to the effects of strain from lattice reconstruction [120]. Meanwhile, $\mathbf{A}_l$ modifies the continuum model through the kinetic term $h_l = -\hbar^2(\hat{\mathbf{k}} - \boldsymbol{\kappa}_l + e\mathbf{A}_l)^2/(2m_l)$, where $\boldsymbol{\kappa}_l = \mathbf{K}$ or $\mathbf{K}'$ denotes the center of the valley in layer $l$ (Fig. 2(a)) and $m_l$ is the effective mass. The strain-induced pseudo-magnetic field $\mathbf{B}_l = \nabla \times \mathbf{A}_l \propto \rho_l \hat{\mathbf{z}}$ can play important roles in determining the topological properties of the system [121-123]. However, $\mathbf{A}_l$ was not explicitly included in the continuum models of most recent works on reconstructed R-type twisted homobilayer $MoTe_2$ and $WSe_2$. By including both $\mathbf{A}_l$ and second harmonics in the continuum model of R-type twisted bilayer $MoTe_2$, it was shown in Ref. [110] that the first four energy bands from DFT calculations can be reasonably well fitted, and the first two or three bands from DFT can be reproduced well by continuum model over a range of twist angles without changing the model parameters.

## 1.6. Continuum models for low-energy **K** valley electrons in a moiré

The stacking registry-modulated interlayer couplings lead to spatially varying potentials and layer-hybridizations for carriers in a moiré superlattice. In this section we discuss the continuum models for band edge electrons in the moiré, with a particular focus on the $\pm\mathbf{K}$ valleys. We will consider the valence band electrons, while the analysis can be carried out similarly for the conduction bands. While the moiré potential and layer-hybridization depend on the TMDs species and their stacking configuration, i.e., R-type or H-type alignment, some general features can be obtained without referring to details of the model.

A minimal model for describing the electronic properties of an R-type homobilayer moiré of TMDs around the valence band edge of $\pm\mathbf{K}$ valley can be constructed in the layer space as

$$H_{\text{R-homo}}(\mathbf{r}) = \begin{pmatrix} h_t(\hat{\mathbf{k}}) + V_t(\mathbf{r}_0) & T(\mathbf{r}_0) \\ T^*(\mathbf{r}_0) & h_b(\hat{\mathbf{k}}) + V_b(\mathbf{r}_0) \end{pmatrix}, \tag{8}$$

where $h_l$ denotes the pristine monolayer Hamiltonian, $\hat{\mathbf{k}}$ is the wave vector, $V_l$ represents the intralayer moiré potential which could contain contributions from the strain-induced band edge modulation and electrostatic contributions from the piezoelectric/ferroelectric charges, and $T$ characterizes the interlayer hopping. For valence band electrons, $h_l(\hat{\mathbf{k}}) \approx -\hbar^2(\hat{\mathbf{k}} - \boldsymbol{\kappa}_l)^2/(2m_l)$. The two layers in a moiré superlattice exhibit spatially varying local stacking registry (Fig. 2(d)), which, in a rigid moiré, can be described by the spatially modulated interlayer translation $\mathbf{r}_0(\mathbf{r}) = (1 - \hat{R}^{-1}\hat{S}^{-1})\mathbf{r}$. The spin-conserved interlayer hopping at **K** valley is therefore a location dependent one, which for spin up can be approximated as (cf. Eq. (2))

$$T(\mathbf{r}) = t_0\left[1 + e^{i\mathbf{b}_1\cdot\mathbf{r}_0} + e^{i(\mathbf{b}_1+\mathbf{b}_2)\cdot\mathbf{r}_0}\right] = t_0\left[1 + e^{i\mathbf{q}_1\cdot\mathbf{r}} + e^{i(\mathbf{q}_1+\mathbf{q}_2)\cdot\mathbf{r}}\right]. \tag{9}$$

$\mathbf{b}_1$ and $\mathbf{b}_2$ are the monolayer reciprocal lattice vectors connecting the three **K** points (Fig. 1(c)), and the value of $t_0$ is to be determined. Here only the lowest harmonics are retained, while higher-order terms can be easily added back. The relation $\mathbf{b}_i \cdot \mathbf{r}_0 = \mathbf{q}_i \cdot \mathbf{r}$ has been employed in the last equality, where $\mathbf{q}_i$ is the moiré reciprocal lattice vector. In a twisted homobilayer, one can write $\mathbf{q}_1 = \left(\frac{1}{2}, \frac{\sqrt{3}}{2}\right)\frac{4\pi}{\sqrt{3}\lambda}$ and $\mathbf{q}_2 = (-1, 0)\frac{4\pi}{\sqrt{3}\lambda}$ (Fig. 2(a)). Note that the commonly employed Eq. (9) in the existing literature (see e.g., Table 4) does not include the momentum-dependence in the vicinity of $\mathbf{K}$ points (cf. Eq. (4)), which might be important for accurate characterization of the electronic and topological properties of the TMDs moiré systems [69, 124-126].

The continuum model in Eq. (8) is most relevant for twisted homobilayer $MoTe_2$ and $WSe_2$ [127-129] in which the valence band edge is located at the $\pm\mathbf{K}$ points, possibly also applicable to heterobilayer $MoTe_2/WSe_2$ whose valence band offset can be experimentally tuned to be small by an interlayer bias [130]. While for other R-type twisted homobilayers, e.g., $MoS_2$, $WS_2$ and $MoSe_2$, the valence band edge is located at the $\mathbf{\Gamma}$ point [131-133] for which continuum models are also proposed [131, 133].

In H-type homobilayer moiré and most other heterobilayer moiré, the models are generally simpler, because effects of the interlayer hopping $T$ can be neglected, due to the large band offset (in heterobilayers) and/or opposite spin alignments (in H-type structures). Meanwhile, the large band offset in heterobilayers dictates that the low-energy carriers are predominantly in one layer. For example, in $WSe_2/MoSe_2$ heterobilayer, the low-energy holes are from the $WSe_2$ layer [134]. In H-type homobilayers, where band-edge holes in the two layers have opposite spin alignments, the spin-conserved interlayer hopping is negligible compared to the large SOC splitting (Fig. 1(d)) and spin-flip interlayer hopping is suppressed by spin conservation. Therefore, to a good approximation

$$H_{\text{hetero/H-homo}}(\mathbf{r}) = -\hbar^2\hat{\mathbf{k}}^2/(2m_l) + V_l(\mathbf{r}_0), \tag{10}$$

where $l$ denotes the contributing layer in a heterobilayer, or $t$ and $b$ in an H-type homobilayer.

The intralayer potential $V_l$ in layer $l$ also follows the moiré periodicity thus can be expanded into the Fourier series. By keeping the lowest harmonics, it can be modeled as

$$V_l(\mathbf{r}) = v_l \sum_{i=1}^{3} \cos(\mathbf{b}_i \cdot \mathbf{r}_0 + \alpha_l) = v_l \sum_{i=1}^{3} \cos(\mathbf{q}_i \cdot \mathbf{r} + \alpha_l). \tag{11}$$

Here $\mathbf{b}_3 = -\mathbf{b}_1 - \mathbf{b}_2$ and $\mathbf{q}_3 = -\mathbf{q}_1 - \mathbf{q}_2$, and the values of $v_l$ and $\alpha_l$ are to be determined. In R-type twisted homobilayers,

$$v_t = v_b \text{ and } \alpha_b = -\alpha_t \quad \text{(R-type homobilayer)}. \tag{12}$$

This is understood as the two layers are interchanged by the out-of-plane mirror operation, and the new structure is related to the original via replacing $\mathbf{r}_0$ by $-\mathbf{r}_0$, thus $V_t(\mathbf{r}_0) = V_b(-\mathbf{r}_0)$. In contrast, in H-type twisted homobilayers, the intralayer potential is layer symmetric with

$$v_t = v_b \text{ and } \alpha_b = \alpha_t \quad \text{(H-type homobilayer)}. \tag{13}$$

For heterobilayer moiré of either H- or R-type, the intralayer potential acting on the contributing layer can also be modeled similarly as that of the H-type homobilayer.

The spatially periodic modulations in the potential terms in the above models turn the massive Dirac cones in $\pm\mathbf{K}$ valleys of monolayer TMDs into moiré minibands of various properties for different TMDs moiré patterns (see e.g., Sec. 4.1). They also endow the moiré excitons with intriguing optical and correlated properties as will be discussed in Secs. 2 and 3.

To obtain the model parameters, $t_0$, $v_l$ and $\alpha_l$, early studies [42, 67, 127, 135] proposed a computationally inexpensive approach by fitting to the DFT band structures of commensurate bilayers of various stacking configurations characterized by constant $\mathbf{r}_0$. Some parameters for R-type twisted homobilayers (hereafter denoted as t$MX_2$) obtained from this local approach are as follows: $t_0 = -8.5$ meV, $v_t = 16$ meV and $\alpha_t = -89.6°$ for $tMoTe_2$ [127]; $t_0 = 9.7$ meV, $v_t = 17.8$ meV and $\alpha_t = 91°$ for $tWSe_2$ [127]; $t_0 = 7.1$ meV, $v_t = 8.586$ meV and $\alpha_t = -88.5°$ for $tMoSe_2$ [42, 43]. The obtained values of $|\alpha_t| = |\alpha_b| \approx 90°$ lead to $V_t(\mathbf{r}) \approx -V_b(\mathbf{r})$ in R-type homobilayers. Such a layer anti-symmetric intralayer potential form is mainly induced by the electrostatic potential from the interlayer charge redistributions [78, 136] that underlies the sliding or interfacial ferroelectricity as discussed in Sec. 1.4 [81-87]. Note that the parameters from this approach depend on the choice of vdW exchange-correlation functional and the reference value for energies in DFT calculations [137] and the effects of spontaneous lattice reconstruction of the moiré including the piezoelectric charge (Sec. 1.5) cannot be captured.

Another approach directly computes the moiré superlattice at a specific twist angle (e.g., $\theta = 3.89°$) from DFT calculations, and fits the moiré energy bands to the above models [68, 110, 111, 114, 129, 138-141]. This approach receives more attention after the experimental observation of integer and fractional quantum anomalous Hall (IQAH and FQAH) effects in $tMoTe_2$ [99-102], which reported larger preferred twist angles for observing the FQAH effect than the initial theoretical predictions [142, 143]. The effects of moiré lattice reconstruction (Sec. 1.5), which were suggested to underlie the discrepancy, can be included in the DFT calculations in this approach. The relaxed lattice geometry and the extracted continuum model parameters, however, also vary with the vdW exchange-correlation functional employed in this approach. Table 4 summarizes some recently obtained model parameters of $tMoTe_2$ from this approach. The parameters are usually obtained by fitting to DFT results at one particular commensurate twist angle (e.g., $\theta = 3.89°$), thus the parameters are quantitatively applicable only in a narrow range of $\theta$ [107, 113, 114].

**Table 4.** Continuum model parameters of R-type $tMoTe_2$ by fitting to moiré bands from DFT. Quantities with primes denote second harmonic contributions in the models. $C_{1,2}^{\uparrow}$ denote the Chern numbers of the first two moiré bands of spin up (cf. Figs. 9(a, b)).

| $\theta$ (∘) | $t_0$ (meV) | $v_t$ (meV) | $\alpha_t$ (∘) | $t_0'$ (meV) | $v_t'$ (meV) | $\alpha_t'$ (∘) | $C_{1,2}^{\uparrow}$ | Ref. |
|---|---|---|---|---|---|---|---|---|
| 4.41 | 13.3 | 22.4 | -91 | 0 | 0 | 0 | (-1, -1) | [138] |
| 3.89 | 11.3 | 15 | -100 | 0 | 0 | 0 | (-1, -1) | [138] |
| 3.89 | -11.2 | 18.4 | -99 | 0 | 0 | 0 | (-1, -1) | [139] |
| 3.89 | -18.8 | 33 | 105.9 | 0 | 0 | 0 | (1, -1) | [111] |
| 3.89 | -16 | 34 | 107.7 | 0 | 0 | 0 | (1, -1) | [144] |
| 3.89 | -23.8 | 41.6 | 107.7 | 0 | 0 | 0 | (1, -1) | [114] |
| 3.89 | -10.77 | 15.88 | 88.43 | 10.21 | 40 | 0 | (1, -1) | [111] |
| 3.89 | -12.2 | 18.9 | 85.23 | 13.12 | 49.98 | 0 | (1, -1) | [107, 145] |
| 3.15 | -7.8 | 20.6 | 75 | 6.9 | 5.8 | 0 | (1, 1) | [110] |

For the low-energy electrons in twisted heterobilayers or H-type homobilayers in the absence of interlayer hopping, the Hamiltonian has the simple form of $H_{\text{hetero/H-homo}} = -\hbar^2\hat{\mathbf{k}}^2/(2m) + v\sum_{i=1}^{3}\cos(\mathbf{b}_i \cdot \mathbf{r}_0 + \alpha)$. The parameters $v$ and $\alpha$ can also be obtained by using the above two approaches. For example, for the contributing $WSe_2$ layer in R-type twisted $WSe_2/MoSe_2$, $v = 13.2$ meV and $\alpha = 94°$ are obtained from the approach based on local commensurate configurations [135]. Several R-type twisted heterobilayers with different chalcogen atoms in the two layers (e.g., $WSe_2/WS_2$) were studied in Ref. [140] by fitting to the DFT bands of commensurate 5.68° twisted bilayer moiré, yielding $v \in [14, 30]$ meV and $\alpha \in [-45°, -35°]$. In Ref. [111], $v = 103$ meV and $\alpha = 56°$ were obtained for H-type $tMoTe_2$ by fitting to a commensurate 3.89° moiré. The potential landscape in the moiré of heterobilayers typically features a triangular array of potential maxima for trapping low-energy holes, making it possible to simulate the Hubbard model physics. While the extra layer pseudospin degree of freedom in H-type twisted homobilayers further allows simulating the bilayer SU(4) Hubbard models that might lead to various spin liquid phases [146-149]. Some of the related developments have been reviewed by Ref. [9, 50, 150], also see Sec. 3.1.

We remark that the continuum models employing the quadratic approximation $h_l \approx -\hbar^2\hat{\mathbf{k}}^2/(2m_l)$ and the lowest harmonics contributions in the moiré potentials have accidental symmetry that do not exist in the lattice geometry. For instance, the combination of time-reversal symmetry and an in-plane mirror symmetry or the effective inversion symmetry exist in the two-band model of twisted homobilayers (Eq. (8)) [151, 152]. Such accidental symmetry could affect the study of certain phenomena, especially those that require the chiral structural symmetry [153-156]. To recover the chiral symmetry, one can instead employ the massive Dirac model for each layer (Eq. (1)) that captures both conduction and valence band edges. The interlayer hopping $T$ and intralayer potentials $V_l$ should also be generalized to include the effects involving the conduction bands (cf. Eq. (3)) [42, 43, 127].

## 2. Moiré excitons in TMDs

### 2.1. Intralayer and interlayer excitons in monolayer and commensurate bilayer TMDs

Excitons in TMDs feature interesting optical properties tied to the spin-valley configurations. We start with monolayer, where interband transition by optical field can be quantified by the dipole transition matrix element $\langle\psi_{\mathrm{c},\mathbf{k},S_z'}|\hat{\mathbf{D}}|\psi_{\mathrm{v},\mathbf{k},S_z}\rangle$, with $\langle\psi_{\mathrm{c},\mathbf{k},S_z'}|\hat{D}_\pm|\psi_{\mathrm{v},\mathbf{k},S_z}\rangle$ and $\langle\psi_{\mathrm{c},\mathbf{k},S_z'}|\hat{D}_z|\psi_{\mathrm{v},\mathbf{k},S_z}\rangle$ describing the coupling strength to photons with $\sigma\pm$ circular and out-of-plane linear polarizations, respectively. Here $\hat{\mathbf{D}} \equiv (\hat{D}_x, \hat{D}_y, \hat{D}_z)$ is the dipole operator and $\hat{D}_\pm \equiv \hat{D}_x \pm i\hat{D}_y$. $|\psi_{n,\mathbf{k},S_z}\rangle$ denotes the conduction ($n$ = c) or valence ($n$ = v) band Bloch state with a wave vector $\mathbf{k}$ and spin $S_z$. The $\hat{C}_3$ and $\hat{\sigma}_\mathrm{h}$ symmetry requirements on the states at the high-symmetry point $\tau\mathbf{K}$ dictate that $\langle\psi_{\mathrm{c},\tau\mathbf{K},S_z'}|\hat{D}_\pm|\psi_{\mathrm{v},\tau\mathbf{K},S_z}\rangle \neq 0$ when $S_z = S_z'$ and $\tau C_{3,\mathrm{c}} - \tau C_{3,\mathrm{v}} = \pm 1 \pmod 3$, whereas $\langle\psi_{\mathrm{c},\tau\mathbf{K},S_z'}|\hat{D}_z|\psi_{\mathrm{v},\tau\mathbf{K},S_z}\rangle \neq 0$ when $S_z \neq S_z'$ and $\tau C_{3,\mathrm{c}} - \tau C_{3,\mathrm{v}} + S_z' - S_z = 0 \pmod 3$. This leads to the optical selection rule in monolayer TMDs: the spin-conserved interband transition at $+\mathbf{K}$ ($-\mathbf{K}$) is coupled to a photon with the $\sigma+$ ($\sigma-$) circular polarization [23], whereas the spin-flip interband transitions involving valence band edges at $+\mathbf{K}$ and $-\mathbf{K}$ are both coupled to photons with the out-of-plane linear polarization [39], see Fig. 5(a).

The optically induced interband transition can create an electron-hole pair. The Coulomb attraction between the electron and hole can then bind them into a hydrogen-like bound state known as the exciton [157, 158], whose $n$-th eigenstate can be written as

$$X^{(n)}_{\tau'\tau,S_z'S_z,\mathbf{Q}} = \sum_{\Delta\mathbf{Q}} \Phi_n(\Delta\mathbf{Q})\psi_{\mathrm{c},\tau'\mathbf{K}+\frac{m_\mathrm{e}}{M}\mathbf{Q}+\Delta\mathbf{Q},S_z'}(\mathbf{r}_\mathrm{e})\psi^*_{\mathrm{v},\tau\mathbf{K}-\frac{m_\mathrm{h}}{M}\mathbf{Q}+\Delta\mathbf{Q},S_z}(\mathbf{r}_\mathrm{h}). \tag{14}$$

Here $\mathbf{r}_{\mathrm{e/h}}$ and $m_{\mathrm{e/h}}$ are the spatial coordinate and effective mass of the electron/hole, respectively, $\mathbf{Q}$ is the center-of-mass (CoM) momentum, and $M \equiv m_\mathrm{e} + m_\mathrm{h}$ is the exciton mass. Similar to the 2D hydrogen model, the electron-hole relative motion forms a series of discrete Rydberg orbitals, which are usually denoted as $n$ = 1$s$, 2$s$, 2$p_{\pm}$, 3$d_{\pm}$, etc. [159-161]. The difference between the quasiparticle band gap and the 1$s$ ground state energy corresponds to the exciton binding energy $E_b$, which characterizes the energy needed to dissociate the exciton into a free electron-hole pair. In monolayer TMDs, the atomically thin geometry of the layered structure greatly reduces the screening efficiency of the material. Combined with the large electron and hole effective masses (~ 0.5$m_0$, with $m_0$ the free electron mass), the exciton binding energy is found to reach several hundred meV [160-165], one order of magnitude larger than those in conventional semiconductor quantum wells. The spatial extension of the electron-hole relative motion, however, is still the Wannier type, with a Bohr radius $a_B$ ~ 1-2 nm much larger than the monolayer lattice constant (~ 0.3 nm) [72, 166-171]. In momentum space, exciton's electron and hole constituents are both well localized in the ±**K** valleys. An exciton with its electron and hole constituents in the same valley (different valleys) is termed as the intra-valley (inter-valley) exciton, whereas that with anti-parallel (parallel) electron and hole spins are termed as the spin-singlet (spin-triplet) exciton (Fig. 5(b)) [157]. Here a spin-up/down hole is obtained by removing a spin-down/up electron in the fully occupied valence band. Optically active bright excitons are in spin-singlet, and the momentum conservation requires them to be in the intra-valley configuration with $\mathbf{Q} \approx 0$. The optical dipole of the exciton is given by

$$\left\langle 0\middle|\hat{\mathbf{D}}\middle|X^{(n)}_{\tau'\tau,S'_zS_z,\mathbf{Q}}\right\rangle = \delta_{\mathbf{Q},0}\delta_{\tau',\tau}\sum_{\Delta\mathbf{Q}}\Phi_n(\Delta\mathbf{Q})\langle\psi_{c,\tau\mathbf{K}+\Delta\mathbf{Q},S'_z}|\hat{\mathbf{D}}|\psi_{v,\tau\mathbf{K}+\Delta\mathbf{Q},S_z}\rangle$$
$$\approx \delta_{\mathbf{Q},0}\delta_{\tau',\tau}\Phi_n(\mathbf{r}=0)\langle\psi_{c,\tau\mathbf{K},S'_z}|\hat{\mathbf{D}}|\psi_{v,\tau\mathbf{K},S_z}\rangle. \quad (15)$$

Here $\Phi_n(\mathbf{r}) \equiv \sum_{\Delta\mathbf{Q}}\Phi_n(\Delta\mathbf{Q})e^{i\Delta\mathbf{Q}\cdot\mathbf{r}}$ is the real-space envelope function of the exciton with $\Phi_n(\mathbf{r}=0) \propto a_B^{-1}$, and an envelope approximation $\langle\psi_{c,\tau\mathbf{K}+\Delta\mathbf{Q},S'_z}|\hat{\mathbf{D}}|\psi_{v,\tau\mathbf{K}+\Delta\mathbf{Q},S_z}\rangle \approx \langle\psi_{c,\tau\mathbf{K},S'_z}|\hat{\mathbf{D}}|\psi_{v,\tau\mathbf{K},S_z}\rangle$ has been used considering that $\Phi_n(\Delta\mathbf{Q})$ is narrowly distributed around $\Delta\mathbf{Q} = 0$ [160, 172]. Note that $|\langle\psi_{c,\tau\mathbf{K},S'_z}|\hat{\mathbf{D}}|\psi_{v,\tau\mathbf{K},S_z}\rangle|$ of the spin-singlet exciton with $S'_z = S_z$ is much larger than that of the spin-triplet exciton with $S'_z = -S_z$, as the latter requires a spin-flip from SOC. The above optical dipole form is finite only for $s$-type Rydberg orbitals with $\Phi_n(\mathbf{r}=0) \neq 0$. The small value of $a_B$ ~ 1-2 nm for the ground state (1$s$) spin-singlet exciton results in its large oscillator strength, which then dominates the optical properties of the monolayer TMDs. Experiments have measured a resonant excitonic absorption for a monolayer TMDs reaching 10-20% [20], and a ~ 30 meV vacuum Rabi splitting of the spin-singlet exciton in a planar cavity [173-185]. On the other hand, if the dependence of $\psi_{n,\tau\mathbf{K}+\Delta\mathbf{Q},S_z}$ on $\Delta\mathbf{Q}$ is incorporated through the $\mathbf{k}\cdot\mathbf{p}$ model, a small but finite optical dipole strength can be obtained for the $p$-type Rydberg orbital [186], which is consistent with *ab initio* GW-BSE calculations [160] and symmetry analysis [187].

A bright exciton in the intra-valley configuration exhibits an excitonic valley pseudospin, which can be described by the Pauli matrices $\hat{\boldsymbol{\tau}} = (\hat{\tau}_x, \hat{\tau}_y, \hat{\tau}_z)$, with $\langle\hat{\tau}_z\rangle = +1$ $(-1)$ corresponding to an exciton in **K** (−**K**) valley. Meanwhile, an in-plane excitonic valley pseudospin corresponds to a coherent superposition of ±**K** valleys. For $s$-orbital bright excitons, its $\hat{C}_3$ quantum number is the same as an electron-hole pair located at ±**K**, allowing the interconversion of the photon polarization with the excitonic valley pseudospin. A $\sigma$+ ($\sigma$−) circularly polarized photon can then interconvert with a spin-singlet bright exciton with $\langle\hat{\tau}_z\rangle$ = +1 (−1), whereas a linearly polarized photon can interconvert with a spin-singlet bright exciton with the in-plane excitonic valley pseudospin. Such optical manipulations of the excitonic valley pseudospin have been demonstrated in a series of experiments [188-192]. Meanwhile, the electron-hole exchange interaction becomes important when the bright exciton has a finite CoM momentum **Q**. It corresponds to a virtual annihilation-creation process of the bright exciton, which can introduce an off-diagonal inter-valley coupling term that depends on **Q** (Fig. 5(c)) [157, 193-196]. This effectively realizes strong valley-orbit coupling that splits the exciton dispersion (Fig. 5(d)), which is responsible for the valley depolarization and decoherence [197], valley-dependent excitonic transport phenomena [198-201] and topological exciton dispersions in moiré potential [202].

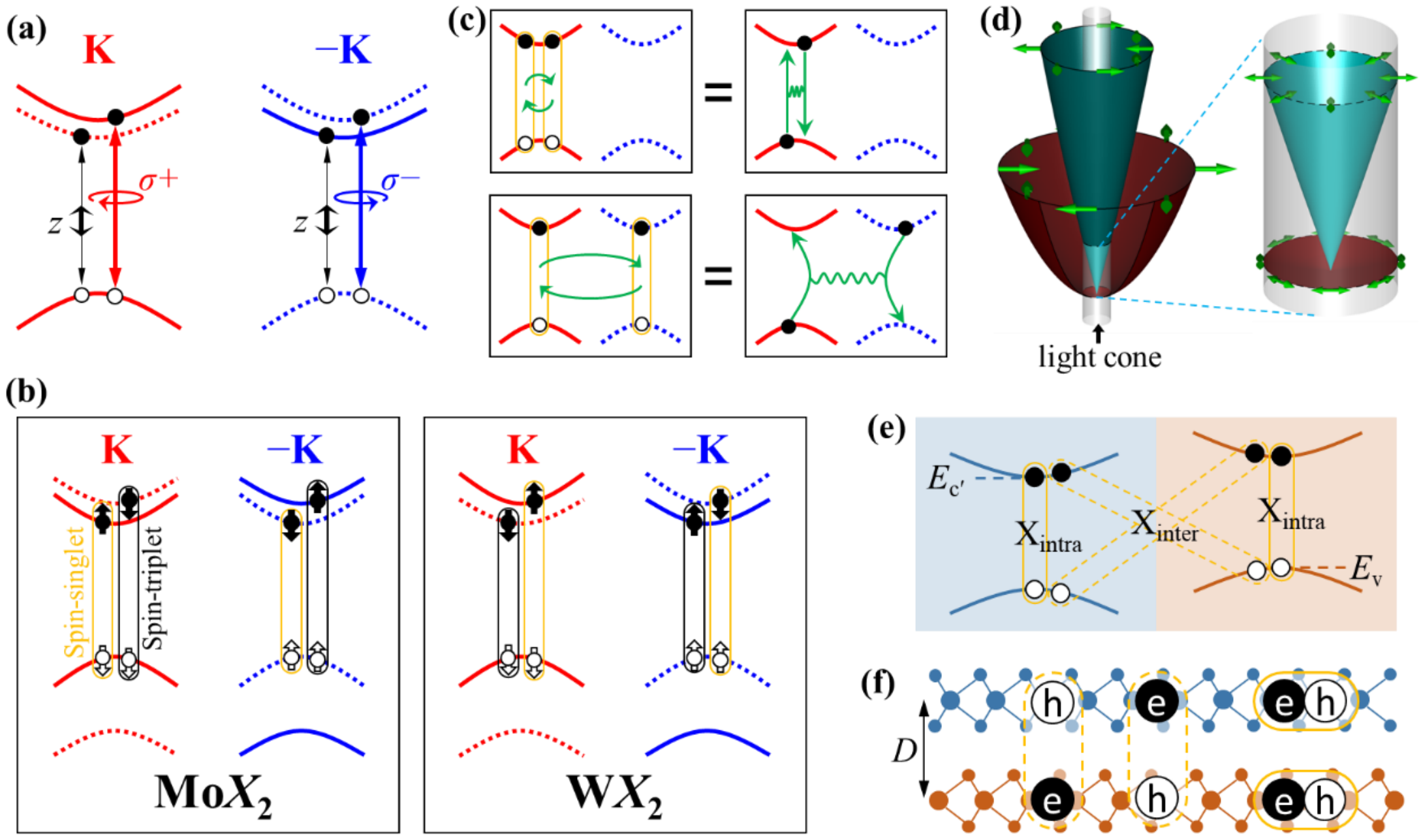

**Figure 5**. (a) The spin- and valley-dependent optical selection rules for near band-edge interband transitions in monolayer TMDs. (b) The spin-singlet exciton is the bound state of an electron (solid dots) and a hole (empty dots) with antiparallel spins, whereas they have parallel spins in a spin-triplet exciton. Only intra-valley excitons are shown here. (c) The electron-hole exchange interaction of the bright exciton. The upper-panel corresponds to the intra-valley process which introduces a diagonal energy shift, whereas the lower-panel corresponds to the inter-valley process which couples two bright excitons in opposite valleys. (d) Dispersions (curved surfaces) and valley pseudospins (single green arrows) of the two bright exciton branches under the electron-hole exchange interaction. Inside the light cone, the upper (lower) branch couples to a linearly polarized photon with the polarization direction longitudinal (transverse) to the CoM momentum. (e) A schematic illustration of the type-II band alignment for a bilayer system, and the corresponding $X_{intra}$ and $X_{inter}$ configurations. $E_{c'}$ and $E_v$ are the conduction and valence band edge energies. Blue and orange backgrounds correspond to two different layers. (f) Real-space pictures for $X_{intra}$ and $X_{inter}$ in a bilayer TMDs with an interlayer distance $D$. (d) Adapted from [157] under the terms of the Creative Commons CC BY license (https://creativecommons.org/licenses/by/4.0/)

In type-II heterobilayer, the Coulomb attraction between the low-energy electron and hole in opposite layers can result in the formation of a tightly bound interlayer exciton ($X_{inter}$) [203], see the illustration in Figs. 5(e, f). Compared to the intralayer exciton ($X_{intra}$) with the electron and hole constituents in the same layer, the ~ 0.6 nm vertical separation between the electron and hole in $X_{inter}$ greatly reduces its oscillator strength (only 0.1-1% of the intralayer one [56, 204-207]). Meanwhile, it also introduces a static out-of-plane electric dipole that brings tunability through an out-of-plane electric field [134, 208]. The binding energy of $X_{inter}$ is weaker than that of $X_{intra}$ by several tens meV [209-214] but still in the order of ~ 0.2 eV [62], and experiments have measured a Bohr radius of $a_B$ ~ 2 nm in $X_{inter}$ [72, 171]. And with the layer separation of electron and hole, their Coulomb exchange is efficiently suppressed, so that $X_{inter}$ has much longer valley lifetime as compared to $X_{intra}$. The low-temperature photoluminescence of type-II heterobilayers is usually dominated by $X_{inter}$ despite their weak oscillator strengths [134, 208], as energies of $X_{intra}$ are above the lowest-energy $X_{inter}$ by several hundred meV. $X_{inter}$ can then be efficiently generated through the excitation of $X_{intra}$ followed by a spin-conserved interlayer charge transfer process [215], which occurs in an ultrafast time scale (< 0.1 ps [74-76]).

In homobilayers, or in heterobilayers with nearly aligned conduction or valence band, interlayer hopping $T_{cc',\mathbf{K}}$ or $T_{vv',\mathbf{K}}$ (ranging from several meV to several tens meV [214, 216-218]) of the electron or hole constituent can lead to a strong hybridization between $X_{inter}$ and $X_{intra}$, as observed in a series of experiments [209-214, 216-222]. The resultant hybrid exciton ($X_{hybrid}$) can feature both large electric and optical dipoles, whose exceptional controllability through external electric and optical fields has attracted great interest. Experiments in naturally stacked 2H bilayer TMDs have shown that the energy separation between $X_{inter}$ and $X_{intra}$ can be continuously varied by an out-of-plane electric field, which can be used to tune their fractions in the formed $X_{hybrid}$ [213, 219-223]. Furthermore, the strong coupling between $X_{hybrid}$ and the cavity mode can lead to the formation of $X_{hybrid}$-polaritons [224-226], which are found to exhibit a largely enhanced nonlinearity (10 times larger than that of conventional $X_{intra}$-polaritons). The underlying mechanism is attributed to the strong interaction between these polaritons and the phase-space filling effect.

## 2.2. Moiré excitons in heterobilayer TMDs in the weak moiré potential limit

To study the effect of a moiré potential on the exciton, below we start from the momentum eigenstate of an electron-hole pair in twisted bilayer under the strong Coulomb binding. The electron and hole in opposite layers are described by Bloch states $\psi_{c',\tau'\mathbf{K}'+\mathbf{k}'}$ and $\psi^*_{v,\tau\mathbf{K}-\mathbf{k}}$, respectively, with $\pm\mathbf{K}'$ ($\pm\mathbf{K}$) the BZ corners of the electron (hole) layer. The interlayer electron-hole Coulomb interaction is found to conserve the quantity $\mathbf{Q} \equiv \mathbf{k} + \mathbf{k}'$ as well as the electron-hole valley indices ($\tau'$, $\tau$), even if $\pm\mathbf{K}'$ and $\pm\mathbf{K}$ are displaced from each other due to the interlayer twist and/or lattice constant mismatch (Fig. 6(a)) [204]. This results in an eigenstate in the form

$$X^{(\mathrm{inter})}_{\tau'\tau,\mathbf{Q}}(\mathbf{r}_\mathrm{e},\mathbf{r}_\mathrm{h}) = \sum_{\Delta\mathbf{Q}} \Phi_{\mathrm{inter}}(\Delta\mathbf{Q})\psi_{c',\tau'\mathbf{K}'+\frac{m_\mathrm{e}}{M}\mathbf{Q}+\Delta\mathbf{Q}}(\mathbf{r}_\mathrm{e})\psi^*_{v,\tau\mathbf{K}-\frac{m_\mathrm{h}}{M}\mathbf{Q}+\Delta\mathbf{Q}}(\mathbf{r}_\mathrm{h}). \qquad (16)$$

$X^{(\mathrm{inter})}_{\tau'\tau,\mathbf{Q}}$ has a kinetic energy $\frac{\hbar^2 Q^2}{2M}$ and a CoM group velocity $\hbar\mathbf{Q}/M$, thus can be termed as the kinematic momentum eigenstate of $X_{inter}$. For the bright $X^{(\mathrm{inter})}_{\tau'\tau,\mathbf{Q}}$ whose direct interconversion with a photon is allowed, the momentum conservation requires its kinematic momentum to be located at the so-called light cone positions $\mathbf{Q} = (\tau\mathbf{K} + \mathbf{G}) - (\tau'\mathbf{K}' + \mathbf{G}')$ with $\mathbf{G}$ ($\mathbf{G}'$) the reciprocal lattice vector of the hole (electron) layer. When the two layers are commensurate with $\mathbf{K} = \pm\mathbf{K}'$, the bright $X^{(\mathrm{inter})}_{\tau'\tau,\mathbf{Q}}$ must have $\mathbf{Q} = 0$ which is the same as that of $X_{intra}$; however, when $\pm\mathbf{K}'$ and $\pm\mathbf{K}$ are displaced from each other, the light cone positions form a series of discrete and finite $\mathbf{Q}$ values with nonzero kinetic energies (Fig. 6(b)) [204]. For low-energy bright $X^{(\mathrm{inter})}_{\tau'\tau,\mathbf{Q}}$ with small $Q$, $\tau\mathbf{K}$ must be close to $\tau'\mathbf{K}'$, which happens for $\tau' = \tau$ when the bilayer stacking is near R-type or $\tau' = -\tau$ when near H-type. The other low-energy $X^{(\mathrm{inter})}_{\tau'\tau,\mathbf{Q}}$ with $\tau' = -\tau$ in near R-type stackings (and $\tau' = \tau$ in near H-type stackings) correspond to inter-valley dark $X_{inter}$ thus will not be considered. The optical dipole $\left\langle 0 \middle| \hat{\mathbf{D}} \middle| X^{(\mathrm{inter})}_{\tau'\tau,\mathbf{Q}} \right\rangle$ decays exponentially with the increase of $Q$, thus only the three main light cones located at $\tau\mathbf{K}_\mathrm{m} \equiv \tau\mathbf{K} - \tau'\mathbf{K}'$ and its $\hat{C}_3$ rotations are important (Fig. 6(b)), whose optical dipoles $\mathbf{D}_0 \equiv \left\langle 0 \middle| \hat{\mathbf{D}} \middle| X^{(\mathrm{inter})}_{\tau'\tau,\tau\mathbf{K}_\mathrm{m}} \right\rangle$, $\mathbf{D}_1 \equiv \left\langle 0 \middle| \hat{\mathbf{D}} \middle| X^{(\mathrm{inter})}_{\tau'\tau,\tau\hat{C}_3\mathbf{K}_\mathrm{m}} \right\rangle$ and $\mathbf{D}_2 \equiv \left\langle 0 \middle| \hat{\mathbf{D}} \middle| X^{(\mathrm{inter})}_{\tau'\tau,\tau\hat{C}_3^2\mathbf{K}_\mathrm{m}} \right\rangle$ are related by $\hat{C}_3$ rotations. Umklapp-type light cones with $Q \geqslant 2K_\mathrm{m}$ have exponentially smaller optical dipole strengths and higher kinetic energies thus can be ignored. Note that all $\hat{D}_+$, $\hat{D}_-$ and $\hat{D}_z$ components of $\left\langle 0\left|\hat{\mathbf{D}}\right|X^{(\text{inter})}_{\tau'\tau,\mathbf{Q}}\right\rangle$ are generally nonzero due to the low symmetry at $\mathbf{Q} \neq 0$, whose magnitudes are about ~ 5% of the $X_{\text{intra}}$ optical dipole [56, 204-207]. The $\hat{D}_z$ component in the spin-singlet $X_{\text{inter}}$ and $\hat{D}_\pm$ component in the spin-triplet $X_{\text{inter}}$, which are both forbidden in the monolayer by the $\hat{\sigma}_\text{h}$ symmetry, now become as significant as other components due to the lack of $\hat{\sigma}_\text{h}$ symmetry in heterobilayers [40]. As a result, both the spin-singlet and spin-triplet $X_{\text{inter}}$ can contribute significantly to the photoluminescence signals [206, 227].

When the interlayer twist angle $\theta$ significantly deviates from 0° or 60°, the kinetic energy $\frac{\hbar^2 K_\text{m}^2}{2M}$ at main light cones becomes large. This explains why the luminescence from $X_{\text{inter}}$ is only observable in bilayers rather close to the R-/H-type stacking. Such bilayer structures feature long-wavelength moiré superlattice patterns with the mBZ corners located at $\pm\mathbf{K}_\text{m}$ and its $\hat{C}_3$ rotations (Fig. 2(a)). It should be emphasized that due to the displacement between $\tau'\mathbf{K}'$ and $\tau\mathbf{K}$, the *crystal momentum* of $X^{(\text{inter})}_{\tau'\tau,\mathbf{Q}}$, which is restricted to the mBZ, is located at $\mathbf{Q} + \tau'\mathbf{K}' - \tau\mathbf{K}$ plus a reciprocal lattice vector of the moiré pattern [40]. In bilayers with vanishing interlayer coupling, the mini-bands of $X^{(\text{inter})}_{\tau'\tau,\mathbf{Q}}$ with respect to its *crystal momentum* can be simply obtained by folding the parabolic dispersion $\frac{\hbar^2 Q^2}{2M}$ into the mBZ (Fig. 6(c)) [228]. Different light cones of $X^{(\text{inter})}_{\tau'\tau,\mathbf{Q}}$ are displaced by reciprocal lattice vectors of the moiré superlattice, thus their *crystal momenta* all correspond to $\mathbf{\Gamma}$ in the mBZ. In the lowest-energy mini-band, the $\mathbf{\Gamma}$-point at an energy $\frac{\hbar^2 K_\text{m}^2}{2M}$ is three-fold degenerate (six if including the valley degeneracy) from the main light cones; the second-lowest-energy $\mathbf{\Gamma}$-point has an energy $\frac{2\hbar^2 K_\text{m}^2}{M}$ and is from 1st-Umklapp light cones (Fig. 6(c)). The relation between the crystal and kinematic momenta can be further tuned by an in-plane magnetic field and heterostrain, which can modify the energies at light cones and lift the degeneracy [44, 229].

After taking into account the interlayer coupling, the electrons and holes experience moiré potentials (Sec. 1.6) [55]. For the strongly bound 1*s* exciton with a Bohr radius $a_\text{B}$ ~ 2 nm [72, 171] smaller than the moiré wavelength, the major effect is a moiré potential experienced by the exciton CoM motion. In a weak potential, where the coupling strength between the main and 1st-Umklapp light cones of $X_{\text{inter}}$ is much smaller than the energy separation $\frac{3\hbar^2 K_\text{m}^2}{2M}$, their hybridization is strongly suppressed. While the three degenerate main light cones are always strongly hybridized by the moiré potential. This splits the original lowest-energy $\mathbf{\Gamma}$-point into three eigenmodes, which are the coherent superpositions of $X^{(\text{inter})}_{\tau'\tau,\tau\mathbf{K}_\text{m}}$, $X^{(\text{inter})}_{\tau'\tau,\tau\hat{C}_3\mathbf{K}_\text{m}}$ and $X^{(\text{inter})}_{\tau'\tau,\tau\hat{C}_3^2\mathbf{K}_\text{m}}$. Given the $\hat{C}_3$ symmetry of the moiré pattern, wavefunctions of the three eigenmodes at $\mathbf{\Gamma}$ show standing wave patterns, with maximum intensities located at the $\hat{C}_3$-symmetric locales $R_h^h$, $R_h^X$ and $R_h^M$ (or $H_h^X$, $H_h^h$ and $H_h^M$), respectively (Fig. 6(d)) [228]. The resultant optical selection rule for each eigenmode then corresponds to an in-plane circularly or out-of-plane linearly polarized photon emission, depending on the valley-spin configuration. A weak moiré potential also strongly hybridizes Umklapp light cones, but the resultant eigenmodes have negligible photoemissions due to their weak optical dipole strengths and higher energies. In this case, one would expect to see photoemission peaks of spin-singlet and spin-triplet $X_{\text{inter}}$ from the main light cones only. On the other hand, in a stronger moiré potential that can lead to the hybridization between the main and Umklapp light cones, additional photoemission peaks from $X_{\text{inter}}$ can emerge at higher energies that correspond to the Umklapp light cones.

Besides $X_{\text{inter}}$, the CoM motion of $X_{\text{intra}}$ in each monolayer can also experience a moiré potential. Bright $X_{intra}$ has only one light cone located at $\mathbf{Q} = 0$, whose optical dipole strength is an order of magnitude larger than that of $X_{inter}$. Nevertheless, a strong moiré potential can also lead to hybridizations between $X_{intra}$ states with the CoM momenta located at finite moiré reciprocal lattice vectors and the one at $\mathbf{Q} = 0$, resulting in the emergence of higher energy Umklapp peaks in the emission/absorption spectrum. In experiments, multiple emission peaks from $X_{inter}$ with different polarization selection rules as well as absorption resonances from $X_{intra}$ have been observed, showing signatures for the formation of moiré-induced exciton minibands [206, 227, 230-232].

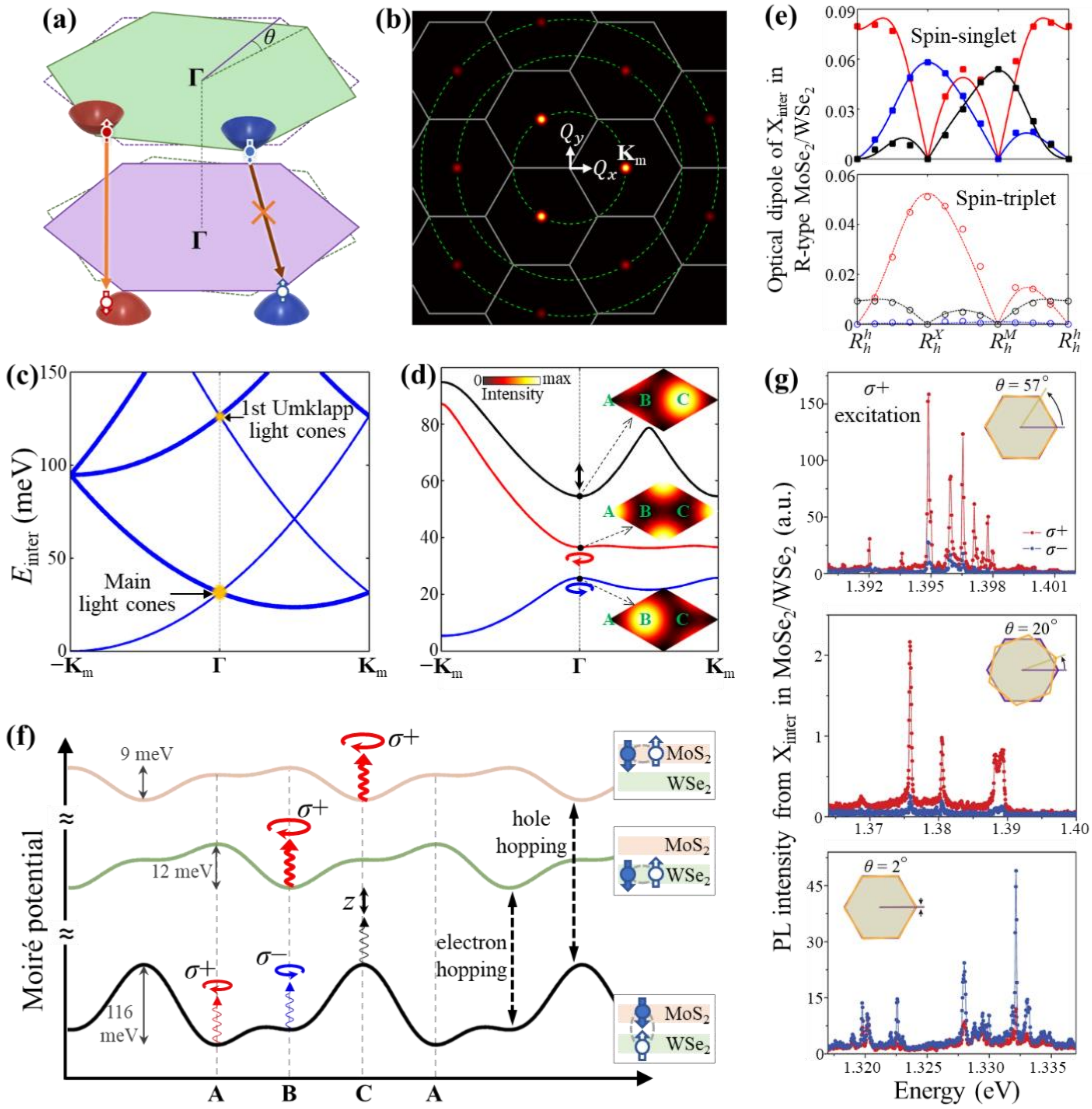

**Figure 6.** (a) The purple and green hexagons correspond to monolayer Brillouin zones of the two layers, whose corners are displaced from one another due to the real space twist angle $\theta$ or lattice constant mismatch. (b) Bright spots denote the light cones in the kinematic momentum space of $X_{inter}$ with $\tau = +$, with the three brightest ones located at $\mathbf{K}_m$ and its $\hat{C}_3$ rotations being the main light cones. Each dashed green circle corresponds to an equal-energy contour. White hexagons show the moiré mBZ. (c) $X_{inter}$ dispersion in the absence of moiré potential, folded into the mBZ of a 5 nm period moiré pattern. Thick lines are doubly degenerate. All light cones are folded to $\mathbf{\Gamma}$ with different energies. (d) The three lowest $X_{inter}$ bands under the effect of a weak moiré potential. The moiré potential leads to three well-separated eigenmodes at $\mathbf{\Gamma}$, which couple to $\sigma-$, $\sigma+$ and $z$ polarized optical fields, respectively. The insets show the spatial distributions of $X_{inter}$ intensities in a moiré supercell, with the maxima located at **B**, **A** and **C**, respectively. (e) Optical dipoles of spin-singlet and spin-triplet $X_{inter}$ with the valley index $(\tau, \tau') = (+, +)$ in R-type $MoSe_2/WSe_2$ as functions of the stacking registry. Red, blue and black symbols correspond to $\widehat{D}_+$, $\widehat{D}_-$ and $\widehat{D}_z$ components, respectively. Curves are analytical fittings. (f) Moiré potential landscapes

for $X_{intra}$ and $X_{inter}$, with optical selection rules of those with ($\tau$, $\tau'$) = (+, +) shown at several energy extrema. (g) Polarization-resolved photoluminescence spectra from trapped $X_{inter}$ in $MoSe_2/WSe_2$ heterobilayers with twist angles of 57° (upper-panel), 20° (middle-panel) and 2° (lower-panel). Under a circularly polarized excitation, the photoluminescence from the samples of 57° and 20° angles is co-circularly polarized (red), whereas that of 2° twist angle is cross-circularly polarized (blue). (a) From [203], reprinted with permission from Springer Nature. (b-d) Reprinted from [228] with permission from Elsevier. (e) From [40], adapted with permission from IOP Publishing. (f) From [56], adapted with permission from AAAS. (g) From [233], reprinted with permission from Springer Nature.

The electron-hole relative motion can also be modulated by the moiré potential, especially for those high-energy Rydberg states with close energies (e.g., 2*s* and $2p_{\pm}$). Taking $X_{intra}$ in monolayer TMDs under the effect of a spatially periodic electrostatic potential $V_e(\mathbf{r})$ as an example, the potential $V_X$ experienced by $X_{intra}$ can be written as

$$V_X = V_e(\mathbf{r}_e) - V_e(\mathbf{r}_h) \approx \mathbf{r}_{eh} \cdot \nabla V_e(\mathbf{R}). \tag{17}$$

Here $\mathbf{R} \equiv \frac{m_e}{M}\mathbf{r}_e + \frac{m_h}{M}\mathbf{r}_h$ and $\mathbf{r}_{eh} \equiv \mathbf{r}_e - \mathbf{r}_h$ are the CoM and electron-hole relative coordinates, respectively. Here we have assumed that the wavelength of $V_e(\mathbf{r})$ is large compared to the exciton Bohr radius, such that $V_e\left(\mathbf{R} \pm \frac{m_{h/e}}{M}\mathbf{r}_{eh}\right) \approx V_e(\mathbf{R}) \pm \frac{m_{h/e}}{M}\mathbf{r}_{eh} \cdot \nabla V_e(\mathbf{R})$. Note that $\mathbf{r}_{eh}$ couples two Rydberg states with different angular momenta, whereas $\nabla V_e(\mathbf{R})$ is a periodic function which couples two different CoM momenta. The periodic electrostatic potential thus introduces a coupling between the exciton CoM and electron-hole relative motions, which can hybridize 2*s* and $2p_{\pm}$ Rydberg states with different CoM momenta [234, 235]. Such $V_e(\mathbf{r})$ can be introduced by the periodic charge distribution in adjacent twisted bilayer hBN or graphene [236-242]. In experiments, the emergence of multiple absorption peaks near the 2*s* resonance of $X_{intra}$ has been observed under a finite $V_e(\mathbf{r})$, showing signatures of strong hybridizations between 2*s* and $2p_{\pm}$ states [239, 242]. When $V_e(\mathbf{r})$ is further strengthened by increasing the doping density in the neighboring twisted bilayer graphene, these absorption peaks red-shift significantly (up to several tens meV), which can become close to the lowest-energy 1*s* state. This suggests that 1*s* exciton can also hybridize strongly with $2p_{\pm}$ and 2*s* states by the periodic electrostatic potential from a neighboring twisted bilayer graphene [234].

## 2.3. Interlayer and hybrid moiré excitons localized in the strong moiré potential limit

In a strong moiré potential, a better starting point to describe $X_{inter}$ will be wavepackets localized around the potential minima [56], which form the basis for tight-binding model to account for the low-energy minibands. A gaussian wavepacket with the valley index ($\tau'$, $\tau$) can be constructed from the kinematic momentum eigenstate (Eq. (16)):

$$\mathcal{X}_{\tau'\tau,\mathbf{R}_c} = \sqrt{\frac{4\pi}{S}}w\sum_{\mathbf{Q}} e^{-i\mathbf{Q}\cdot\mathbf{R}_c - \frac{Q^2w^2}{2}} X^{(\text{inter})}_{\tau'\tau,\mathbf{Q}}(\mathbf{r}_e, \mathbf{r}_h). \tag{18}$$

Here $S$ is the 2D area of the system that ensures the normalization of $\mathcal{X}_{\tau'\tau,\mathbf{R}_c}$, and $\mathbf{R}_c$ is the wavepacket's real-space center position. Due to the strong confinement from the moiré potential, the experimentally measured wavepacket width $w$ is ~ 2 nm [170, 171], much smaller than the moiré wavelength $\lambda$. This results in a momentum-space distribution width $2/w$ much larger than $K_m = \frac{4\pi}{3\lambda}$, implying that the wavepacket's momentum-space distribution covers all the three main light cones [56, 203]. The optical dipole $\langle 0|\hat{\mathbf{D}}|\mathcal{X}_{\tau'\tau,\mathbf{R}_c}\rangle$ of the wavepacket is thus the coherent sum of those from the three main light cones:

$$\langle 0|\hat{\mathbf{D}}|\mathcal{X}_{\tau'\tau,\mathbf{R}_c}\rangle \propto e^{-i\tau\mathbf{K}_m\cdot\mathbf{R}_c}\mathbf{D}_0 + e^{-i\tau\hat{C}_3\mathbf{K}_m\cdot\mathbf{R}_c}\mathbf{D}_1 + e^{-i\tau\hat{C}_3^2\mathbf{K}_m\cdot\mathbf{R}_c}\mathbf{D}_2. \tag{19}$$

Note that the interference between the main light cones leads to the modulation of $\langle 0|\hat{\mathbf{D}}|\mathcal{X}_{\tau'\tau,\mathbf{R}_c}\rangle$ with $\mathbf{R}_c$. A wavepacket $\mathcal{X}_{\tau'\tau,\mathbf{R}_c}$ at a location $\mathbf{R}_c$ then has finite $\hat{D}_+$, $\hat{D}_-$ and $\hat{D}_z$ components in its optical dipole (Fig. 6(e)). Considering the $\hat{C}_3$ symmetry of the moiré pattern, moiré potential extrema should be located at regions with $R_h^h$, $R_h^X$ and $R_h^M$ (or $H_h^X$, $H_h^h$ and $H_h^M$) stacking registries. A wavepacket $\mathcal{X}_{\tau'\tau,\mathbf{R}_c}$ localized at these regions has the same $\hat{C}_3$ quantum number as $\psi_{c,\tau'\mathbf{K}'}(\mathbf{r}_e)\psi^*_{v,\tau\mathbf{K}}(\mathbf{r}_h)$ in a commensurate bilayer with the corresponding stacking registry, which is given in Table 1. This enforces $\langle 0|\hat{\mathbf{D}}|\mathcal{X}_{\tau'\tau,\mathbf{R}_c}\rangle$ to have only the $\hat{D}_+$, $\hat{D}_-$ or $\hat{D}_z$ component, with the corresponding optical selection rules summarized in Table 5 (also see Fig. 6(f)). Considering the tens of meV splitting between the spin-singlet and spin-triplet, as well as the potential difference between $R_h^h$, $R_h^X$ and $R_h^M$ (or $H_h^X$, $H_h^h$ and $H_h^M$), it is possible to distinguish the spin, valley and position configurations of $X_{inter}$ wavepackets through their emission energies and polarizations [56, 206, 227, 230, 231].

**Table 5.** The polarization selection rules for the spin-singlet and spin-triplet bright $X_{inter}$ localized at $R_h^h$, $R_h^X$ and $R_h^M$ ($H_h^X$, $H_h^h$ and $H_h^M$) stacking configurations. The hole is assumed at **K** valley ($\tau = +$).

| | $R_h^h$ & $H_h^X$ | $R_h^X$ & $H_h^h$ | $R_h^M$ & $H_h^M$ |
|---|---|---|---|
| Spin-singlet with $\tau = +$ | $\sigma+$ | $\sigma-$ | $z$ |
| Spin-triplet with $\tau = +$ | $z$ | $\sigma+$ | $\sigma-$ |

When the inter-site hopping between nearest-neighbor potential minima is suppressed by a small wavepacket width $w \ll \lambda$, the low-energy miniband of $X_{inter}$ gets flattened [44, 243]. In this case the $X_{inter}$ state is trapped at the potential minimum, and its photoemission is expected to exhibit the single-photon nature [56]. In high-quality moiré-patterned $MoSe_2/WSe_2$ heterobilayers, ultranarrow (~ 0.1 meV) $X_{inter}$ emission peaks have been observed under a rather low temperature (~ 1 K) and low excitation power (Fig. 6(g)) [233, 244]. These emission peaks exhibit a series of remarkable properties including strong circular polarizations under a circularly polarized photoexcitation (cross-polarized in near R-type and co-polarized in near H-type heterobilayers), uniform g-factors among emitters in the same and different samples ($\approx 6.7$ in all near R-type samples and $\approx -15.8$ in all near H-type samples), suggesting that they are from $X_{inter}$ localized at the $\hat{C}_3$-symmetric moiré potential minima. A photon correlation measurement has revealed the anti-bunching behavior of these narrow emission peaks, indicating their single-photon nature [245].

In twisted homobilayer TMDs, the sensitive dependence of $T_{nn',\mathbf{K}}$ on the stacking registry results in position-dependent hybridizations between localized $X_{inter}$ and $X_{intra}$, which can also result in intriguing phenomena related to $X_{hybrid}$ [246]. In a near R-type homobilayer $MoTe_2$ moiré pattern, the lowest-energy bright $X_{inter}$ with $(\tau', \tau) = (+, +)$ is expected to be a gaussian wavepacket localized at $R_h^X$ or $R_h^M$, which has a $\hat{C}_3$ quantum number $-1$ thus can emit a $\sigma-$ circularly polarized photon. In contrast, the bright $X_{intra}$ wave packet in **K** valley always has a $\hat{C}_3$ quantum number +1 and can emit a $\sigma+$ circularly polarized photon, irrespective of its center position. The hybridization between $X_{inter}$ and $X_{intra}$ at $R_h^X$ or $R_h^M$ is thus forbidden by the $\hat{C}_3$ symmetry. However, the $\hat{C}_3$ symmetry is broken when the $X_{inter}$ wavepacket center is slightly displaced from the $R_h^X$ or $R_h^M$ region, e.g., by the dipolar force from a neighboring $X_{inter}$. This gives rise to a finite interlayer hopping, and the hybridization between $X_{inter}$ and $X_{intra}$ wavepackets becomes possible. With opposite circular polarizations of the $X_{inter}$ and $X_{intra}$ contributions, the photoemission of the $X_{hybrid}$ wavepacket exhibits a finite linear polarization, whose strength and direction are determined by the displacement vector. Also note that the optical dipole of $X_{inter}$ is much weaker (~ 5%) than that of $X_{intra}$, it is thus suggested that a tiny displacement of the wavepacket can substantially change the polarization and magnitude of $X_{hybrid}$'s optical dipole [246].

## 2.4. Dipole-dipole and exchange interactions between localized excitons

Exciton-exciton interactions in quantum well systems are usually analyzed in the basis of momentum eigenstates [247-254]. In the moiré confinement, as low-energy $X_{inter}$ are usually strongly localized at the moiré potential minima, it is more suitable to discuss interactions between two strongly localized $X_{inter}$ in real space. For two $X_{inter}$ wavepackets at different sites separated by an inter-site distance (~ 10 nm) much larger than the wavepacket width, their interaction is dominated by the spin/valley-independent dipole-dipole interaction $V_{\mathrm{dd}}$. On the other hand, for two $X_{inter}$ wavepackets located at the same site, the exchange-related interaction due to the indistinguishability of the two electrons/holes becomes rather important, which is sensitive to the electron and hole spin-valley indices. Below we analyze their forms.

In experiments, when the excited $X_{inter}$ density is low (less than one $X_{inter}$ per moiré supercell), different $X_{inter}$ tend to occupy different sites and their interaction is dominated by $V_{\mathrm{dd}}$, which leads to a continuous increase of the photoluminescence energy with the excitation density [134, 255, 256]. The resultant density-dependent energy shift underlies the intriguing diffusion dynamics of $X_{inter}$ [208, 257-259]. Note that $V_{\mathrm{dd}}$ is affected by the atomically-thin geometry of layered materials [260]. Similar to the widely adopted Rytova-Keldysh form of the Coulomb interaction in monolayer TMDs [261-265], one can model the TMDs bilayer by two infinitely-thin 2D layers separated by a vertical distance $D$ [266], where dielectric screening of each layer is characterized by the screening length $s_0 \approx 4.5$ nm. By solving the Poisson's equation, the derived dipole-dipole interaction between two $X_{inter}$ wavepackets separated by $\mathbf{r}$ has the form

$$V_{\mathrm{dd}} \approx 2V_{\mathrm{intra}}(\mathbf{r}) - 2V_{\mathrm{inter}}(\mathbf{r}) = 2\int_0^{\infty} \frac{(1-e^{-Dq})J_0(qr)}{1+qs_0(1-e^{-Dq})} dq, \tag{20}$$

with $V_{\mathrm{intra}}$ ($V_{\mathrm{inter}}$) the intralayer (interlayer) Coulomb interaction (see Fig. 7(a)), and $J_0$ the Bessel function of the first kind. Compared to the traditional dipole-dipole interaction $V_{\mathrm{dd}}^{(3\mathrm{D})} = 2/r - 2/\sqrt{r^2+D^2}$, the calculated $V_{\mathrm{dd}}$ value in bilayer TMDs is found to be anomalously enhanced by several times at an inter-exciton distance of several nm or larger, whereas encapsulation by thick hBN layers only slightly affects $V_{\mathrm{dd}}$ [260]. The calculated value agrees well with that extracted from the experiment in Ref. [267], see Fig. 7(b). This is in sharp contrast to Coulomb interactions $V_{\mathrm{intra}}$ and $V_{\mathrm{inter}}$, which are greatly weakened by $s_0$ and the screening of the encapsulating hBN layers [264, 265]. The underlying mechanism can be attributed to the distribution of 2D induced-charge densities in TMDs layers by the excitonic dipolar field. Such behaviors of dipole-dipole interactions in bilayer systems signify the importance of inter-site interactions between $X_{inter}$ wavepackets, which can facilitate the study of novel quantum states in condensed matter physics, e.g., the excitonic charge-density wave and supersolid phases [268].

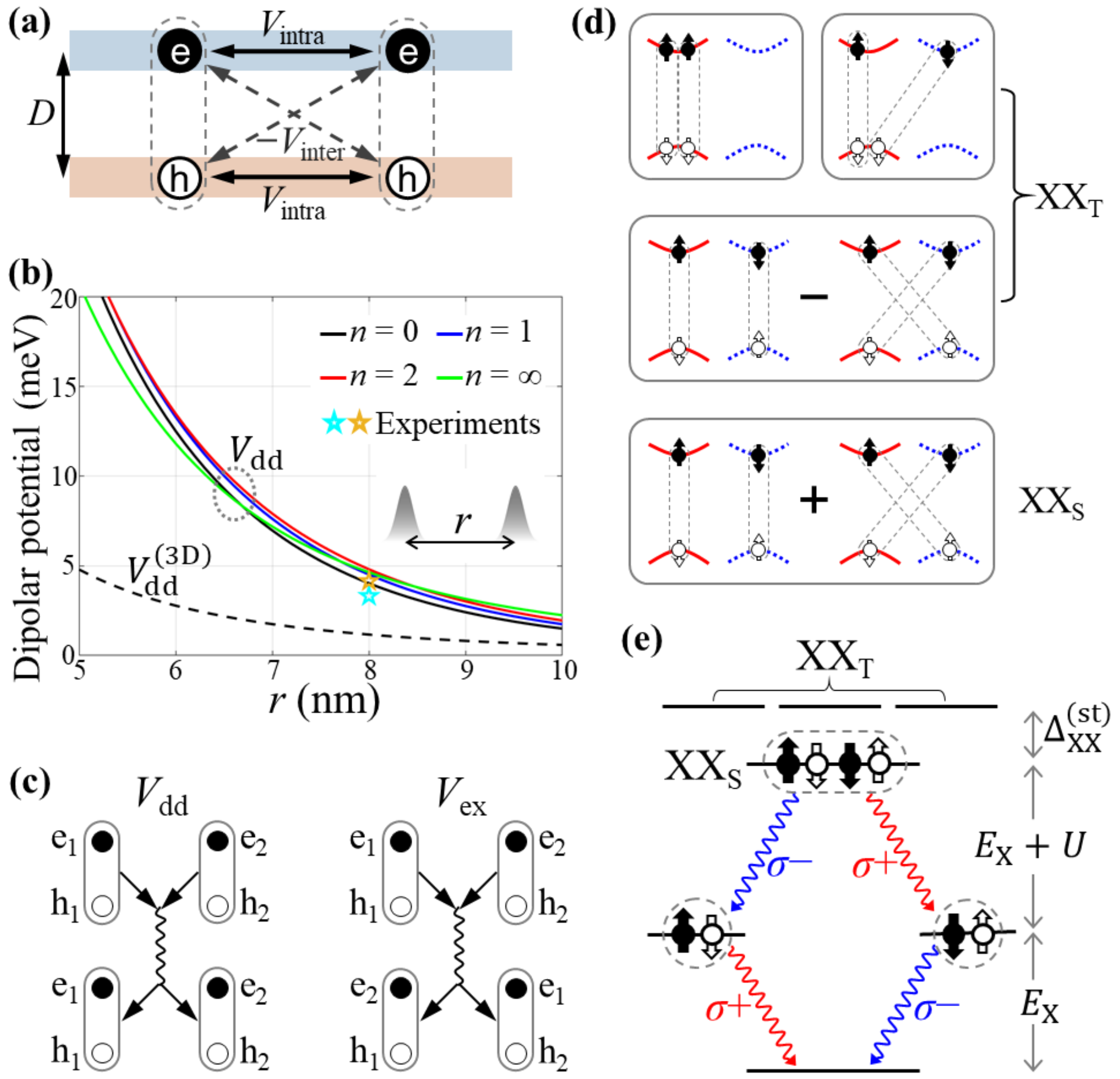

**Figure 7.** (a) An illustration of the exciton-exciton interaction $\hat{V}_{\text{XX}}$. (b) The calculated dipole-dipole interaction strength as a function of the inter-exciton distance $r$. Solid lines are theoretical results with $n$ denoting the layer number of the encapsulating multilayer hBN, and star symbols are extracted from the experiment in Ref. [267]. Dashed line corresponds to the traditional dipole-dipole interaction form $V_{\text{dd}}^{(3D)} = 2/r - 2/\sqrt{r^2 + D^2}$, which is much weaker than the solid lines as well as the experimental results. (c) Diagrams of the dipole-dipole ($V_{\text{dd}}$) and exchange ($V_{\text{ex}}$) interaction terms between two excitons. (d) Two-exciton spin-triplet ($XX_T$) and spin-singlet ($XX_S$) forms under various spin-valley configurations of electrons and holes. Red solid and blue dashed lines correspond to $+\mathbf{K}$ and $-\mathbf{K}$ valleys, respectively. (e) The lowest-energy exciton-pair at a potential minimum in $XX_S$ state corresponds to two $X_{\text{inter}}$ in opposite valleys, whose cascaded emission can generate a polarization-entangled photon pair. (b) From [260], reprinted with permission from IOP Publishing.

For two $X_{\text{inter}}$ wavepackets located at the same site, we denote their single-exciton wave functions as $\psi(\mathbf{r}_e, \mathbf{r}_h)|s_e\rangle|s_h\rangle$ and $\psi(\mathbf{r}'_e, \mathbf{r}'_h)|s'_e\rangle|s'_h\rangle$, with $s_e, s'_e = \uparrow, \downarrow$ ($s_h, s'_h = \Downarrow, \Uparrow$) the spin-valley indices for the electron (hole). Due to the permutation antisymmetry of two electrons or two holes, even if the exciton-exciton interaction is not considered, the two-exciton wave function is not in the simple form $\psi(\mathbf{r}_e, \mathbf{r}_h)\psi(\mathbf{r}'_e, \mathbf{r}'_h)|s_e s'_e\rangle|s_h s'_h\rangle$ but contains various permutation terms. Depending on whether the spin part of the two-electron or two-hole wave function is in a spin-triplet or spin-singlet form, the orbital part is permutation antisymmetric or symmetric. We then construct the following orbital wave function forms for the two-exciton state

$$\psi_{\mathrm{XX}}^{(\mathrm{anti})}=\frac{\psi(\mathbf{r}_{\mathrm{e}},\mathbf{r}_{\mathrm{h}})\psi(\mathbf{r}_{\mathrm{e}}',\mathbf{r}_{\mathrm{h}}')-\psi(\mathbf{r}_{\mathrm{e}},\mathbf{r}_{\mathrm{h}}')\psi(\mathbf{r}_{\mathrm{e}}',\mathbf{r}_{\mathrm{h}})}{\sqrt{2-2\alpha}}, \tag{21}$$
$$\psi_{\mathrm{XX}}^{(\mathrm{sym})}=\frac{\psi(\mathbf{r}_{\mathrm{e}},\mathbf{r}_{\mathrm{h}})\psi(\mathbf{r}_{\mathrm{e}}',\mathbf{r}_{\mathrm{h}}')+\psi(\mathbf{r}_{\mathrm{e}},\mathbf{r}_{\mathrm{h}}')\psi(\mathbf{r}_{\mathrm{e}}',\mathbf{r}_{\mathrm{h}})}{\sqrt{2+2\alpha}}.$$

Here $\alpha\equiv\int d\mathbf{r}_{\mathrm{e}}d\mathbf{r}_{\mathrm{h}}d\mathbf{r}_{\mathrm{e}}'d\mathbf{r}_{\mathrm{h}}'\psi^*(\mathbf{r}_{\mathrm{e}}',\mathbf{r}_{\mathrm{h}}')\psi^*(\mathbf{r}_{\mathrm{e}},\mathbf{r}_{\mathrm{h}})\psi(\mathbf{r}_{\mathrm{e}},\mathbf{r}_{\mathrm{h}}')\psi(\mathbf{r}_{\mathrm{e}}',\mathbf{r}_{\mathrm{h}})$ is a real value corresponding to the wave function overlap. The total two-exciton wave function forms either a spin-triplet ($\psi_{\mathrm{XX}}^{(\mathrm{anti})}|\sigma_{\mathrm{e}}\sigma_{\mathrm{e}}\rangle\frac{|\Downarrow\Uparrow\rangle+|\Uparrow\Downarrow\rangle}{\sqrt{2}}$, $\psi_{\mathrm{XX}}^{(\mathrm{anti})}\frac{|\downarrow\uparrow\rangle+|\uparrow\downarrow\rangle}{\sqrt{2}}|\sigma_{\mathrm{h}}\sigma_{\mathrm{h}}\rangle$, $\psi_{\mathrm{XX}}^{(\mathrm{anti})}\frac{|\uparrow\downarrow\rangle+|\downarrow\uparrow\rangle}{\sqrt{2}}\frac{|\Downarrow\Uparrow\rangle+|\Uparrow\Downarrow\rangle}{\sqrt{2}}$, denoted as $\mathrm{XX_T}$) or a spin-singlet ($\psi_{\mathrm{XX}}^{(\mathrm{sym})}\frac{|\uparrow\downarrow\rangle-|\downarrow\uparrow\rangle}{\sqrt{2}}\frac{|\Downarrow\Uparrow\rangle-|\Uparrow\Downarrow\rangle}{\sqrt{2}}$, denoted as $\mathrm{XX_S}$). For simplicity, the exciton-exciton interaction $\hat{V}_{\mathrm{XX}}=V_{\mathrm{intra}}(\mathbf{r}_{\mathrm{e}}-\mathbf{r}_{\mathrm{e}}')+V_{\mathrm{intra}}(\mathbf{r}_{\mathrm{h}}-\mathbf{r}_{\mathrm{h}}')-V_{\mathrm{inter}}(\mathbf{r}_{\mathrm{e}}-\mathbf{r}_{\mathrm{h}}')-V_{\mathrm{inter}}(\mathbf{r}_{\mathrm{h}}-\mathbf{r}_{\mathrm{e}}')$ can be treated as a perturbation. The energy corrections to the two-exciton states up to the lowest-order are

$$\left\langle\psi_{\mathrm{XX}}^{(\mathrm{anti})}\middle|\hat{V}_{\mathrm{XX}}\middle|\psi_{\mathrm{XX}}^{(\mathrm{anti})}\right\rangle=\frac{V_{\mathrm{dd}}+V_{\mathrm{ex}}}{1-\alpha}, \tag{22}$$
$$\left\langle\psi_{\mathrm{XX}}^{(\mathrm{sym})}\middle|\hat{V}_{\mathrm{XX}}\middle|\psi_{\mathrm{XX}}^{(\mathrm{sym})}\right\rangle=\frac{V_{\mathrm{dd}}-V_{\mathrm{ex}}}{1+\alpha}.$$

Here, $V_{\mathrm{dd}}\equiv\int d\mathbf{r}_{\mathrm{e}}d\mathbf{r}_{\mathrm{h}}d\mathbf{r}_{\mathrm{e}}'d\mathbf{r}_{\mathrm{h}}'\psi^*(\mathbf{r}_{\mathrm{e}}',\mathbf{r}_{\mathrm{h}}')\psi^*(\mathbf{r}_{\mathrm{e}},\mathbf{r}_{\mathrm{h}})\hat{V}_{\mathrm{XX}}\psi(\mathbf{r}_{\mathrm{e}},\mathbf{r}_{\mathrm{h}})\psi(\mathbf{r}_{\mathrm{e}}',\mathbf{r}_{\mathrm{h}}')$ corresponds to the direct Coulomb or dipole-dipole interaction term and $V_{\mathrm{ex}}\equiv-\int d\mathbf{r}_{\mathrm{e}}d\mathbf{r}_{\mathrm{h}}d\mathbf{r}_{\mathrm{e}}'d\mathbf{r}_{\mathrm{h}}'\psi^*(\mathbf{r}_{\mathrm{e}}',\mathbf{r}_{\mathrm{h}}')\psi^*(\mathbf{r}_{\mathrm{e}},\mathbf{r}_{\mathrm{h}})\hat{V}_{\mathrm{XX}}\psi(\mathbf{r}_{\mathrm{e}},\mathbf{r}_{\mathrm{h}}')\psi(\mathbf{r}_{\mathrm{e}}',\mathbf{r}_{\mathrm{h}})$ is the exchange interaction terms [247-252], which are schematically illustrated in Fig. 7(c). These interaction terms together with the finite $\alpha$ lead to energy separations between different two-exciton states.

Due to the spin-valley locking of band edge carriers, whether the two-exciton state is in $\mathrm{XX_T}$ or $\mathrm{XX_S}$ depends on the electron and hole valley indices (Fig. 7(d)). Only when the two electrons (as well as the two holes) fall in opposite valleys, can the two-exciton state be in $\mathrm{XX_S}$. Both the strengths of $V_{\mathrm{dd}}$ and $V_{\mathrm{ex}}$, as well as the wave function overlap $\alpha$, become increasingly significant with the reduction of the wavepacket's width. In bilayer TMDs moiré patterns, the typical width of $\mathrm{X_{inter}}$ wavepackets is ~ 2 nm [170, 171]. For such parameters, a rough estimation indicates that the energy of $\mathrm{XX_T}$ is above $\mathrm{XX_S}$ by several tens meV [269]. Meanwhile, $\mathrm{XX_S}$ has an energy $U$ ~ 30 meV above that of a non-interacting $\mathrm{X_{inter}}$ pair, which agrees well with the experimentally measured emission energies from two-exciton occupied sites [269-273]. The lowest-energy exciton-pair localized in a potential minimum thus is expected to emit a polarization-entangled photon pair (Fig. 7(e)) [56]. Although the entanglement hasn't been confirmed experimentally yet, the absence of circular polarization from the emissions of such two-exciton states is fully consistent with the theoretical prediction [269, 273].

In the above analysis, the single-exciton energy splitting $\Delta_{\mathrm{X}}^{(\mathrm{st})}$ between a spin-singlet exciton with $(s_{\mathrm{e}},s_{\mathrm{h}})=(\uparrow,\Downarrow)$ or $(\downarrow,\Uparrow)$ and a spin-triplet exciton with $(s_{\mathrm{e}},s_{\mathrm{h}})=(\uparrow,\Uparrow)$ or $(\downarrow,\Downarrow)$ is not included. $\Delta_{\mathrm{X}}^{(\mathrm{st})}$ is generally finite due to the electron-hole exchange energy inside the spin-singlet exciton [274, 275], which can reach ~ 10 meV for $\mathrm{X_{intra}}$ but becomes much weaker for $\mathrm{X_{inter}}$ due to the spatial separation between the electron and hole. When referring to two-exciton states, $\Delta_{\mathrm{X}}^{(\mathrm{st})}$ couples $\mathrm{XX_S}$ with a spin wave function $\frac{|\uparrow\downarrow\rangle-|\downarrow\uparrow\rangle}{\sqrt{2}}\frac{|\Downarrow\Uparrow\rangle-|\Uparrow\Downarrow\rangle}{\sqrt{2}}$ and $\mathrm{XX_T}$ with a spin wave function $\frac{|\uparrow\downarrow\rangle+|\downarrow\uparrow\rangle}{\sqrt{2}}\frac{|\Downarrow\Uparrow\rangle+|\Uparrow\Downarrow\rangle}{\sqrt{2}}$. The two-exciton eigenstate is thus determined by the competition between the single-exciton splitting $\Delta_{\mathrm{X}}^{(\mathrm{st})}$ and the two-exciton splitting $\Delta_{\mathrm{XX}}^{(\mathrm{st})}=\frac{V_{\mathrm{dd}}+V_{\mathrm{ex}}}{1-\alpha}-\frac{V_{\mathrm{dd}}-V_{\mathrm{ex}}}{1+\alpha}$. When $2\Delta_{\mathrm{X}}^{(\mathrm{st})}\gg\Delta_{\mathrm{XX}}^{(\mathrm{st})}$, the eigenstates of two excitons with opposite electron spins and opposite hole spins should be a pair of intravalley excitons and a pair of intervalley excitons, which are neither $\mathrm{XX_T}$ nor $\mathrm{XX_S}$ with an energy separation of $2\Delta_{\mathrm{X}}^{(\mathrm{st})}$. On the other hand, when $\Delta_{\mathrm{XX}}^{(\mathrm{st})} \gg 2\Delta_{\mathrm{X}}^{(\mathrm{st})}$, the two-exciton eigenstates should be $\psi_{\mathrm{XX}}^{(\mathrm{anti})} \frac{|\uparrow\downarrow\rangle+|\downarrow\uparrow\rangle}{\sqrt{2}} \frac{|\Downarrow\Uparrow\rangle+|\Uparrow\Downarrow\rangle}{\sqrt{2}}$ and $\psi_{\mathrm{XX}}^{(\mathrm{sym})} \frac{|\uparrow\downarrow\rangle-|\downarrow\uparrow\rangle}{\sqrt{2}} \frac{|\Downarrow\Uparrow\rangle-|\Uparrow\Downarrow\rangle}{\sqrt{2}}$. As $\Delta_{\mathrm{XX}}^{(\mathrm{st})}$ decays with the increase of the wavepacket width $w$ whereas $\Delta_{\mathrm{X}}^{(\mathrm{st})}$ corresponds to a single-exciton property largely independent on $w$, $2\Delta_{\mathrm{X}}^{(\mathrm{st})} \gg \Delta_{\mathrm{XX}}^{(\mathrm{st})}$ ( $\Delta_{\mathrm{XX}}^{(\mathrm{st})} \gg 2\Delta_{\mathrm{X}}^{(\mathrm{st})}$ ) can be realized for two excitons in plane-wave forms (for two $X_{\mathrm{inter}}$ wavepackets localized at the same site).

## 3. Correlation phenomena of moiré charge carriers and moiré excitons

### 3.1. Correlated electronic crystals in the moiré potentials

In most heterobilayer moiré superlattices, low-energy carriers are localized at the moiré potential minima that form a triangular lattice. The inter-site hopping between different potential minima decays exponentially with the moiré wavelength, which can be tuned through the twist angle. Meanwhile, the reduced screening in the 2D geometry leads to strong Coulomb interactions between carriers. Moiré patterns thus can serve as a platform for simulating correlated phenomena with tunable filling factor $v$ (defined as the average number of doped carriers per moiré unit cell). At $v$=1, the inter-site hopping can be suppressed by the strong on-site repulsion $U$, resulting in a Mott insulator state in the triangular lattice (see the illustration in Fig. 8(a)). A series of experiments have reported the observation of Mott insulator states in heterobilayer $WS_2$/$WSe_2$ and homobilayer TMDs either by transport measurements [276-280], or by monitoring the exciton absorption/photoluminescence which shows an abrupt change when the carriers become an insulator [147, 267, 281-286], or through a real-space imaging using the STS [287, 288]. The magnitude of $U$ is estimated to be several tens meV from the measured critical temperature of the insulating behavior [277, 279]. Moreover, insulating behaviors are also observed under fractional filling factors ($v$ = 1/3, 1/2, 2/3, etc.), where carriers form ordered arrays with periodicities larger than the moiré wavelength thus break the discrete-translation symmetry and sometimes also point-group symmetry of the moiré pattern [267, 279-283, 287, 289]. This corresponds to the formation of a series of generalized Wigner crystal states: including a triangular lattice under $v$ = 1/3, a honeycomb lattice under $v$ = 2/3 (Fig. 8(b)), and stripe lattices under $v$ = 1/2 which spontaneously break the $\hat{C}_3$ symmetry (Fig. 8(c)). The real-space lattice configurations of such generalized Wigner crystals have been directly visualized through STS (Fig. 8(d)) [287]. Conventional Hubbard model with on-site interaction only should be conducting under $v$ < 1. The above observations of correlated insulators at fractional filling indicate that inter-site interactions between carriers separated by ~ 10 nm can also be stronger than the single-particle bandwidth [288, 290]. It then suggests that the long-range nature of the Coulomb potential plays important roles in the moiré systems, and a more appropriate model for the carriers in the bilayer moiré pattern is the extended Hubbard model.

In monolayer TMDs and TMDs/hBN/TMDs systems where the moiré potential is absent, the Coulomb interaction can still dominate over the kinetic energy under a low-enough density and low enough temperature, leading to the spontaneous breaking of the continuous translation symmetry and formation of a Wigner crystal [291-293]. Early experiments in GaAs and AlAs semiconductor quantum wells usually rely on strong magnetic fields to quench the carrier's kinetic energy and enhance the localization. In layered TMDs systems, the combination of weak screening from the 2D geometry and large carrier effective masses can facilitate the realization of Wigner crystals without magnetic fields. In monolayer $MoSe_2$, experimental signatures of the Wigner crystal have been reported under an electron doping density lower than $3\times10^{11}$ $cm^{-2}$, which corresponds to a triangular lattice with a periodicity > 20 nm [291]. In $MoSe_2$/hBN/$MoSe_2$ vdW structures with an hBN spacer of 1 nm thickness, electrons in the two $MoSe_2$ layers can crystallize into triangular Wigner crystals that are interlocked by the interlayer Coulomb interaction to form a bilayer Wigner crystal [294]. Depending on the ratio between electron densities in the two layers, the bilayer Wigner crystal can have various lattice types with different thermodynamic stabilities.

As charge carriers' kinetic and interaction energies favor delocalization and localization respectively, their competition can lead to a phase transition between the crystal and liquid states. The increase of the kinetic energy with temperature results in the thermal melting of Wigner crystals. Meanwhile, the kinetic energy of the carriers increases faster than the Coulomb interaction when the density becomes higher, which leads to the quantum melting of Wigner crystals. A widely adopted rule to determine the critical temperature/density is the modified Lindemann criterion, which states that the melting occurs when the modified Lindemann ratio $\eta$ (the ratio between the fluctuation of the relative displacement for a pair of nearest-neighbor sites and the lattice period $\lambda$) exceeds some critical value $\eta_c$ [295-297]. Note that $\eta_c$ has significantly different values for the monolayer triangular and bilayer honeycomb Wigner crystals, as in the latter case the nearest-neighbor sites locate in opposite layers without spatial overlap. The numerical estimation is $\eta_c \approx 0.3$ for the triangular lattice [296], whereas the experimentally obtained modified Lindemann ratio is $\eta \approx 0.56$ for the bilayer honeycomb Wigner crystal [294]. Besides the modified Lindemann criterion which could be oversimplified for some systems, microscopic numerical calculations have also been carried out to understand the solid-liquid phase transition in TMDs moiré patterns [298-302].

The modified Lindemann ratio is determined by the spatial fluctuation of the crystal site, which is related to its low-energy excitations. In the above-mentioned various electronic crystals (or Wigner-Mott insulators), carriers localized at different sites are strongly correlated through the long-range Coulomb interaction, whose low-energy excitations are thus dominated by the collective vibration of the crystal sites (i.e., phonon modes). In early works on the triangular-type Wigner crystal in GaAs and AlAs semiconductor quantum wells, the emergence of such phonon modes is often viewed as a signature of the crystalline order for the carriers [303-305]. In moiré patterned TMDs systems, a series of electronic crystals not restricted to the triangular-type are stabilized by the confinement from the moiré potential, whose phonon modes are distinct from the well-studied triangular-type Wigner crystal [306]. Compared to ionic crystals, electronic crystals have much weaker interaction strengths but much smaller effective masses, resulting in the same order of magnitudes (tens meV) for their phonon frequencies. Furthermore, the small carrier effective mass brings great tunability to phonons of electronic crystals through external fields. It has been suggested that chiral phonons with finite magnetizations and large Berry curvatures can emerge in triangular- and honeycomb-type electronic crystals under time-reversal symmetry breaking by a magnetic field, or in honeycomb crystals under inversion symmetry breaking by an out-of-plane electric field [306]. These chiral phonons are located at the superlattice mBZ center $\mathbf{\Gamma}_m$ or corners $\pm\mathbf{K}_m$, and are optically addressable through an infrared absorption or two-photon Raman process.

Bringing a moiré superlattice and a pristine monolayer adjacent to each other also leads to an interesting possibility of realizing interlayer excitonic insulators. This has been explored in layered structures consisting of a $WSe_2$ monolayer separated from a $WS_2/WSe_2$ moiré heterobilayer by thin hBN spacers [284, 285]. One can start with doping the moiré bilayer at hole filling factor 1 which is then in a Mott insulator state, while leaving the $WSe_2$ monolayer empty. With the application of an out-of-plane electric field, holes can be moved from the moiré bilayer to the $WSe_2$ monolayer. This is equivalent to adding equal number of electrons and holes respectively to the moiré bilayer and $WSe_2$ monolayer, which can bound together in the moiré landscape and can be described as interlayer excitons. Alternatively, such states can also form in angle-aligned monolayer $WS_2$/multilayer $WSe_2$ moiré superlattices [286, 307], where interlayer coupling at the $WS_2/WSe_2$ interface induces a moiré superlattice only in the first $WSe_2$ layer adjacent to the $WS_2$ layer, leaving other $WSe_2$ layers nearly unaffected. In both types of systems, formation of interlayer exciton insulators is observed up to a critical exciton density, where the system is electronically insulating while interlayer excitons can flow as in a liquid [284-286, 307].

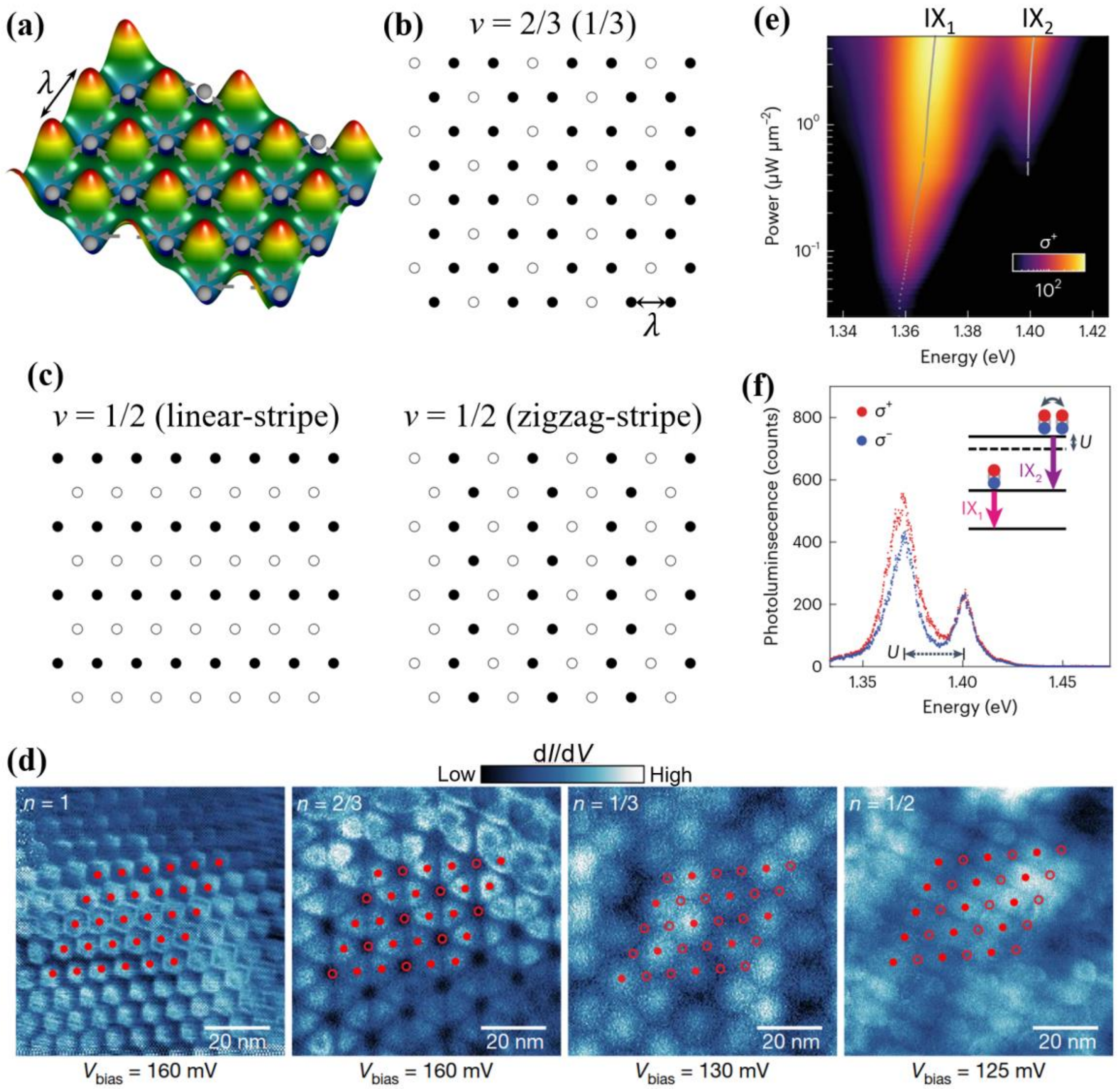

**Figure 8.** (a) A schematic illustration of a moiré pattern under a filling factor $\nu = 1$, which forms a triangular lattice with a wavelength $\lambda$. The colored surface shows the moiré potential landscape, and gray balls correspond to carriers trapped at the potential minima. The strong on-site repulsion leads to the formation of a Mott insulator, which is further stabilized by inter-site Coulomb repulsions (gray arrows). (b) The generalized Wigner crystal in a honeycomb lattice configuration under a fractional filling $\nu$ = 2/3. Solid (empty) dots correspond to occupied (unoccupied) moiré potential minima. Under $\nu = 1/3$, the notation of the occupied and unoccupied sites is switched, resulting in a triangular lattice. (c) Charge-order configurations of linear- and zigzag-stripe lattices under $\nu = 1/2$. (d) The local STM d$I$/d$V$ images of various generalized Wigner crystals, visualizing real-space lattice configurations under electron fillings $n$ = 1, 2/3, 1/3 and 1/2 (from left to right). Solid (empty) red dots denote moiré sites occupied (unoccupied) by electrons. (e) The photoluminescence spectra from $X_{inter}$ under a continuously increased excitation power. The lower-energy peak $IX_1$ is the photoluminescence from $X_{inter}$ at singly occupied moiré sites. The higher-energy peak $IX_2$, which appears only above a threshold excitation power, is from two-exciton spin-singlet states in doubly occupied moiré sites. (f) Polarization-resolved photoluminescence spectra under $\sigma$+-polarized optical excitation. The emission circular polarization is finite for peak $IX_1$, but vanishes for peak $IX_2$. The inset shows the origins of the two peaks. (a) From [297], reprinted with permission from Chinese Physical Society and IOP Publishing. (d) From [287], reprinted with permission from Springer Nature. (e, f) From [269], reprinted with permission from Springer Nature.

## 3.2. Excitonic crystals from the exciton-exciton interaction

Besides the correlated electronic crystals, the repulsive interaction between $X_{inter}$ can also lead to the formation of correlated excitonic crystals. The dipole-dipole interaction between two $X_{inter}$ (scales as $\propto 1/r^3$ when separated by a large distance $r$) is much weaker than the Coulomb interaction between charged carriers. Thus the excitonic crystal is expected to form under a high-enough density when the moiré potential is absent, in sharp contrast to the electronic Wigner crystal. In experiments, due to the exciton-exciton annihilation whose efficiency increases with the exciton density [308-310] and the fact that a phase transition from bound excitons into electron/hole plasmas occurs when the density is above the so-called Mott threshold (in the order of $10^{12}$-$10^{13}$ cm$^{-2}$ [311, 312]), a high exciton density becomes hard to realize. The moiré potential thus plays an essential role for suppressing the kinetic energy and realizing the excitonic crystal phase. Experiments in heterobilayer TMDs moiré patterns have shown that additional higher-energy emission peaks from $X_{inter}$ emerge when the exciton filling factor $\nu_X$ reaches an integer (1, 2, …), signifying the realization of excitonic Mott insulators (Fig. 8(e)) [269-273, 313, 314]. The ~ 30 meV peak splitting corresponds to the on-site repulsion $U$ between two $X_{inter}$ occupying the same moiré potential minimum, which is in a two-exciton spin-singlet state with the two electrons (and two holes) in opposite valleys. This results in the vanishing circular polarization of emissions from doubly occupied moiré sites (Fig. 8(f)). Besides these optical signatures, the formed excitonic Mott insulator also shows a strongly suppressed diffusion length, which further confirms its insulating behavior [272, 314].

Unlike the fermionic electrons and holes, excitons can be viewed as bosons when the inter-exciton distance is much larger than the Bohr radius $a_B$. $X_{inter}$ in the moiré pattern thus can be used to simulate the Bose-Hubbard model, with the exciton density tunable through the optical excitation power. Furthermore, a two-component Hubbard model with both fermionic and bosonic species can be realized when carriers and $X_{inter}$ are both injected into the moiré patterned bilayer, with the filling factor $\nu$ of the fermion ($\nu_X$ of the boson) controllable by doping through a gate voltage (optical pumping). It is shown that the strong on-site repulsion between charge carriers, between $X_{inter}$, as well as between a carrier and an $X_{inter}$ can give rise to a mixed correlated insulator when the total filling factor $\nu + \nu_X = 1$ [270, 272, 315].

# 4. Topological and quantum geometric properties of R-type twisted homobilayer TMDs and quantum anomalous Hall effects

In most heterobilayer TMDs where the interlayer hopping of carriers is suppressed by the large band offset, the layer-hybridization is negligible and the moiré potential usually serves as a modulated electrostatic landscape. While in R-type twisted homobilayer TMDs, carrier interlayer hopping and the resultant layer-hybridization become significant, which underlies the emergence of quantum geometric properties for the low-energy carriers. Importantly, topological ‘flat’ bands can emerge [127], leading to the groundbreaking experimental observations of the integer and fractional quantum anomalous Hall effects in R-type twisted bilayer $MoTe_2$ at zero external magnetic fields [99-102].

### 4.1. Topological properties of R-type twisted homobilayer TMDs and real-space quantum geometry

Figure 9(a, b) present some representative energy bands of spin-up electrons in R-type twisted bilayer $MoTe_2$ ($tMoTe_2$) using the continuum model and parameters from Ref. [127] (cf. Table 4), where isolated topological bands with small band widths and nonzero Chern numbers emerge. The electric and topological properties can be understood from the interplay of an effective scalar potential and a real-space Berry curvature (pseudo-magnetic field) in the adiabatic limit, which is applicable to long-period moiré with smooth spatial variations [42, 43].

The Hamiltonian for spin up electrons in the top valence bands reads $H = -\frac{\hat{\mathbf{p}}^2}{2m} + \mathcal{U} = -\frac{\hat{\mathbf{p}}^2}{2m} + \begin{pmatrix} V_t & \tilde{T} \\ \tilde{T}^* & V_b \end{pmatrix}$, where $\tilde{T} = Te^{i(\boldsymbol{\kappa}_b - \boldsymbol{\kappa}_t)\cdot\mathbf{r}}$ with $T$ defined in Eq. (9), and $V_{t,b}$ are given by Eq. (11). This Hamiltonian is equivalent to Eq. (8) by a unitary transformation, the non-moiré periodic (due to the $e^{\pm i(\boldsymbol{\kappa}_b - \boldsymbol{\kappa}_t)\cdot\mathbf{r}}$ factors in $\tilde{T}$ and $\tilde{T}^*$) but $\hat{C}_3$-symmetric potential $\mathcal{U}$ is important to preserve the $\hat{C}_3$ symmetry of the pseudo-magnetic field. $\mathcal{U}$ can be rewritten as an effective Zeeman term $\mathcal{U} = \mathcal{V}_0 + \hat{\boldsymbol{\sigma}} \cdot \vec{\mathcal{V}}$, where $\hat{\boldsymbol{\sigma}} = (\hat{\sigma}_x, \hat{\sigma}_y, \hat{\sigma}_z)$ denotes the layer pseudospin operator. In R-type twisted homobilayer TMDs, the direction vector $\vec{n} = \vec{\mathcal{V}}/\mathcal{V}$ forms a skyrmion texture with vortex and antivortex in the moiré unit cell (Fig. 9(c)), which points out-of-plane oppositely at $R_h^X$ and $R_h^M$ stackings consistent with the electric polarization discussed in Sec. 1.4.

The eigenstate of electrons can be expanded as $|\Psi\rangle = \psi_+|\chi_+\rangle + \psi_-|\chi_-\rangle$, where $\psi_\pm$ and $|\chi_\pm\rangle$ characterizes the CoM and layer internal degree of freedom, respectively. Here $|\chi_\pm\rangle$ are the local eigenvectors of $\mathcal{U}$ with energy $\mathcal{V}_0 \pm \mathcal{V}$, satisfying $\langle\chi_\pm|\hat{\boldsymbol{\sigma}}|\chi_\pm\rangle = \pm\vec{n}$. In the adiabatic limit, the eigenstate of low-energy electrons is approximated as $|\Psi\rangle \approx \psi_+|\chi_+\rangle$, i.e., the layer pseudospin of the electrons aligns with $\vec{n}$ in the adiabatic motion (Fig. 9(c)). Meanwhile, the CoM motion is governed by

$$\left[-\frac{1}{2m}(\hat{\mathbf{p}} + e\mathbf{A})^2 + \mathcal{V}_0 + \mathcal{V} + \mathcal{G}\right]\psi_+ = E\psi_+, \tag{23}$$

which describes electrons moving in an emergent magnetic field

$$\boldsymbol{\mathcal{B}} = \nabla \times \mathbf{A} = \frac{\hbar}{2e}\vec{n} \cdot (\partial_x\vec{n} \times \partial_y\vec{n})\hat{\mathbf{z}}, \tag{24}$$

and an overall scalar potential $\mathcal{V}_0 + \mathcal{V} + \mathcal{G}$ with $\mathcal{G} = -\frac{\hbar^2}{8m}(\nabla\vec{n})^2$ the geometric scalar potential [316]. The $\boldsymbol{\mathcal{B}}$ field is opposite in the two valleys due to time-reversal symmetry, its flux in a unit cell is determined by the solid angle enclosed by $\vec{n}$. One also identifies that the real-space geometric quantities $\mathcal{G}$ and $\boldsymbol{\mathcal{B}}$ are respectively connected to the real (metric) and imaginary (curvature) parts of the quantum geometric tensor of the internal degree of freedom: $T_{ij} = \langle\partial_i\chi_+|(1 - |\chi_+\rangle\langle\chi_+|)|\partial_j\chi_+\rangle = [\partial_i\vec{n}\cdot\partial_j\vec{n} + \mathrm{i}\,\vec{n}\cdot(\partial_i\vec{n} \times \partial_j\vec{n})]/4$ , where $i, j = x, y$ [317]. This adiabatic approximation provides an intuitive understanding for the topological properties of the moiré bands as detailed next. And it has also been employed [318, 319] to explain the presence of a 'magic' angle ($\sim 1.4°$) in the continuum model of twisted homobilayer TMDs [129, 142], where the first moiré band has a minimal bandwidth.

Figure 9(d, e) shows the distribution of $\mathcal{B}$ and $\mathcal{G}$, which share similar profiles. There exists exactly one magnetic flux quantum in each unit cell, as determined by the $4\pi$ solid angle of the $\vec{n}$ texture (Fig. 9(c)). The integral of $\mathcal{G}$ in the unit cell is also found to be independent of $\theta$. Thus

the intensity of $\mathcal{B}$ and $\mathcal{G}$ scales inversely with the unit cell area. Figure 9(f) shows the profile of $\mathcal{V}_0 + \mathcal{V}$ with the largest values around $R_h^X$ and $R_h^M$ stackings. Being independent of spatial gradients, the intensity of $\mathcal{V}_0 + \mathcal{V}$ is independent of $\theta$ in a rigid moiré (typically $\mathcal{O}(10)$ meV). Consequently, $\mathcal{G}$ is negligible compared to $\mathcal{V}_0 + \mathcal{V}$ when $\theta$ is small, while $\mathcal{G}$ increases rapidly with $\theta$ and modifies the overall potential landscape $\mathcal{V}_0 + \mathcal{V} + \mathcal{G}$ when $\theta$ is large [42, 43].

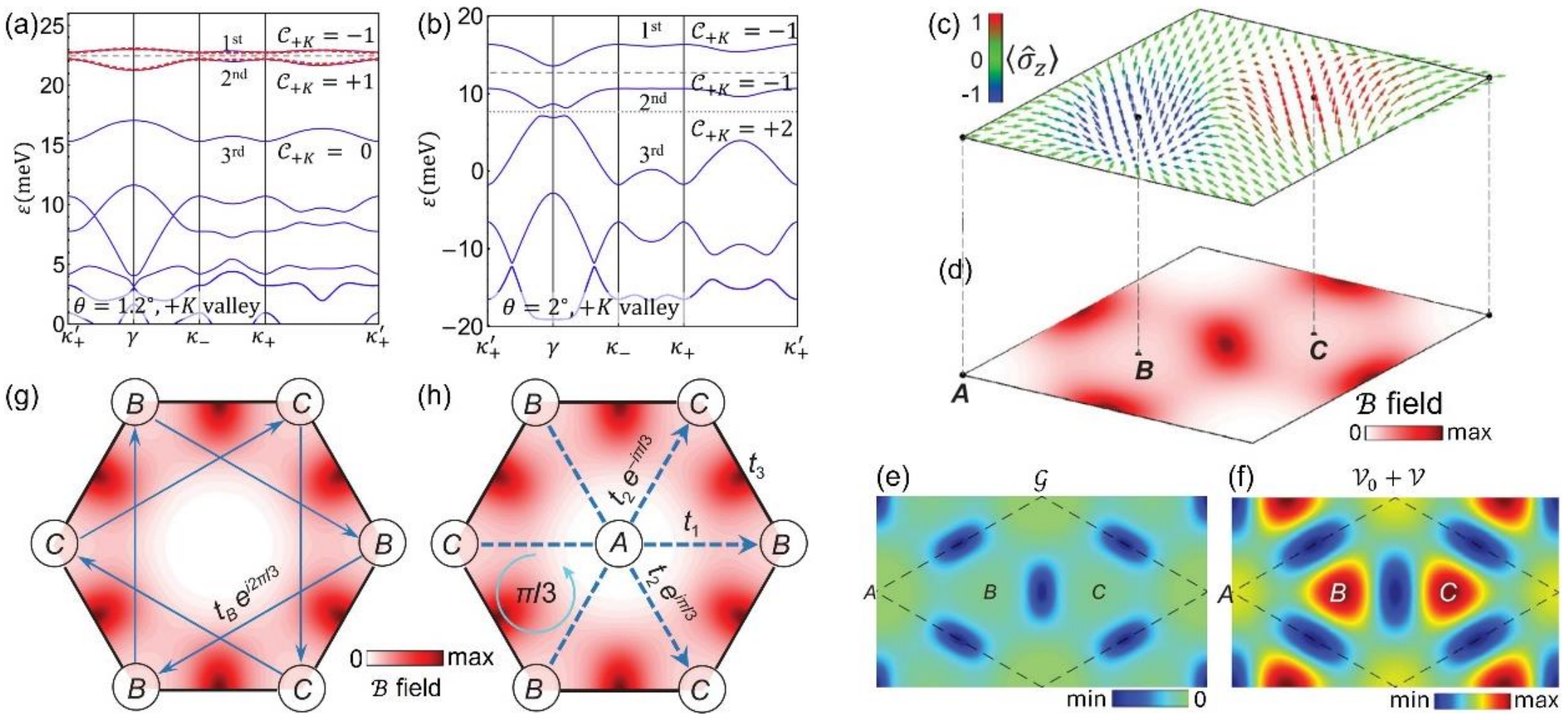

**Figure 9.** (a, b) Energy bands of spin-up electrons in R-type twisted bilayer $MoTe_2$ with $\theta = 1.2°$ and $\theta = 2°$. (c) Schematics of layer pseudospin texture $\langle \chi_+ | \hat{\boldsymbol{\sigma}} | \chi_+ \rangle = \vec{n}$ for spin-up electrons in R-type t$MoTe_2$. (d–f) Distribution of the pseudo-magnetic field $\mathcal{B}$ (d), geometric scalar potential $\mathcal{G}$ (e), and $\mathcal{V}_0 + \mathcal{V}$ (f). (g, h) Schematics of the three-orbital (g) and two-orbital (h) flux lattice model. Here A, B and C high-symmetry points correspond to $R_h^h$, $R_h^X$ and $R_h^M$, respectively. (a) Reproduced with permission from Ref. [127]. (c-h) adapted from Ref. [42].

The coexistence of the pseudo-magnetic field $\boldsymbol{\mathcal{B}}$ and the scalar potential $\mathcal{V}_0 + \mathcal{V} + \mathcal{G}$ realizes an effective flux lattice with lattice sites located at the scalar potential maxima that can trap holes. When $\theta$ is small, an effective honeycomb lattice with sites at $R_h^X$ and $R_h^M$ stacking emerges (Fig. 9(g)), electrons on adjacent sites are localized in opposite layers as determined by the layer pseudospin texture. Recent STS measurements have found signatures of such layer pseudospin texture [320, 321]. In the case of small $\theta$, the first two moiré bands of spin-up electrons exhibit Chern numbers of $\pm 1$ respectively (Fig. 9(a)). Such topological properties can be attributed to an effective Haldane model [322] realized on the fluxed honeycomb lattice [42, 127]. As $\theta$ is enlarged, the effective lattice becomes triangular (due to the increase of $\mathcal{G}$) with three inequivalent sites at A, B and C with $R_h^h$, $R_h^X$ and $R_h^M$ stacking registries, respectively (Fig. 9(h)). This structural change is accompanied by a topological phase transition, where the first three moiré bands of spin-up have Chern numbers of $-1$, $-1$, and 2 respectively (Fig. 9(b)).

A three-orbital tight-binding model for low-energy holes on a fluxed lattice (Fig. 9(h)) can be constructed to characterize the electronic and topological properties of the first three moiré bands [42]:

$$\hat{H}_{3O} = \sum_l \left( \varepsilon_A \hat{A}_l^\dagger \hat{A}_l + \varepsilon_B \hat{B}_l^\dagger \hat{B}_l + \varepsilon_C \hat{C}_l^\dagger \hat{C}_l \right) - \sum_{\langle l,m \rangle} \left( t_1 e^{i\phi_1^{l,m}} \hat{A}_l^\dagger \hat{B}_m + t_2 e^{i\phi_2^{l,m}} \hat{A}_l^\dagger \hat{C}_m + t_3 \hat{C}_l^\dagger \hat{B}_m + h.c. \right), \tag{25}$$

where $\hat{A}^l$, $\hat{B}^l$ and $\hat{C}^l$ are annihilation operators for the orbital at A, B and C sites in the $l$-th unit cell, $\phi_{1,2}^{l,m}$ is the hopping phase along the path of nearest neighbors $\langle l, m\rangle$ as determind by the underlying pseudo-magnetic field. The potential landscape dictates $\varepsilon_B = \varepsilon_C$ and $t_1 = t_2 > t_3$, in which a smaller $t_3$ is caused by the scalar potential barrier between B and C sites and the large and opposite layer-polarizations of orbitals at B and C. The relative intensity of $\varepsilon_{B,C}$ and $\varepsilon_A$ depends on $\theta$, as it changes the magnitude of $\mathcal{G}$. This three-orbital model can well reproduce the first three moiré bands and their Chern numbers from the continuum model or DFT calculations. The validity of such a three-orbital tight-binding model is also corroborated by some recent studies by constructing the Wannier orbitals [139, 323].

In the case of small twist angles, as $\varepsilon_A > \varepsilon_{B,C}$, a reduced two-orbital tight-binding model based on the B and C orbitals (Fig. 9(g)) can be constructed to account for the first two moiré bands [42, 127]

$$\hat{H}_{2O} = \sum_l \left(\varepsilon_B \hat{B}_l^\dagger \hat{B}_l + \varepsilon_C \hat{C}_l^\dagger \hat{C}_l\right) - \sum_{\langle l,m\rangle} \left(t_3 \hat{C}_l^\dagger \hat{B}_m + h.c.\right) \\ - \sum_{\langle\langle l,m\rangle\rangle} \left(t_B e^{i\phi_B^{l,m}} \hat{B}_l^\dagger \hat{B}_m + t_C e^{i\phi_C^{l,m}} \hat{C}_l^\dagger \hat{C}_m + h.c.\right), \tag{26}$$

where hopping processes up to the next nearest neighbors $\langle\langle l, m\rangle\rangle$ have been considered and $\phi_{B,C}^{l,m}$ takes values of $\pm 2\pi/3$ depending on the hopping direction. This model can capture the first two moiré bands and their Chern numbers reasonably well, although long-range hopping processes beyond the next-nearest neighbors might be needed to satisfactorily reproduce the moiré bands from the continuum model or DFT when the twist angle is moderate (e.g., $\gtrsim 1°$) [129]. This two-orbital model effectively realizes the Haldane model [322] in each valley, and the Kane-Mele model [324] with two valleys. Recent experiments have reported signatures of the quantum spin Hall in tMoTe$_2$ [325] and tWSe$_2$ [326].

The layer pseudospin texture, which determines the real-space geometric quantities thus the properties of the moiré, can be tuned with an out-of-plane electric field $E_\perp$ and an in-plane magnetic field $\mathbf{B}_\parallel$. The $E_\perp$ field shifts the intralayer potentials $V_t$ and $V_b$, while the $\mathbf{B}_\parallel$ field affects the phase of interlayer hopping as $\tilde{T}' = \tilde{T} e^{i\zeta}$, where $\zeta = \frac{eD}{\hbar} \mathbf{B}_\parallel \cdot (\hat{\mathbf{z}} \times \mathbf{r})$ with $D$ the interlayer distance approximated as constant. As such, $E_\perp$ changes the polar angle of the layer pseudospin texture $\vec{n}$, while $\mathbf{B}_\parallel$ affects the azimuthal angle [43, 327]. Thus both geometric quantities $\boldsymbol{\mathcal{B}}$ and $\mathcal{G}$ can be tuned with $E_\perp$ and $\mathbf{B}_\parallel$ in the adiabatic limit. There exists a critical value of $E_\perp$, beyond which $\vec{n}$ encloses a vanishing solid angle thus null magnetic flux of $\boldsymbol{\mathcal{B}}$ in a moiré unit cell [42, 43]. The change of scalar potential landscape, $\boldsymbol{\mathcal{B}}$ field and the magnetic flux by tuning $E_\perp$ can lead to drastic changes in the electronic and topological properties of the moiré. Especially, the first moiré band can become topologically trivial in the presence of a large $E_\perp$ with the electrons strongly localized in one layer on the triangular sites composed of either $R_h^X$ or $R_h^M$ stacking depending on the direction of $E_\perp$ [328]. On the other hand, the emergence of an effective out-of-plane magnetic field induced by $\mathbf{B}_\parallel$ shows explicitly that $\mathbf{B}_\parallel$ can affect the orbital motion of electrons in 2D layered materials through interlayer coupling. A Hall current then can be expected when $\mathbf{B}_\parallel$ is applied to a twisted bilayer, realizing the planar Hall effect in the 2D system even in the absence of spin-orbit coupling [329].

It is interesting to notice that the layer pseudospin texture $\vec{n}(\mathbf{r}, t)$ can gain both spatial and temporal dependence when a temporally varying $E_\perp(t)$ is applied [327]. This leads to a pseudo-electric field $E_i = \frac{\hbar}{2e} \vec{n} \cdot (\partial_i \vec{n} \times \partial_t \vec{n})$ satisfying the Faraday's law of induction $\nabla \times \mathbf{E} + \partial_t \mathbf{B} =$ 0, where $i = x, y$. The emergent electromagnetic fields can be employed to manipulate local valley/spin currents in the moiré.

We remark that the adiabatic limit can be generalized to systems with a larger dimension in the internal degree of freedom (e.g., multilayer structures) and achieve non-Abelian pseudo-magnetic fields by projecting onto a subspace spanned by more than one internal states [330, 331]. In some twisted bilayers with non-uniform strain, the interplay of strain pseudo-magnetic field and interlayer coupling can also behave as an effective non-Abelian gauge field [332].

## 4.2. Quantum geometry in hybrid space and chirality-enabled time-reversal-even Hall effects

Apart from layer pseudospin texture in the real space, interlayer hybridization in a twisted bilayer also introduces layer pseudospin textures in the momentum space. The latter can be characterized by $l^z_{n,m}(k) = \langle u_{n\boldsymbol{k}} | \hat{\sigma}_z | u_{m\boldsymbol{k}} \rangle$, where $|u_{n\boldsymbol{k}}\rangle$ denotes the periodic part of the Bloch state on band $n$ with crystal wave vector $\mathbf{k}$. A nontrivial distribution of $l^z_{n,m}(\mathbf{k})$ due to interlayer hybridization in the chiral geometry makes twisted homobilayers unique platforms for exploring novel Hall effects of quantum geometric origins in the presence of time-reversal symmetry. Here we discuss two examples: (i) Extrinsic layer-counterflow linear Hall effect; (ii) Intrinsic crossed-nonlinear dynamical Hall effect.

### 4.2.1. Extrinsic linear dipole Hall effect

Generally, the current flow in a bilayer can be decomposed into a layer-symmetric (or monopole) part and a layer-antisymmetric (or interlayer dipole) part. The former corresponds to the overall charge current, whereas the latter is a pure electric dipole current with charge counterflows in the two layers. Here we focus on the Hall currents. The Onsager reciprocal relation for the electric conductivity [333] dictates a null charge Hall current in time-reversal symmetric systems within the linear response theory. However, no such restriction exists for the dipole Hall current (Fig. 10(a)), which can exist even in a non-magnetic environment [153, 334, 335].

In time-reversal symmetric bilayers, the dipole Hall current can only come from extrinsic origins within the linear response theory. Symmetry analysis reveals that the dipole Hall current is allowed only if the bilayer geometry is chiral, whose Hamiltonian $H$ lacks inversion and mirror symmetries [153, 334, 335]. From semiclassical transport theory, the dipole Hall conductivity reads $\sigma_H^{\mathrm{dH}} = \frac{e^2}{\hbar}\tau W^{\mathrm{dH}}$, where $\tau$ is the relaxation time (not to be confused with the valley index), and

$$W^{\mathrm{dH}} = -\frac{\hbar}{2}\sum_n \int \frac{d\mathbf{k}}{(2\pi)^2}\frac{\partial f_0}{\partial \varepsilon_n}\left[\mathbf{v}_n(\mathbf{k}) \times \mathbf{v}_n^{\mathrm{dH}}(\mathbf{k})\right]_z \tag{27}$$

is time-reversal-even and intrinsic to the electronic structure of the bilayer [153, 336]. Here $\mathbf{v}_n = \langle u_{n\mathbf{k}} | \hat{\mathbf{v}} | u_{n\mathbf{k}} \rangle$ is the group velocity on the $n$-th band with $\hat{\mathbf{v}} = \partial_{\mathbf{k}} H(\mathbf{k})/\hbar$ and $\varepsilon_n$ is the band energy, $\mathbf{v}_n^{\mathrm{dH}} = \langle u_{n\mathbf{k}} | \frac{1}{2}\{\hat{\mathbf{v}}, \hat{\sigma}_z\} | u_{n\mathbf{k}} \rangle$ is the dipole velocity, $f_0$ is the equilibrium Fermi-Dirac distribution function. The dependence on $\partial f_0/\partial \varepsilon_n$ implies that the dipole Hall current is a Fermi surface property, and it vanishes if the interlayer hybridization is absent. One can also rewrite the dipole Hall conductivity as

$$\sigma_H^{\mathrm{dH}} = \frac{e^2}{\hbar}\tau \sum_n \int \frac{d\mathbf{k}}{(2\pi)^2} f_0\, \omega_n^{\mathrm{dH}}(\mathbf{k}), \tag{28}$$

which measures the geometric quantity

$$\omega_n^{\mathrm{dH}}(\mathbf{k}) = \frac{1}{2}\left[\partial_{\mathbf{k}} \times \mathbf{v}_n^{\mathrm{dH}}(\mathbf{k})\right]_z = \hbar\,\mathrm{Re}\sum_{n'\neq n} \frac{\left[\mathbf{v}_{nn'}(\mathbf{k}) \times \mathbf{v}_{n'n}^{\mathrm{dH}}(\mathbf{k})\right]_z}{\varepsilon_n(\mathbf{k}) - \varepsilon_{n'}(\mathbf{k})} \tag{29}$$

of the occupied states, and interband matrix elements of the velocity and dipole velocity operators are involved in the last equality. Equation (28) formally resembles the intrinsic $k$-space Berry curvature contribution to the Hall conductivity [337]. A large $\omega_n^{\mathrm{dH}}(\mathbf{k})$ requires strongly layer-hybridized electronic states (i.e., $|\langle \mathrm{u}_{n\mathbf{k}}|\hat{\sigma}_z|\mathrm{u}_{n\mathbf{k}}\rangle| \ll 1$) in a chiral bilayer, and it tends to concentrate around regions on the bands with small gaps.

The dipole Hall current manifests as opposite Hall currents in the two layers. The Hall conductivity that quantifies the charge Hall current in the top or bottom layer is given by $\sigma_H^t = -\sigma_H^b = \sigma_H^{\mathrm{dH}}/2$. It is predicted to be sizeable in twisted R-stacking homobilayer TMDs (Fig. 10(b-d)), where the magnitude can be tuned by an out-of-plane electric field and the current direction can be flipped by reversing the twist angle. The Hall counterflows in the two layers also induce an in-plane magnetic moment, which underlies the emergence of circular dichroism in twisted bilayers [334, 338, 339].

Note that the above discussions focus on time-reversal symmetric scenarios, where no net linear charge Hall current flows in the bilayer, and the dipole Hall current can only be of extrinsic origin (Eq. (28)). One might, however, notice that electron interaction can spontaneously break time-reversal symmetry at certain filling factors when the moiré bands of a twisted bilayer are narrow (see Sec. 4.3). In this latter case, a net intrinsic charge Hall current contributed by the $k$-space Berry curvature can emerge; furthermore, there might also exist an intrinsic dipole Hall current originated from the so-called **k**-space dipole Berry curvature [340]. This intrinsic dipole Hall effect, however, has different symmetry requirements compared to the extrinsic one (in addition to time-reversal symmetry breaking), and its presence sometimes requires applying an out-of-plane electric field to the moiré.

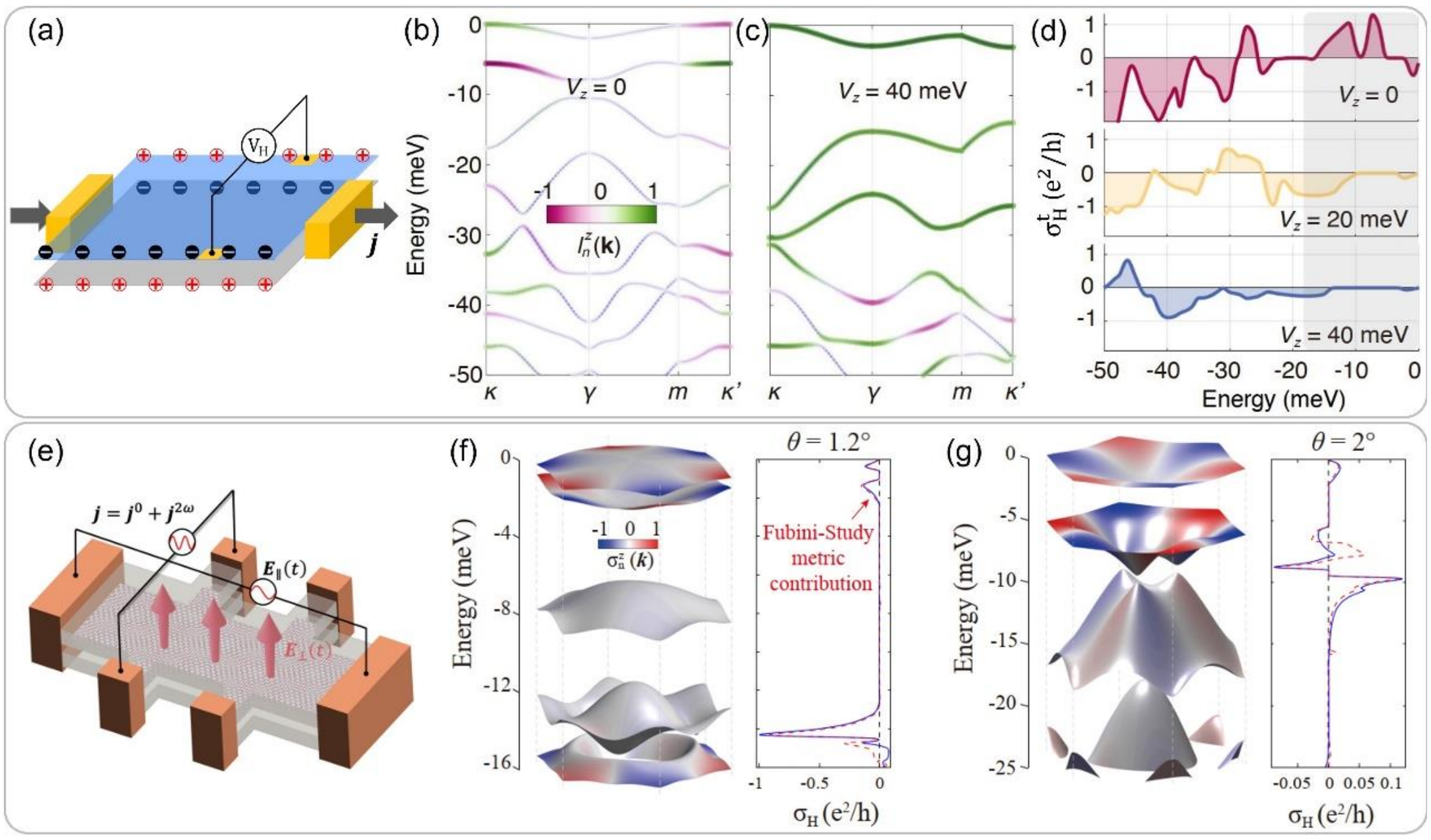

**Figure 10.** (a–d) Interlayer dipole Hall effect. (a) Schematics of the dipole Hall current in the two layers when an in-plane electric field is applied to the twisted bilayer. (b, c) Valence bands for spin-up electrons in R-type 2° $tMoTe_2$ in the presence of different interlayer bias ($V_z$). The color denotes value of $\langle u_{nk}|\sigma_z|u_{nk}\rangle$. (d) Hall conductivity in the top layer for different $V_z$. (e-g) Crossed-nonlinear dynamical Hall effect. (e) Schematics of the experimental setup. (f, g) Valence band structure and intrinsic Hall

conductivity with respect to the in-plane field for spin-up electrons in R-type $tMoTe_2$ at twist angle (f) $\theta = 1.2°$ and (g) $\theta = 2°$. The color on the energy bands denotes value of $\langle u_{n\mathbf{k}}|\sigma_z|u_{n\mathbf{k}}\rangle$, and the red dashed curve in the conductivity plots represents the Fubini-Study metric contribution. (a-d) Reprinted from Ref. [153]. (e-g) Reprinted from Ref. [155].

### 4.2.2. Intrinsic crossed-nonlinear dynamical Hall effect

Another novel transport phenomenon that is unique to chiral layered structures with strong interlayer coupling is the so-called crossed-nonlinear dynamical Hall effect – an intrinsic Hall effect that is linear in an in-plane $\mathbf{E}_{\parallel}$ and an out-of-plane AC electric field $\mathbf{E}_{\perp}(t)$, see schematics in Fig. 10(e) [155]. From the semiclassical transport theory, the crossed-nonlinear charge Hall current density in a bilayer is given by

$$\mathbf{j} = \chi^{\text{int}} \dot{\mathbf{E}}_{\perp} \times \mathbf{E}_{\parallel}, \tag{30}$$

where the response coefficient reads

$$\begin{aligned} \chi^{\text{int}} &= \frac{e^2}{\hbar} \sum_n \int \frac{d\mathbf{k}}{(2\pi)^2} f_0 [\partial_{\mathbf{k}} \times \boldsymbol{\mathcal{P}}_n(\mathbf{k})]_z \\ &= -e^2 \sum_n \int \frac{d\mathbf{k}}{(2\pi)^2} \frac{\partial f_0}{\partial \varepsilon_n} [\mathbf{v}_n \times \boldsymbol{\mathcal{P}}_n(\mathbf{k})]_z \end{aligned}. \tag{31}$$

Here $\boldsymbol{\mathcal{P}}_n$ is a band geometric quantity, i.e., the interlayer Berry connection polarizability (BCP)

$$\boldsymbol{\mathcal{P}}_n(\mathbf{k}) = 2\hbar^2 \text{Re} \sum_{m \neq n} \frac{p^{nm}(\mathbf{k}) \mathbf{v}^{mn}(\mathbf{k})}{[\varepsilon_n(\mathbf{k}) - \varepsilon_m(\mathbf{k})]^3}. \tag{32}$$

Its numerator involves the interband matrix elements of the velocity operator $\hat{\mathbf{v}}$ and the interlayer dipole operator $\hat{p} = eD\hat{\sigma}_z$, where $e$ is the electron charge and $D$ is the interlayer distance taken as constant. All the other quantities are the same as those defined in the previous section. A large interlayer BCP, thus $\chi^{\text{int}}$, requires strongly interlayer hybridized electronic states in a chiral geometry.

The expression of $\boldsymbol{\mathcal{P}}_n$ resembles that of the Fubini-Study metric (FSM) [341], it is thus noted that the effect is closely related to the hybrid quantum metric in the $(\mathbf{k}, E_{\perp})$ space. Specifically, $\boldsymbol{\mathcal{P}}_n$ can be rewritten as

$$\boldsymbol{\mathcal{P}}_n = -2\hbar \frac{\boldsymbol{g}^n}{\varepsilon_n - \varepsilon_{\bar{n}}} - 2\hbar \sum_{m \neq n, \bar{n}} \frac{\mathbf{g}^{nm}(\varepsilon_m - \varepsilon_{\bar{n}})}{(\varepsilon_n - \varepsilon_m)(\varepsilon_n - \varepsilon_{\bar{n}})}, \tag{33}$$

where $\bar{n}$ labels the nearest neighbor to band $n$, $\mathbf{g}^{nm} = \text{Re}\langle \partial_{E_\perp} u_n | u_m \rangle \langle u_m | \partial_{\boldsymbol{k}} u_n \rangle$ and $\boldsymbol{g}^n = \sum_{m \neq n} \mathbf{g}^{nm} = \text{Re}\langle \partial_{E_\perp} u_n | (1 - |u_n\rangle\langle u_n|) | \partial_{\mathbf{k}} u_n \rangle$ is the FSM in the $(\mathbf{k}, E_{\perp})$ space. The first term of Eq. (33) can be considered as the FSM contribution, while the second term contains effects from additional remote bands (ARB) on $n$. Similarly, $\chi^{\text{int}}$ can be separated into an FSM term and ARB contribution

$$\chi^{\text{int}} = 2\hbar e^2 \sum_n \int \frac{d\mathbf{k}}{(2\pi)^2} \frac{\partial f_0}{\partial \varepsilon_n} \frac{(\mathbf{v}_n \times \boldsymbol{g}^n)_z}{\varepsilon_n - \varepsilon_{\bar{n}}} + \chi_{\text{ARB}}^{\text{int}}, \tag{34}$$

where the expression of the latter [155] is not given explicitly here. In two-band systems, only the FSM term exists; while in multi-band cases, the FSM contribution dominates if band $n$ is close to another band but well-separated from the others.

Interestingly, when $\mathbf{E}_\parallel$ is also AC and has the same frequency as $E_\perp$, e.g., $\mathbf{E}_\parallel = \mathbf{E}_\parallel^0 \cos\omega t$ and $E_\perp = E_\perp^0 \cos(\omega t + \varphi)$ with $\varphi$ denoting a phase difference between the two electric fields, rectification and frequency doubling effects can be achieved. With such AC electric field configuration, the Hall current becomes

$$\mathbf{j} = \sigma_\mathrm{H}(\hat{\mathbf{z}} \times \mathbf{E}_\parallel^0)\sin\varphi + \sigma_\mathrm{H}(\hat{\mathbf{z}} \times \mathbf{E}_\parallel^0)\sin(2\omega t + \varphi), \tag{35}$$

where $\sigma_\mathrm{H} = \omega E_\perp^0 \chi^{\mathrm{int}}/2$ has the dimension of conductivity and quantifies the response with respect to $\mathbf{E}_\parallel$. Figure 10(f, g) present some numerical results by considering R-type $tMoTe_2$, the dominating FSM contribution is shown in red curves. Note that the on/off, direction, and magnitude of the rectified current can be controlled by varying $\varphi$. The direction of the total current can also be switched by reversing the direction of $E_\perp$ or the twist angle of the bilayer.

We remark that similar transport phenomena exist in magnetic materials. For example, the layer Hall effect in antiferromagnetic $MnBi_2Te_2$ [342-344] and the intrinsic dipole Hall effect in twisted bilayer $MoTe_2$ with spontaneous time-reversal symmetry breaking [340] are the counterparts of the extrinsic dipole Hall effect in Sec. 4.2.1; while the intrinsic nonlinear Hall effect in antiferromagnets [345-349] can be regarded as the counterpart of the crossed-nonlinear dynamical Hall effect. Thanks to the interlayer hybridization in the chiral geometry, twisted homobilayers allow the various Hall effects to emerge in a non-magnetic environment with band geometric origins unique to chiral layered structures (Secs. 4.2.1 and 4.2.2).

In Secs. 4.2.1 and 4.2.2, an in-plane electric field is employed for driving the Hall currents. It is interesting to note that one can replace the in-plane electric field by a temperature gradient to achieve the dipole Nernst effect [350] or the dynamical Nernst effect [156]. In addition to the novel Hall effects enabled by the structural chirality, the layer pseudospin degree of freedom also makes possible other intriguing phenomena in layered vdW materials. One example is the anomalous out-of-plane electric dipole and multipole generation induced by an in-plane electric field [351]. This effect also has band geometric origins in the Berry curvature and quantum metric of the hybrid space spanned by $(\mathbf{k}, E_\perp)$. The electronic and quantum geometric properties of the moiré superlattices can also manifest in various other transport behaviors, e.g., the giant nonlinear Hall effect in R-type $tWSe_2$ [352, 353] enabled by the Berry curvature dipole [354, 355]. Various Berry curvature related transport phenomena in moiré materials have been recently discussed in Ref. [356] and some nonlinear electronic and optical physics in moiré superlattices are reviewed in Ref. [357].

### 4.3. Quantum anomalous Hall effects in the moiré of TMDs

Upon filling the topological minibands in R-type twisted homobilayer TMDs (Sec. 4.1), the small bandwidth makes possible spontaneous spin/valley polarization by the Coulomb exchange interaction. Flat Chern band can form and lead to integer and fractional quantum anomalous Hall (denoted as IQAH and FQAH hereafter) effects in the absence of applied magnetic field. In this section, we will focus on some of the recent experimental developments in these directions.

#### 4.3.1. IQAH effect at $\nu = -1$ of H-type $MoTe_2/WSe_2$ heterobilayer moiré

The first experimental evidence of nontrivial band topology in the moiré of TMDs was reported in H-type nearly aligned $MoTe_2/WSe_2$ heterobilayers [130]. Specifically, the IQAH effect was observed at filling factor $\nu = -1$ (one hole per moiré unit cell) in the presence of a large $E_\perp$ field, as signaled by a quantized Hall resistance of $h/e^2$ accompanied by vanishing longitudinal resistance at zero magnetic field (Fig. 11(a, b)). Reducing $E_\perp$ turns the system into a Mott insulator, during which a charge gap closure was not observed. This observation of the IQAH effect is intriguing as the interlayer hopping is expected to be suppressed in a H-type $MoTe_2/WSe_2$ bilayer (Sec. 1.6): charge carriers near the band edges of the two layers have opposite spin alignments; meanwhile, the band alignment is type-I with both conduction and valence band edges from $MoTe_2$ and the interlayer valence band offset is large ($\gg$ 100 meV). The absence of IQAH effect under similar conditions in the R-type aligned $MoTe_2/WSe_2$ heterobilayers [278] is also intriguing.

The experimental observations stimulated a series of theory works that proposed different possible origins of the IQAH effect [68, 122, 358-366]. Ref. [68] introduced a continuum model for valence-band electrons of opposite spin in the two layers similar to that of the homobilayers (Eq. (8)), with a weak spin-flip interlayer hopping term. An effective monolayer model that only involves the valence band edge of $MoTe_2$ but incorporates the lattice relaxation effects as a periodic pseudo-magnetic field and a scalar potential was proposed in Ref. [122]. Other models that consider the massive Dirac fermions of $MoTe_2$ in a periodic scalar potential are also proposed [360, 366]. Despite starting from different models, many of the works found the IQAH state to be valley-polarized in nature [122, 358, 360-362, 364]. Later experimental results, on the other hand, suggested the observed IQAH to have a valley-coherent character [367]. And some alternative explanations based on intervalley excitonic states have been put forward [361, 363, 365].

Besides the IQAH, spin Hall effect [368, 369] and evidences of quantum spin Hall insulator at filling $\nu = -2$ with phase transition to Chern insulator in a small out-of-plane magnetic field [370] are reported in the H-type $MoTe_2/WSe_2$. This moiré system also provides a unique and versatile platform to explore the Kondo-related physics [371-377].

#### 4.3.2. FQAH effects in the first moiré band of $tMoTe_2$

FQAH effect is the counterpart of fractional quantum Hall effect at zero applied magnetic field, where a flat Chern band plays the role of a Landau level. The first experimental observations of this long quested topological phenomenon are recently reported in $tMoTe_2$ moiré superlattices at fractional fillings (e.g., $\nu = -2/3$ and $-3/5$) [99-102].

The ferromagnetism resulted from spontaneous valley polarization in R-type $tMoTe_2$ was first discovered in a 3.9° twisting sample with trion photoluminescence and reflective magnetic circular dichroism (RMCD) measurements [328]. Specifically, a drop in the photoluminescence intensity was spotted at filling factors $\nu = \pm 1$ (one electron/hole per moiré unit cell), signaling the formation of correlated insulators; and importantly, pronounced RMCD signal – signature of ferromagnetism – was observed in the vicinity of $\nu = -1$, which is accompanied by magnetic hysteresis loops when an out-of-plane magnetic field is applied (Figs. 11(c–e)). This ferromagnetic behavior is tunable with an out-of-plane electric field, which polarizes the charge carriers towards a specific layer and eventually eliminates the ferromagnetism (Fig. 11(d)).

The first experimental signatures of FQAH effects in $tMoTe_2$ are from optical measurements of the incompressibility. Following the discovery of ferromagnetism [328], further experiments on these systems in Ref. [99] reported that the coercive field and Curie temperature from the RMCD signals are clearly enhanced around filling $\nu = -1$ and $\nu = -2/3$ (Fig. 11(e)), and these two correlated states and another weaker one at $\nu = -3/5$ are incompressible as identified by the suppressed trion photoluminescence (Fig. 11(f)). More importantly, Ref. [99] found these states are topological with Chern number $C = -1, -2/3$ and $-3/5$ respectively, which are extracted from the optical fan diagram (Fig. 11(f)) with the charge gaps fitted to the Streda formula [378]. In another independent experiment [100], optical signatures of IQAH and FQAH states are also identified at $\nu = -1$ and $\nu = -2/3$ in the electronic incompressibility based on an optical readout technique for measuring the chemical potential. In both works, it is found that an out-of-plane electric field can drive a phase transition from the topological Chern insulators to some trivial insulating states (e.g., Mott insulator or charge density waves).

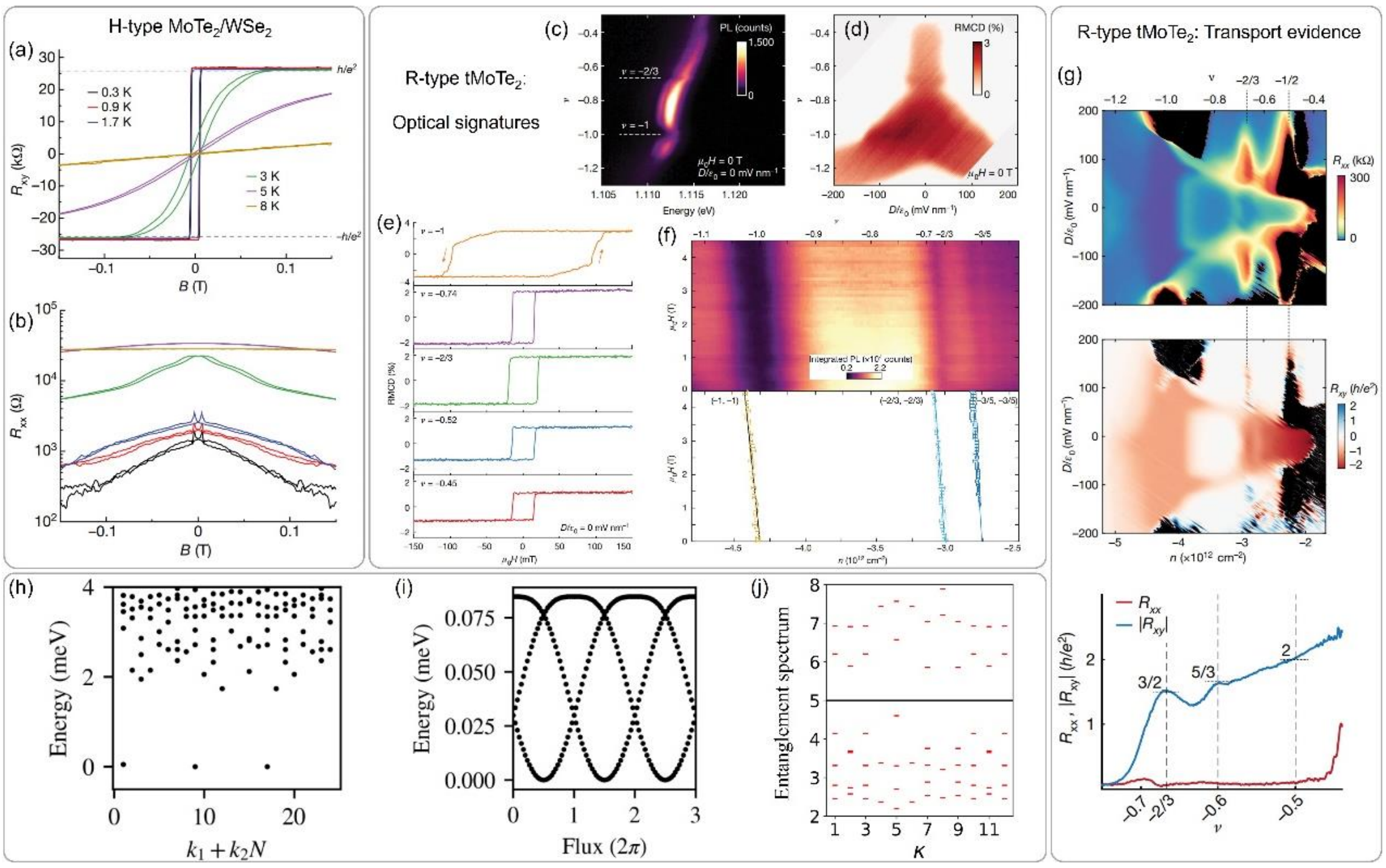

**Figure 11.** (a, b) IQAH effect in the moiré of H-type $MoTe_2/WSe_2$. (a) Hall and (b) longitudinal resistance as a function of out-of-plane magnetic field (B) at various temperatures. (c-f) Optical signatures of IQAH and FQAH effects in R-type $tMoTe_2$. (c) PL intensity as a function of filling factor ($\nu$) and photon energy. (d) RMCD signal versus $\nu$ and out-of-plane electric field ($D/\epsilon_0$). (e) RMCD signal versus applied magnetic field ($\mu_0 H$) swept back and forth at selected fillings. (f) Integrated PL intensity versus magnetic field ($\mu_0 H$) and carrier density ($n$), and Wannier diagrams showing $C=-1$ IQAH state at $\nu=-1$, $C=-2/3$ FQAH state at $\nu=-2/3$ and $C=-3/5$ FQAH state at $\nu=-3/5$. (g) Transport evidence of FQAH effects in R-type $tMoTe_2$. Longitudinal (top panel) and Hall (middle panel) resistance as a function of electric field ($D/\varepsilon_0$) and carrier density ($n$). Top axis shows the filling factor $\nu$. (Bottom panel) Longitudinal and Hall resistance versus $\nu$ at $D/\varepsilon_0=0$. (h-j) Signatures of FQAH effect in ED calculations. (h) An example of the many-body spectrum of $tMoTe_2$ as a function of total crystalline momentum at $\nu=-2/3$. (i) Evolution of the ground states under flux insertion along the $k_2$ direction. (j) An example of the entanglement spectrum of $tMoTe_2$. (a, b) Reproduced with permission from Ref. [130]. (c–f) Reproduced from Ref. [99]. (g) Reproduced from Ref. [101]. (h, i) Reproduced with permission from Ref. [114]. (j) Adapted with permission from Ref. [379].

Following the progress in optical studies, direct evidence of IQAH and FQAH effects in $tMoTe_2$ from electrical transport measurements was also discovered. In Ref. [101], clearly quantized Hall resistance plateaus $R_{xy}=C^{-1}h/e^2$ with $C=-1$, $-2/3$, and $-3/5$ were observed at filling $\nu=-1$, $-2/3$, and $-3/5$ respectively, which are accompanied by suppressed longitudinal Hall resistance $R_{xx}$ (Fig. 11(g)). The phase transition from Chern insulators to trivial insulating states driven by an out-of-plane electric field is also observed. In an independent experiment [102], integer and fractional Hall resistances at $\nu=-1$ and $-2/3$ were also observed.

Ref. [101] also reported interesting transport features around $\nu=-1/2$ (Fig. 11(g)). Specifically, $R_{xy}$ scales linearly and passes through $2h/e^2$ while $R_{xx}$ remains small, indicating that the state is compressible, which is consistent with the absence of gap features in the trion photoluminescence at $\nu=-1/2$ (Fig. 11(c)). Such features are similar to those of the composite Fermi liquid state observed at half-filling of the conventional lowest Landau level in a strong magnetic field, implying the existence of a zero-field analog of the composite Fermi liquid state in $tMoTe_2$ [380, 381]. Further experimental evidence in support of the zero-field composite Fermi liquid state is very recently reported using trion sensing spectroscopy [382].

As the field rapidly evolves, signatures of the FQAH effects from imaging techniques [383, 384] and transient optical spectroscopy [385] have also been reported. Ref. [383] applied the newly developed exciton resonant-microwave impedance microscopy on $tMoTe_2$. The imaginary component of the reflected microwave signal, which correlates with the local conductivity, exhibits signatures of the IQAH and FQAH effects with insulating bulk and conductive edges. A nanoscale superconducting sensor was utilized in Ref. [384] to map the magnetic fringe fields in $tMoTe_2$. Oscillations in the local magnetic field were observed at integer and fractional fillings, which are consistent with the formation of the IQAH and FQAH states. Using pump–probe spectroscopy, Ref. [385] identified a large number of correlated states at fractional fillings of holes as well as electrons that were absent in earlier static spectroscopic sensing and transport measurements, and also revealed the dynamics of the correlated states.

Many of the experimental observations in $tMoTe_2$ are supported by numerical studies through exact diagonalization (ED) of the many-body Hamiltonian with Coulomb interaction projected to the flat Chern band(s) [114, 138, 139, 142, 143, 152, 379, 386]. Based on the continuum model and parameters proposed in Ref. [127], FQAH states in $tMoTe_2$ were first numerically predicted at filling $\nu = -1/3$ and $-2/5$ around the 'magic' twist angle $\theta \approx 1.4°$ [142], where the first moiré band is near exactly flat. The discrepancies between theoretical predictions and experimental observations ($\theta \approx 3° \sim 4°$, and $\nu = -2/3$ and $-3/5$) were attributed to lattice reconstruction effects by later theoretical studies (e.g., Table 4). With the updated models, numerical studies found a considerably small spin gap at $\nu = -1/3$ [114, 138, 152], indicating a much weaker ferromagnetism at $\nu = -1/3$, in line with the RMCD results and absence of FQAH effect at $\nu = -1/3$ in experiments; while the FQAH states at $\nu = -2/3$ were consistently predicted for $\theta \approx 3° \sim 4°$.

Meanwhile, theoretical studies of the correlated phenomena in $tMoTe_2$ are complicated by a few factors. Details in accurately modeling the non-interacting electronic properties of the moiré, including the form of the Hamiltonian and its parameter values that vary with twist angle, remain controversial (Sec. 1.6). Distinct models could lead to correlated behaviors with quantitative differences and/or completely distinct natures. Limitations from computational capability could also affect reliably identifying the correlated phases: ED calculations oftentimes are performed by projecting the Coulomb interaction onto the first moiré band only, which neglects the effects of band mixing that could be important in $tMoTe_2$ [139, 152, 379].

Given the above subtleties, interested readers might refer to the various works for specific numerical results. Here we simply summarize some of the signatures of the FQAH states at fractional filling $\nu = p/q$ from ED calculations [387-391]: (i) The ground states are $q$-fold degenerate under periodic boundary condition and well separated from excited states [e.g., Fig. 11(h)]; (ii) Under magnetic flux insertion, the ground states evolve into each other after $2\pi$ flux and return to the initial configuration with a $2q\pi$ periodicity, during which they remain gapped from the excited states [e.g., Fig. 11(i)]; (iii) The quasiparticle excitations obey the generalized Pauli exclusion principle [387, 392], which sets constraints on the number of low-energy excitations. The (particle) entanglement spectrum [393] extracted from the reduced density matrix of the ground states exhibit a clear gap, and the counting of states below the gap satisfies the exclusion statistics [e.g., Fig. 11(j)] [387, 388].

We remark that the search of candidate systems for the FQAH effects can be guided by scrutinizing the properties of quantum geometry (e.g., uniformity of Berry curvature and quantum metric, trace or ideal-band conditions) and mimicking the lowest Landau level physics. These topics were covered in some earlier reviews (e.g., Ref. [390]), and some new understandings have been developed recently [394-398] as motivated by the studies of moiré superlattices. The adiabatic approximation approach (Sec. 4.1), which predicts the emergence of a pseudo-magnetic field, offers a natural starting point to make some analytical connections to the Landau level paradigm for understanding some of the correlated behaviors in $tMoTe_2$ [318, 319, 399-402].

It is noteworthy to point out that FQAH states can also emerge in unconventional scenarios, where e.g., the models are not continuously connected to Landau levels or the quantum geometry is far from ideal [403-407]. Ref. [403] considered the bipartite limit of the three-orbital tight-binding model of $tMoTe_2$ ($t_3 = 0$ in Eq. (25)), where an exact flat band with rather uniform quantum geometry emerges but having a touching point with a dispersive band. Numerical simulations have found FQAH states at 1/3 and 2/3 filling of such a singular flat band with nearest-neighbor repulsion over a broad range surpassing the bandwidth of the two touching bands. Interestingly, maintaining the band touching was found to be important for stabilizing the FQAH effect. By introducing anisotropy in the model [404], the band touching can be lifted, the exact flat band becomes isolated but with Chern number equals zero. Surprisingly, FQAH states were found numerically in such a topologically trivial flat band at 2/3 filling in the small interaction strength regime far below the band gap size, where inter-band mixing effects are negligible and the trivial flat band remains well isolated [404].

Finally, we note that related experimental advancements have also been achieved in $tWSe_2$ with significant differences compared to $tMoTe_2$. The IQAH effects were reported in $tWSe_2$, at a much smaller twist angle ($\sim 1.23°$) and at integer fillings of both $\nu = -1$ and $-3$ [408]. In contrast, experimental signatures of the IQAH effect at $\nu = -3$ have been elusive in $tMoTe_2$, but ferromagnetism and unquantized anomalous Hall resistance have been observed in samples with $2.5° < \theta < 4°$ [409-411]. Theoretical studies have predicted FQAH effects in $tWSe_2$ [138, 143], but their experimental signatures are still lacking. At intermediate twist angles ($2° \lesssim \theta \lesssim 3°$) of $tWSe_2$, Ref. [412] reported spontaneous Stoner ferromagnetism around $\nu = -0.8$ and the emergence of Chern insulator at $\nu = -1$ after applying a magnetic field. In $tWSe_2$ with larger twist angles ($4° \lesssim \theta \lesssim 5°$), earlier works reported correlated insulating behaviors at $\nu = -1$ [276] with continuous metal–insulator transitions tunable by carrier density and displacement field [413]. Hints of the existence of a superconducting state were also reported in Ref. [276], while two recent works reported robust signatures of superconductivity (but no ferromagnetism) in $\theta = 3.65°$ (also 3.5°) [414] and $\theta = 5°$ [415] $tWSe_2$ within a finite region near $\nu = -1$ and an out-of-plane electric field. Some intriguing observations from these two studies are as follows: Ref. [414] found that the superconducting phase boarders on two distinct metallic states at larger and smaller carrier densities and transitions to a correlated insulator when the displacement field is enlarged; Ref. [415] also observed some metallic state in proximity to the superconductor (but no proximate insulating phases) with a Fermi surface reconstruction possibly due to antiferromagnetic ordering. The maximum critical temperature was found to be about 0.2 K and 0.4 K respectively in the two works. These experiments have triggered theoretical interests in identifying the origin and properties of the superconductivity [416-436], for instance, theories based on band topology-induced quantum fluctuations [434], spinon pairing in a parent insulating phase [433], phenomenological boson-mediated BCS model [436], Kohn-Luttinger mechanism [429], intervalley-coherent antiferromagnetic spin fluctuations [426] etc. have been proposed. Very recently, experimental signatures of superconductivity have also been reported in $tMoTe_2$ with $\theta \approx 3.83°$ and $-0.76 < \nu < -0.71$ [437]. The superconducting state has a maximum critical temperature of 1.2 K and is most pronounced at zero out-of-plane electric field. Notably, the normal state of the superconductor exhibits anomalous Hall effects. The drastically contrasted behaviors in $tMoTe_2$ and $tWSe_2$ suggest that there exist subtle differences between the two moiré superlattices and call for a better and more accurate theoretical description of them (Sec. 1.6).

## 5. Outlook

With the exciting progresses made, some aspects of the existing studies require further experimental and theoretical endeavors. On the theoretical side, the computational limitations [107, 111, 113] have restricted the capability of unambiguously identifying models of various moiré superlattices (Secs. 1.6 and 4.3.1) and quantitative power in the predictions of correlated phases [152] for comparison with the observations and guiding future experimental studies. On the experimental aspect, characterization of the structural properties of the moiré superlattices (Sec. 1.5) can help to resolve some of debates on the models of moiré.

Meanwhile, there exist exciting possibilities for future experimental exploration. The FQAH phases that emerge at fractionally filling the first moiré band of $tMoTe_2$ discussed in Sec. 4.3.2 are Abelian ones, analogous to their counterparts in the lowest Landau level. In quantum Hall systems, non-Abelian topological phases, e.g., the Moore-Read Pfaffian/anti-Pfaffian state [438-441] or the Read-Rezayi state [442] were predicted in the first (excited) Landau level. It is thus interesting to search for their analogues in the Chern bands. Recent DFT studies found a sequence of narrow bands with identical Chern number $C^{\uparrow} = 1$ in $tMoTe_2$ at twist angle around 2° [107, 113], which are promising as counterparts of the higher Landau levels. Indeed, the calculations suggested that the second moiré band at such twist angles share similar characters as the first Landau level, and ED calculations based on the extracted parameters found signatures of Moore-Read type states at its half filling [113, 145, 396, 397, 443-445]. In addition to FQAH phases, the moiré bands could also host topologically ordered phases that preserve the time-reversal symmetry, i.e., fractional topological (or fractional quantum spin Hall, FQSH) insulators [446]. Experimentally, a triple quantum spin Hall insulator with three pairs of helical edge states was observed in a 2.1° $tMoTe_2$ at filling $\nu = -6$ [325], suggesting that the first three moiré bands share identical Chern numbers. Signatures of the FQSH insulator were also reported in the same study at $\nu = -3$, which has stimulated many theoretical studies for understanding the experimental results and proposing intriguing phases [447-456].

In addition to the FQAH and FQSH states, theoretical studies have also explored other correlated topological states at fractional fillings such as composite fermi liquids and the so-called quantum anomalous Hall crystals [457, 458], as well as various intriguing phases and transport phenomena at integer fillings (e.g., $\nu = -1$ and $-2$) in $tMoTe_2$ [144, 323, 340, 459-462]. It would be interesting to experimentally explore these various theoretical predictions. We note that there exist other practical superlattice platforms for exploring the FQAH effects, including the moiré of rhombohedral multilayer graphene on hBN (where zero-field FQAH effects have been observed [463-465]) and periodically strained graphene or TMDs [123, 466, 467]. Experimental efforts in these systems are also highly desirable, especially in the periodically strained systems where experimental progress has been very limited [468].

In the long run, recall that one of the motivations for studying the fractional quantum Hall and FQSH states is to utilize them for topological quantum computing [469, 470], where some of the proposals require additional superconducting and ferromagnetic components [471-473]. Moiré materials free of external magnetic fields are promising candidates for experimentally implementing these schemes. By employing an on-chip 2D growth method [474], progress in introducing superconducting junctions into $tMoTe_2$ has been achieved recently [475, 476] and further developments are highly appealing. And towards a practical platform for implementing topological quantum computing, higher critical temperatures of the correlated topological phases are important, which are limited by several factors including the moiré band dispersion and quantum geometry. A desirable pursuit is to raise the Curie temperature of the intrinsic ferromagnetism in the twisted bilayers, which may also help to improve the stability of the FQAH states.

The moiré of bilayers has been the focus until now, while richer structures with distinct properties and additional tunability can be achieved by increasing the number of layers. For instance, in the case of twisted trilayers, out-of-plane mirror symmetric configurations with moiré landscapes similar to that of twisted bilayers can be achieved if the outer layers are identical and perfectly aligned [477, 478], while supermoiré patterns with quasi-periodicity much larger than that of twisted bilayers can be realized if the outer layers are mis-orientated [136, 479-481]. Given the recognized complexities in quantitative modeling the twisted bilayer TMDs, great care should be taken in the theoretical modeling of multilayer structures. Nevertheless, the new moiré landscapes in multilayers likely will allow one to explore novel variants of the topics discussed here. For instance, heterotrilayers with identical outer layers can host quadrupolar excitons in addition to dipolar excitons, which can lead to interesting optical features and correlated phases [482-485]; while supermoiré patterns might exhibit real-space mosaics of topological and correlated phases of different characters.

## 6. Acknowledgments

D.Z. and W.Y. acknowledge support by Research Grant Council of Hong Kong SAR (HKU SRFS2122-7S05, AoE/P-701/20, A-HKU705/21), the National Key R&D Program of China (2020YFA0309600), National Natural Science Foundation of China (No. 12425406), and New Cornerstone Science Foundation. H.Y. acknowledges support by NSFC under grant No. 12274477, and the Department of Science and Technology of Guangdong Province in China (2019QN01X061).

## 7. References

[1] G. Li, A. Luican, J.M.B. Lopes dos Santos, A.H. Castro Neto, A. Reina, J. Kong, E.Y. Andrei, Observation of Van Hove singularities in twisted graphene layers, Nat. Phys., 6 (2010) 109-113.

[2] M. Yankowitz, J. Xue, D. Cormode, J.D. Sanchez-Yamagishi, K. Watanabe, T. Taniguchi, P. Jarillo-Herrero, P. Jacquod, B.J. LeRoy, Emergence of superlattice Dirac points in graphene on hexagonal boron nitride, Nat. Phys., 8 (2012) 382-386.

[3] L.A. Ponomarenko, R.V. Gorbachev, G.L. Yu, D.C. Elias, R. Jalil, A.A. Patel, A. Mishchenko, A.S. Mayorov, C.R. Woods, J.R. Wallbank, M. Mucha-Kruczynski, B.A. Piot, M. Potemski, I.V. Grigorieva, K.S. Novoselov, F. Guinea, V.I. Fal'ko, A.K. Geim, Cloning of Dirac fermions in graphene superlattices, Nature, 497 (2013) 594-597.

[4] C.R. Dean, L. Wang, P. Maher, C. Forsythe, F. Ghahari, Y. Gao, J. Katoch, M. Ishigami, P. Moon, M. Koshino, T. Taniguchi, K. Watanabe, K.L. Shepard, J. Hone, P. Kim, Hofstadter's butterfly and the fractal quantum Hall effect in moiré superlattices, Nature, 497 (2013) 598-602.

[5] B. Hunt, J.D. Sanchez-Yamagishi, A.F. Young, M. Yankowitz, B.J. LeRoy, K. Watanabe, T. Taniguchi, P. Moon, M. Koshino, P. Jarillo-Herrero, R.C. Ashoori, Massive Dirac Fermions and Hofstadter Butterfly in a van der Waals Heterostructure, Science, 340 (2013) 1427-

1430.
[6] Y. Cao, V. Fatemi, S. Fang, K. Watanabe, T. Taniguchi, E. Kaxiras, P. Jarillo-Herrero, Unconventional superconductivity in magic-angle graphene superlattices, Nature, 556 (2018) 43-50.
[7] Y. Cao, V. Fatemi, A. Demir, S. Fang, S.L. Tomarken, J.Y. Luo, J.D. Sanchez-Yamagishi, K. Watanabe, T. Taniguchi, E. Kaxiras, R.C. Ashoori, P. Jarillo-Herrero, Correlated insulator behaviour at half-filling in magic-angle graphene superlattices, Nature, 556 (2018) 80-84.
[8] E.Y. Andrei, D.K. Efetov, P. Jarillo-Herrero, A.H. MacDonald, K.F. Mak, T. Senthil, E. Tutuc, A. Yazdani, A.F. Young, The marvels of moiré materials, Nat. Rev. Mater., 6 (2021) 201-206.
[9] K.P. Nuckolls, A. Yazdani, A microscopic perspective on moiré materials, Nat. Rev. Mater., 9 (2024) 460-480.
[10] E.Y. Andrei, A.H. MacDonald, Graphene bilayers with a twist, Nat. Mater., 19 (2020) 1265-1275.
[11] J. Liu, X. Dai, Orbital magnetic states in moiré graphene systems, Nat. Rev. Phys., 3 (2021) 367-382.
[12] P. Törmä, S. Peotta, B.A. Bernevig, Superconductivity, superfluidity and quantum geometry in twisted multilayer systems, Nat. Rev. Phys., 4 (2022) 528-542.
[13] C.N. Lau, M.W. Bockrath, K.F. Mak, F. Zhang, Reproducibility in the fabrication and physics of moiré materials, Nature, 602 (2022) 41-50.
[14] S. Bhowmik, A. Ghosh, U. Chandni, Emergent phases in graphene flat bands, Rep. Prog. Phys., 87 (2024) 096401.
[15] Q.H. Wang, K. Kalantar-Zadeh, A. Kis, J.N. Coleman, M.S. Strano, Electronics and optoelectronics of two-dimensional transition metal dichalcogenides, Nature Nanotech., 7 (2012) 699-712.
[16] K.F. Mak, J. Shan, Photonics and optoelectronics of 2D semiconductor transition metal dichalcogenides, Nat. Photon., 10 (2016) 216-226.
[17] J.R. Schaibley, H. Yu, G. Clark, P. Rivera, J.S. Ross, K.L. Seyler, W. Yao, X. Xu, Valleytronics in 2D materials, Nat. Rev. Mater., 1 (2016) 16055.
[18] A. Ciarrocchi, F. Tagarelli, A. Avsar, A. Kis, Excitonic devices with van der Waals heterostructures: valleytronics meets twistronics, Nat. Rev. Mater., 7 (2022) 449-464.
[19] G.-B. Liu, D. Xiao, Y. Yao, X. Xu, W. Yao, Electronic structures and theoretical modelling of two-dimensional group-VIB transition metal dichalcogenides, Chem. Soc. Rev., 44 (2015) 2643-2663.
[20] K.F. Mak, C. Lee, J. Hone, J. Shan, T.F. Heinz, Atomically Thin MoS2: A New Direct-Gap Semiconductor, Phys. Rev. Lett., 105 (2010) 136805.
[21] A. Splendiani, L. Sun, Y. Zhang, T. Li, J. Kim, C.-Y. Chim, G. Galli, F. Wang, Emerging Photoluminescence in Monolayer MoS2, Nano Lett., 10 (2010) 1271-1275.
[22] X. Xu, W. Yao, D. Xiao, T.F. Heinz, Spin and pseudospins in layered transition metal dichalcogenides, Nat. Phys., 10 (2014) 343-350.
[23] D. Xiao, G.-B. Liu, W. Feng, X. Xu, W. Yao, Coupled Spin and Valley Physics in Monolayers of MoS2 and Other Group-VI Dichalcogenides, Phys. Rev. Lett., 108 (2012) 196802.
[24] A. Kormányos, G. Burkard, M. Gmitra, J. Fabian, V. Zólyomi, N.D. Drummond, V. Fal'ko, k·p theory for two-dimensional transition metal dichalcogenide semiconductors, 2D

Mater., 2 (2015) 022001.
[25] Z.Y. Zhu, Y.C. Cheng, U. Schwingenschlögl, Giant spin-orbit-induced spin splitting in two-dimensional transition-metal dichalcogenide semiconductors, Phys. Rev. B, 84 (2011) 153402.
[26] A. Ramasubramaniam, Large excitonic effects in monolayers of molybdenum and tungsten dichalcogenides, Phys. Rev. B, 86 (2012) 115409.
[27] P.V. Nguyen, N.C. Teutsch, N.P. Wilson, J. Kahn, X. Xia, A.J. Graham, V. Kandyba, A. Giampietri, A. Barinov, G.C. Constantinescu, N. Yeung, N.D.M. Hine, X. Xu, D.H. Cobden, N.R. Wilson, Visualizing electrostatic gating effects in two-dimensional heterostructures, Nature, 572 (2019) 220-223.
[28] G.-B. Liu, W.-Y. Shan, Y. Yao, W. Yao, D. Xiao, Three-band tight-binding model for monolayers of group-VIB transition metal dichalcogenides, Phys. Rev. B, 88 (2013) 085433.
[29] A. Kormányos, V. Zólyomi, N.D. Drummond, P. Rakyta, G. Burkard, V.I. Fal'ko, Monolayer MoS2: Trigonal warping, the Γ valley, and spin-orbit coupling effects, Phys. Rev. B, 88 (2013) 045416.
[30] A. Kormányos, V. Zólyomi, N.D. Drummond, G. Burkard, Spin-Orbit Coupling, Quantum Dots, and Qubits in Monolayer Transition Metal Dichalcogenides, Phys. Rev. X, 4 (2014) 011034.
[31] E.S. Kadantsev, P. Hawrylak, Electronic structure of a single MoS2 monolayer, Solid State Commun., 152 (2012) 909-913.
[32] K. Kośmider, J.W. González, J. Fernández-Rossier, Large spin splitting in the conduction band of transition metal dichalcogenide monolayers, Phys. Rev. B, 88 (2013) 245436.
[33] T. Cheiwchanchamnangij, W.R.L. Lambrecht, Quasiparticle band structure calculation of monolayer, bilayer, and bulk MoS2, Phys. Rev. B, 85 (2012) 205302.
[34] K. Kośmider, J. Fernández-Rossier, Electronic properties of the MoS2-WS2 heterojunction, Phys. Rev. B, 87 (2013) 075451
[35] R. Roldán, M.P. López-Sancho, F. Guinea, E. Cappelluti, J.A. Silva-Guillén, P. Ordejón, Momentum dependence of spin-orbit interaction effects in single-layer and multi-layer transition metal dichalcogenides, 2D Mater., 1 (2014) 034003.
[36] T. Cheiwchanchamnangij, W.R.L. Lambrecht, Y. Song, H. Dery, Strain effects on the spin-orbit-induced band structure splittings in monolayer MoS2 and graphene, Phys. Rev. B, 88 (2013) 155404.
[37] H. Ochoa, R. Roldán, Spin-orbit-mediated spin relaxation in monolayer MoS2, Phys. Rev. B, 87 (2013) 245421.
[38] R. Pisoni, A. Kormanyos, M. Brooks, Z. Lei, P. Back, M. Eich, H. Overweg, Y. Lee, P. Rickhaus, K. Watanabe, T. Taniguchi, A. Imamoglu, G. Burkard, T. Ihn, K. Ensslin, Interactions and Magnetotransport through Spin-Valley Coupled Landau Levels in Monolayer MoS_2, Phys Rev Lett, 121 (2018) 247701.
[39] G. Wang, C. Robert, M.M. Glazov, F. Cadiz, E. Courtade, T. Amand, D. Lagarde, T. Taniguchi, K. Watanabe, B. Urbaszek, X. Marie, In-Plane Propagation of Light in Transition Metal Dichalcogenide Monolayers: Optical Selection Rules, Phys. Rev. Lett., 119 (2017) 047401.
[40] H. Yu, G.-B. Liu, W. Yao, Brightened spin-triplet interlayer excitons and optical selection rules in van der Waals heterobilayers, 2D Mater., 5 (2018) 035021.

[41] Z. Bi, N.F.Q. Yuan, L. Fu, Designing flat bands by strain, Phys. Rev. B, 100 (2019) 035448.
[42] H. Yu, M. Chen, W. Yao, Giant magnetic field from moiré induced Berry phase in homobilayer semiconductors, Natl. Sci. Rev., 7 (2020) 12-20.
[43] D. Zhai, W. Yao, Theory of tunable flux lattices in the homobilayer moiré of twisted and uniformly strained transition metal dichalcogenides, Phys. Rev. Materials, 4 (2020) 094002.
[44] H. Zheng, D. Zhai, W. Yao, Twist versus heterostrain control of optical properties of moiré exciton minibands, 2D Mater., 8 (2021) 044016.
[45] M. Kögl, P. Soubelet, M. Brotons-Gisbert, A.V. Stier, B.D. Gerardot, J.J. Finley, Moiré straintronics: a universal platform for reconfigurable quantum materials, npj 2D Mater. Appl., 7 (2023) 32.
[46] F. Escudero, A. Sinner, Z. Zhan, P.A. Pantaleón, F. Guinea, Designing moiré patterns by strain, Phys. Rev. Research, 6 (2024) 023203.
[47] Y. Bai, L. Zhou, J. Wang, W. Wu, L.J. McGilly, D. Halbertal, C.F.B. Lo, F. Liu, J. Ardelean, P. Rivera, N.R. Finney, X.-C. Yang, D.N. Basov, W. Yao, X. Xu, J. Hone, A.N. Pasupathy, X.-Y. Zhu, Excitons in strain-induced one-dimensional moiré potentials at transition metal dichalcogenide heterojunctions, Nat. Mater., 19 (2020) 1068-1073.
[48] P. Wang, G. Yu, Y.H. Kwan, Y. Jia, S. Lei, S. Klemenz, F.A. Cevallos, R. Singha, T. Devakul, K. Watanabe, T. Taniguchi, S.L. Sondhi, R.J. Cava, L.M. Schoop, S.A. Parameswaran, S. Wu, One-dimensional Luttinger liquids in a two-dimensional moire lattice, Nature, 605 (2022) 57-62.
[49] G. Yu, P. Wang, A.J. Uzan-Narovlansky, Y. Jia, M. Onyszczak, R. Singha, X. Gui, T. Song, Y. Tang, K. Watanabe, T. Taniguchi, R.J. Cava, L.M. Schoop, S. Wu, Evidence for two dimensional anisotropic Luttinger liquids at millikelvin temperatures, Nat Commun, 14 (2023) 7025.
[50] K.F. Mak, J. Shan, Semiconductor moiré materials, Nat. Nanotech., 17 (2022) 686-695.
[51] L. Du, M.R. Molas, Z. Huang, G. Zhang, F. Wang, Z. Sun, Moiré photonics and optoelectronics, Science, 379 (2023) eadg0014.
[52] D. Huang, J. Choi, C.-K. Shih, X. Li, Excitons in semiconductor moiré superlattices, Nat. Nanotech., 17 (2022) 227-238.
[53] N.P. Wilson, W. Yao, J. Shan, X. Xu, Excitons and emergent quantum phenomena in stacked 2D semiconductors, Nature, 599 (2021) 383-392.
[54] M. Kapfer, B.S. Jessen, M.E. Eisele, M. Fu, D.R. Danielsen, T.P. Darlington, S.L. Moore, N.R. Finney, A. Marchese, V. Hsieh, P. Majchrzak, Z. Jiang, D. Biswas, P. Dudin, J. Avila, K. Watanabe, T. Taniguchi, S. Ulstrup, P. Bøggild, P.J. Schuck, D.N. Basov, J. Hone, C.R. Dean, Programming twist angle and strain profiles in 2D materials, Science, 381 (2023) 677-681.
[55] C. Zhang, C.-P. Chuu, X. Ren, M.-Y. Li, L.-J. Li, C. Jin, M.-Y. Chou, C.-K. Shih, Interlayer couplings, Moiré patterns, and 2D electronic superlattices in MoS2/WSe2 hetero-bilayers, Sci. Adv., 3 (2017) e1601459.
[56] H. Yu, G.-B. Liu, J. Tang, X. Xu, W. Yao, Moiré excitons: From programmable quantum emitter arrays to spin-orbit-coupled artificial lattices, Sci. Adv., 3 (2017) e1701696.
[57] K. Liu, L. Zhang, T. Cao, C. Jin, D. Qiu, Q. Zhou, A. Zettl, P. Yang, S.G. Louie, F. Wang, Evolution of interlayer coupling in twisted molybdenum disulfide bilayers, Nat. Commun., 5 (2014) 4966.

[58] A.M.v.d. Zande, J. Kunstmann, A. Chernikov, D.A. Chenet, Y. You, X. Zhang, P.Y. Huang, T.C. Berkelbach, L. Wang, F. Zhang, M.S. Hybertsen, D.A. Muller, D.R. Reichman, T.F. Heinz, J.C. Hone, Tailoring the Electronic Structure in Bilayer Molybdenum Disulfide via Interlayer Twist, Nano Lett., 14 (2014) 3869-3875.
[59] Y. Pan, S. Fölsch, Y. Nie, D. Waters, Y.-C. Lin, B. Jariwala, K. Zhang, K. Cho, J.A. Robinson, R.M. Feenstra, Quantum-Confined Electronic States Arising from the Moiré Pattern of MoS2-WSe2 Heterobilayers, Nano Lett., 18 (2018) 1849-1855.
[60] V.O. Özçelik, J.G. Azadani, C. Yang, S.J. Koester, T. Low, Band alignment of two-dimensional semiconductors for designing heterostructures with momentum space matching, Phys. Rev. B, 94 (2016) 035125.
[61] J. Kang, S. Tongay, J. Zhou, J. Li, J. Wu, Band offsets and heterostructures of two-dimensional semiconductors, Appl. Phys. Lett., 102 (2013) 012111.
[62] N.R. Wilson, P.V. Nguyen, K.L. Seyler, P. Rivera, A.J. Marsden, Z.P.L. Laker, G.C. Constantinescu, V. Kandyba, A. Barinov, N.D.M. Hine, X. Xu, D.H. Cobden, Determination of band offsets, hybridization, and exciton binding in 2D semiconductor heterostructures, Sci. Adv., 3 (2017) e1601832.
[63] M.-H. Chiu, C. Zhang, H.W. Shiu, C.-P. Chuu, C.-H. Chen, C.-Y.S. Chang, C.-H. Chen, M.-Y. Chou, C.-K. Shih, L.-J. Li, Determination of band alignment in the single-layer MoS2/WSe2 heterojunction, Nat. Commun., 6 (2015) 7666.
[64] J.M.B.L.d. Santos, N.M.R. Peres, A.H.C. Neto, Graphene Bilayer with a Twist: Electronic Structure, Phys. Rev. Lett., 99 (2007) 256802.
[65] R. Bistritzer, A.H. MacDonald, Moiré bands in twisted double-layer graphene, Proc. Natl. Acad. Sci. USA, 108 (2011) 12233-12237.
[66] S. Shallcross, S. Sharma, E. Kandelaki, O.A. Pankratov, Electronic structure of turbostratic graphene, Phys. Rev. B, 81 (2010) 165105
[67] Y. Wang, Z. Wang, W. Yao, G.-B. Liu, H. Yu, Interlayer coupling in commensurate and incommensurate bilayer structures of transition metal dichalcogenides, Phys. Rev. B, 95 (2017) 115429.
[68] Y. Zhang, T. Devakul, L. Fu, Spin-textured Chern bands in AB-stacked transition metal dichalcogenide bilayers, PNAS, 118 (2021) e2112673118.
[69] Q. Tong, H. Yu, Q. Zhu, Y. Wang, X. Xu, W. Yao, Topological mosaics in moiré superlattices of van der Waals heterobilayers, Nat. Phys., 13 (2017) 356-362.
[70] Q. Zhu, M.W.-Y. Tu, Q. Tong, W. Yao, Gate tuning from exciton superfluid to quantum anomalous Hall in van der Waals heterobilayer, Sci. Adv., 5 eaau6120.
[71] F. Liu, Q. Li, X.-Y. Zhu, Direct determination of momentum-resolved electron transfer in the photoexcited van der Waals heterobilayer WS2/MoS2, Phys. Rev. B, 101 (2020) 201405.
[72] D. Schmitt, J.P. Bange, W. Bennecke, A. AlMutairi, G. Meneghini, K. Watanabe, T. Taniguchi, D. Steil, D.R. Luke, R.T. Weitz, S. Steil, G.S.M. Jansen, S. Brem, E. Malic, S. Hofmann, M. Reutzel, S. Mathias, Formation of moiré interlayer excitons in space and time, Nature, 608 (2022) 499-503.
[73] A. Sood, J.B. Haber, J. Carlström, E.A. Peterson, E. Barre, J.D. Georgaras, A.H.M. Reid, X. Shen, M.E. Zajac, E.C. Regan, J. Yang, T. Taniguchi, K. Watanabe, F. Wang, X. Wang, J.B. Neaton, T.F. Heinz, A.M. Lindenberg, F.H.d. Jornada, A. Raja, Bidirectional phonon emission in two-dimensional heterostructures triggered by ultrafast charge transfer, Nat. Nanotechnol., 18 (2022) 29-35.

[74] X. Hong, J. Kim, S.-F. Shi, Y. Zhang, C. Jin, Y. Sun, S. Tongay, J. Wu, Y. Zhang, F. Wang, Ultrafast charge transfer in atomically thin MoS2/WS2 heterostructures, Nat. Nanotechnol., 9 (2014) 682-686.
[75] Y. Yu, S. Hu, L. Su, L. Huang, Y. Liu, Z. Jin, A.A. Purezky, D.B. Geohegan, K.W. Kim, Y. Zhang, L. Cao, Equally Efficient Interlayer Exciton Relaxation and Improved Absorption in Epitaxial and Nonepitaxial MoS2/WS2 Heterostructures, Nano Lett., 15 (2015) 486-491.
[76] A.F. Rigosi, H.M. Hill, Y. Li, A. Chernikov, T.F. Heinz, Probing Interlayer Interactions in Transition Metal Dichalcogenide Heterostructures by Optical Spectroscopy: MoS2/WS2 and MoSe2/WSe2, Nano Lett., 15 (2015) 5033-5038.
[77] H. Yu, Z. Zhou, W. Yao, Distinct moiré textures of in-plane electric polarizations for distinguishing moiré origins in homobilayers, Sci. China-Phys. Mech. Astron., 66 (2023) 107711.
[78] L. Li, M. Wu, Binary Compound Bilayer and Multilayer with Vertical Polarizations: Two-Dimensional Ferroelectrics, Multiferroics, and Nanogenerators, ACS Nano, 11 (2017) 6382-6388.
[79] F. Ferreira, V.V. Enaldiev, V.I. Fal'ko, S.J. Magorrian, Weak ferroelectric charge transfer in layer-asymmetric bilayers of 2D semiconductors, Sci. Rep., 11 (2021) 13422.
[80] V.V. Enaldiev, F. Ferreira, S.J. Magorrian, V.I. Fal'ko, Piezoelectric networks and ferroelectric domains in twistronic superlattices in WS2/MoS2 and WSe2/MoSe2 bilayers, 2D Mater., 8 (2021) 025030.
[81] M.V. Stern, Y. Waschitz, W. Cao, I. Nevo, K. Watanabe, T. Taniguchi, E. Sela, M. Urbakh, O. Hod, M.B. Shalom, Interfacial ferroelectricity by van der Waals sliding, Science, 372 (2021) 1462-1466.
[82] C.R. Woods, P. Ares, H. Nevison-Andrews, M.J. Holwill, R. Fabregas, F. Guinea, A.K. Geim, K.S. Novoselov, N.R. Walet, L. Fumagalli, Charge-polarized interfacial superlattices in marginally twisted hexagonal boron nitride, Nat. Commun., 12 (2021) 347.
[83] K. Yasuda, X. Wang, K. Watanabe, T. Taniguchi, P. Jarillo-Herrero, Stacking-engineered ferroelectricity in bilayer boron nitride, Science, 372 (2021) 1458-1462.
[84] X. Wang, K. Yasuda, Y. Zhang, S. Liu, K. Watanabe, T. Taniguchi, J. Hone, L. Fu, P. Jarillo-Herrero, Interfacial ferroelectricity in rhombohedral-stacked bilayer transition metal dichalcogenides, Nat. Nanotech., 17 (2022) 367-371.
[85] A. Weston, E.G. Castanon, V. Enaldiev, F. Ferreira, S. Bhattacharjee, S. Xu, H. Corte-León, Z. Wu, N. Clark, A. Summerfield, T. Hashimoto, Y. Gao, W. Wang, M. Hamer, H. Read, L. Fumagalli, A.V. Kretinin, S.J. Haigh, O. Kazakova, A.K. Geim, V.I. Fal'ko, R. Gorbachev, Interfacial ferroelectricity in marginally twisted 2D semiconductors, Nat. Nanotech., 17 (2022) 390-395.
[86] L. Rogée, L. Wang, Y. Zhang, S. Cai, P. Wang, M. Chhowalla, W. Ji, S.P. Lau, Ferroelectricity in untwisted heterobilayers of transition metal dichalcogenides, Science, 376 (2022) 973-978.
[87] J. Liang, D. Yang, J. Wu, J.I. Dadap, K. Watanabe, T. Taniguchi, Z. Ye, Optically Probing the Asymmetric Interlayer Coupling in Rhombohedral-Stacked MoS2 Bilayer, Phys. Rev. X, 12 (2022) 041005.
[88] D. Bennett, G. Chaudhary, R.-J. Slager, E. Bousquet, P. Ghosez, Polar meron-antimeron networks in strained and twisted bilayers, Nat. Commun., 14 (2023) 1629.
[89] J. Ji, G. Yu, C. Xu, H.J. Xiang, General Theory for Bilayer Stacking Ferroelectricity, Phys. Rev. Lett., 130 (2023) 146801.

[90] D. Bennett, W.J. Jankowski, G. Chaudhary, E. Kaxiras, R.-J. Slager, Theory of polarization textures in crystal supercells, Phys. Rev. Research, 5 (2023) 033216.
[91] T. Akamatsu, T. Ideue, L. Zhou, Y. Dong, S. Kitamura, M. Yoshii, D. Yang, M. Onga, Y. Nakagawa, K. Watanabe, T. Taniguchi, J. Laurienzo, J. Huang, Z. Ye, T. Morimoto, H. Yuan, Y. Iwasa, A van der Waals interface that creates in-plane polarization and a spontaneous photovoltaic effect, Science, 372 (2021) 68-72.
[92] T.-H.-Y. Vu, D. Bennett, G.N. Pallewella, M.H. Uddin, K. Xing, W. Zhao, S.H. Lee, Z. Mao, J.B. Muir, L. Jia, J.A. Davis, K. Watanabe, T. Taniguchi, S. Adam, P. Sharma, M.S. Fuhrer, M.T. Edmonds, Imaging topological polar structures in marginally twisted 2D semiconductors, arXiv:2405.15126, DOI.
[93] S. Chiodini, G. Venturi, J. Kerfoot, J. Zhang, E.M. Alexeev, T. Taniguchi, K. Watanabe, A.C. Ferrari, A. Ambrosio, Electromechanical Response of Saddle Points in Twisted hBN Moire Superlattices, ACS Nano, 19 (2025) 16297-16306.
[94] Y. Li, S. Wan, H. Liu, H. Huang, Z. Li, X. Weng, M. Zhu, Y. Zhou, Topological Polar Networks in Twisted Rhombohedral-Stacked Bilayer WSe2 Moiré Superlattices, Nano Lett., 24 (2024) 13349-13355.
[95] C.S. Tsang, X. Zheng, T. Yang, Z. Yan, W. Han, L.W. Wong, H. Liu, S. Gao, K.H. Leung, C.-S. Lee, S.P. Lau, M. Yang, J. Zhao, T.H. Ly, Polar and quasicrystal vortex observed in twisted-bilayer molybdenum disulfide, Science, 386 (2024) 198-205.
[96] E. Pan, Z. Li, F. Yang, K. Niu, R. Bian, Q. Liu, J. Chen, B. Dong, R. Wang, T. Zhou, A. Zhou, X. Luo, J. Chu, J. Lin, W. Li, F. Liu, Observation and manipulation of two-dimensional topological polar texture confined in moiré interface, Nat. Commun., 16 (2025) 3026.
[97] J.J.M. Sangers, A. Brokkelkamp, S. Conesa-Boj, Strain-Induced Moiré Polarization Vortices in Twisted-Multilayer WSe2, Small, n/a (2025) 2503363.
[98] V.V. Enaldiev, V. Zólyomi, C. Yelgel, S.J. Magorrian, V.I. Fal'ko, Stacking Domains and Dislocation Networks in Marginally Twisted Bilayers of Transition Metal Dichalcogenides, Phys. Rev. Lett., 124 (2020) 206101.
[99] J. Cai, E. Anderson, C. Wang, X. Zhang, X. Liu, W. Holtzmann, Y. Zhang, F. Fan, T. Taniguchi, K. Watanabe, Y. Ran, T. Cao, L. Fu, D. Xiao, W. Yao, X. Xu, Signatures of Fractional Quantum Anomalous Hall States in Twisted MoTe2, Nature, 622 (2023) 63-68.
[100] Y. Zeng, Z. Xia, K. Kang, J. Zhu, P. Knüppel, C. Vaswani, K. Watanabe, T. Taniguchi, K.F. Mak, J. Shan, Thermodynamic evidence of fractional Chern insulator in moiré MoTe2, Nature, 622 (2023) 69-73.
[101] H. Park, J. Cai, E. Anderson, Y. Zhang, J. Zhu, X. Liu, C. Wang, W. Holtzmann, C. Hu, Z. Liu, T. Taniguchi, K. Watanabe, J.-h. Chu, T. Cao, L. Fu, W. Yao, C.-Z. Chang, D. Cobden, D. Xiao, X. Xu, Observation of Fractionally Quantized Anomalous Hall Effect, Nature, 622 (2023) 74-79.
[102] F. Xu, Z. Sun, T. Jia, C. Liu, C. Xu, C. Li, Y. Gu, K. Watanabe, T. Taniguchi, B. Tong, J. Jia, Z. Shi, S. Jiang, Y. Zhang, X. Liu, T. Li, Observation of Integer and Fractional Quantum Anomalous Hall Effects in Twisted Bilayer MoTe2, Phys. Rev. X, 13 (2023) 031037.
[103] S. Carr, D. Massatt, S.B. Torrisi, P. Cazeaux, M. Luskin, E. Kaxiras, Relaxation and domain formation in incommensurate two-dimensional heterostructures, Phys. Rev. B, 98 (2018) 224102.
[104] M.R. Rosenberger, H.-J. Chuang, M. Phillips, V.P. Oleshko, K.M. McCreary, S.V. Sivaram, C.S. Hellberg, B.T. Jonker, Twist Angle-Dependent Atomic Reconstruction and Moiré Patterns in Transition Metal Dichalcogenide Heterostructures, ACS Nano, 14 (2020)

4550-4558.
[105] A. Weston, Y. Zou, V. Enaldiev, A. Summerfield, N. Clark, V. Zólyomi, A. Graham, C. Yelgel, S. Magorrian, M. Zhou, J. Zultak, D. Hopkinson, A. Barinov, T.H. Bointon, A. Kretinin, N.R. Wilson, P.H. Beton, V.I. Fal'ko, S.J. Haigh, R. Gorbachev, Atomic reconstruction in twisted bilayers of transition metal dichalcogenides, Nat. Nanotech., 15 (2020) 592-597.
[106] F. Ferreira, S.J. Magorrian, V.V. Enaldiev, D.A. Ruiz-Tijerina, V.I. Fal'ko, Band energy landscapes in twisted homobilayers of transition metal dichalcogenides, Appl. Phys. Lett., 118 (2021) 241602.
[107] X.-W. Zhang, C. Wang, X. Liu, Y. Fan, T. Cao, D. Xiao, Polarization-driven band topology evolution in twisted MoTe2 and WSe2, Nat. Commun., 15 (2024) 4223.
[108] M.H. Naik, M. Jain, Ultraflatbands and Shear Solitons in Moiré Patterns of Twisted Bilayer Transition Metal Dichalcogenides, Phys. Rev. Lett., 121 (2018) 266401.
[109] M.H. Naik, I. Maity, P.K. Maiti, M. Jain, Kolmogorov-Crespi Potential For Multilayer Transition-Metal Dichalcogenides: Capturing Structural Transformations in Moiré Superlattices, J. Phys. Chem. C, 123 (2019) 9770-9778.
[110] N. Mao, C. Xu, J. Li, T. Bao, P. Liu, Y. Xu, C. Felser, L. Fu, Y. Zhang, Transfer learning relaxation, electronic structure and continuum model for twisted bilayer MoTe2, Commun. Phys., 7 (2024) 262.
[111] Y. Jia, J. Yu, J. Liu, J. Herzog-Arbeitman, Z. Qi, H. Pi, N. Regnault, H. Weng, B.A. Bernevig, Q. Wu, Moiré fractional Chern insulators. I. First-principles calculations and continuum models of twisted bilayer MoTe2, Phys. Rev. B, 109 (2024) 205121.
[112] S.J. Magorrian, V.V. Enaldiev, V. Zólyomi, F. Ferreira, V.I. Fal'ko, D.A. Ruiz-Tijerina, Multifaceted moiré superlattice physics in twisted WSe2 bilayers, Phys. Rev. B, 104 (2021) 125440.
[113] C. Xu, N. Mao, T. Zeng, Y. Zhang, Multiple Chern Bands in Twisted MoTe_2 and Possible Non-Abelian States, Phys Rev Lett, 134 (2025) 066601.
[114] C. Wang, X.-W. Zhang, X. Liu, Y. He, X. Xu, Y. Ran, T. Cao, D. Xiao, Fractional Chern Insulator in Twisted Bilayer MoTe2, Phys. Rev. Lett., 132 (2024) 036501.
[115] M.A.H. Vozmediano, M.I. Katsnelson, F. Guinea, Gauge fields in graphene, Phys. Rep., 496 (2010) 109-148.
[116] B. Amorim, A. Cortijo, F. de Juan, A.G. Grushin, F. Guinea, A. Gutiérrez-Rubio, H. Ochoa, V. Parente, R. Roldán, P. San-Jose, J. Schiefele, M. Sturla, M.A.H. Vozmediano, Novel effects of strains in graphene and other two dimensional materials, Phys. Rep., 617 (2016) 1-54.
[117] G.G. Naumis, S. Barraza-Lopez, M. Oliva-Leyva, H. Terrones, Electronic and optical properties of strained graphene and other strained 2D materials: a review, Rep. Prog. Phys., 80 (2017) 096501.
[118] S. Shabani, D. Halbertal, W. Wu, M. Chen, S. Liu, J. Hone, W. Yao, D.N. Basov, X. Zhu, A.N. Pasupathy, Deep moiré potentials in twisted transition metal dichalcogenide bilayers, Nat. Phys., 17 (2021) 720-725.
[119] M.H. Naik, E.C. Regan, Z. Zhang, Y.-H. Chan, Z. Li, D. Wang, Y. Yoon, C.S. Ong, W. Zhao, S. Zhao, M.I.B. Utama, B. Gao, X. Wei, M. Sayyad, K. Yumigeta, K. Watanabe, T. Taniguchi, S. Tongay, F.H. da Jornada, F. Wang, S.G. Louie, Intralayer charge-transfer moiré excitons in van der Waals superlattices, Nature, 609 (2022) 52-57.
[120] H. Li, S. Li, M.H. Naik, J. Xie, X. Li, J. Wang, E. Regan, D. Wang, W. Zhao, S. Zhao, S. Kahn, K. Yumigeta, M. Blei, T. Taniguchi, K. Watanabe, S. Tongay, A. Zettl, S.G. Louie, F.

Wang, M.F. Crommie, Imaging moiré flat bands in three-dimensional reconstructed WSe2/WS2 superlattices, Nat. Mater., 20 (2021) 945-950.
[121] M.A. Cazalilla, H. Ochoa, F. Guinea, Quantum Spin Hall Effect in Two-Dimensional Crystals of Transition-Metal Dichalcogenides, Phys. Rev. Lett., 113 (2014).
[122] Y.-M. Xie, C.-P. Zhang, J.-X. Hu, K.F. Mak, K.T. Law, Valley-Polarized Quantum Anomalous Hall State in Moiré MoTe2/WSe2 Heterobilayers, Phys. Rev. Lett., 128 (2022) 026402.
[123] D. Zhai, Z. Lin, W. Yao, Supersymmetry dictated topology in periodic gauge fields and realization in strained and twisted 2D materials, Rep. Prog. Phys., 87 (2024) 108004.
[124] S. Carr, S. Fang, Z. Zhu, E. Kaxiras, Exact continuum model for low-energy electronic states of twisted bilayer graphene, Phys. Rev. Research, 1 (2019) 013001.
[125] F. Guinea, N.R. Walet, Continuum models for twisted bilayer graphene: Effect of lattice deformation and hopping parameters, Phys. Rev. B, 99 (2019) 205134.
[126] M. Xie, A.H. MacDonald, Weak-Field Hall Resistivity and Spin-Valley Flavor Symmetry Breaking in Magic-Angle Twisted Bilayer Graphene, Phys. Rev. Lett., 127 (2021) 196401.
[127] F. Wu, T. Lovorn, E. Tutuc, I. Martin, A.H. MacDonald, Topological Insulators in Twisted Transition Metal Dichalcogenide Homobilayers, Phys. Rev. Lett., 122 (2019) 086402.
[128] H. Pan, F. Wu, S. Das Sarma, Band topology, Hubbard model, Heisenberg model, and Dzyaloshinskii-Moriya interaction in twisted bilayer WSe2, Phys. Rev. Research, 2 (2020).
[129] T. Devakul, V. Crépel, Y. Zhang, L. Fu, Magic in twisted transition metal dichalcogenide bilayers, Nat. Commun., 12 (2021) 6730.
[130] T. Li, S. Jiang, B. Shen, Y. Zhang, L. Li, Z. Tao, T. Devakul, K. Watanabe, T. Taniguchi, L. Fu, J. Shan, K.F. Mak, Quantum anomalous Hall effect from intertwined moiré bands, Nature, 600 (2021) 641-646.
[131] M. Angeli, A.H. MacDonald, Γ valley transition metal dichalcogenide moiré bands, Proc. Natl. Acad. Sci. U.S.A., 118 (2021) e2021826118.
[132] L. Xian, M. Claassen, D. Kiese, M.M. Scherer, S. Trebst, D.M. Kennes, A. Rubio, Realization of nearly dispersionless bands with strong orbital anisotropy from destructive interference in twisted bilayer MoS2, Nat. Commun., 12 (2021) 5644.
[133] Y. Zhang, T. Liu, L. Fu, Electronic structures, charge transfer, and charge order in twisted transition metal dichalcogenide bilayers, Phys. Rev. B, 103 (2021) 155142.
[134] P. Rivera, J.R. Schaibley, A.M. Jones, J.S. Ross, S. Wu, G. Aivazian, P. Klement, K. Seyler, G. Clark, N.J. Ghimire, J. Yan, D.G. Mandrus, W. Yao, X. Xu, Observation of long-lived interlayer excitons in monolayer MoSe2-WSe2 heterostructures, Nat. Commun., 6 (2015) 6242.
[135] F. Wu, T. Lovorn, E. Tutuc, A.H. MacDonald, Hubbard Model Physics in Transition Metal Dichalcogenide Moiré Bands, Phys. Rev. Lett., 121 (2018) 026402.
[136] Q. Tong, M. Chen, F. Xiao, H. Yu, W. Yao, Interferences of electrostatic moiré potentials and bichromatic superlattices of electrons and excitons in transition metal dichalcogenides, 2D Mater., 8 (2020) 025007.
[137] N. Morales-Durán, J. Wang, G.R. Schleder, M. Angeli, Z. Zhu, E. Kaxiras, C. Repellin, J. Cano, Pressure-enhanced fractional Chern insulators along a magic line in moiré transition metal dichalcogenides, Phys. Rev. Research, 5 (2023) L032022.

[138] A.P. Reddy, F. Alsallom, Y. Zhang, T. Devakul, L. Fu, Fractional quantum anomalous Hall states in twisted bilayer MoTe2 and WSe2, Phys. Rev. B, 108 (2023) 085117.
[139] C. Xu, J. Li, Y. Xu, Z. Bi, Y. Zhang, Maximally localized Wannier functions, interaction models, and fractional quantum anomalous Hall effect in twisted bilayer MoTe2, PNAS, 121 (2024) e2316749121.
[140] Y. Zhang, N.F.Q. Yuan, L. Fu, Moir\'e quantum chemistry: Charge transfer in transition metal dichalcogenide superlattices, Phys. Rev. B, 102 (2020) 201115.
[141] S. Kundu, M.H. Naik, H.R. Krishnamurthy, M. Jain, Moir\'e induced topology and flat bands in twisted bilayer ${\mathrm{WSe}}_{2}$: A first-principles study, Phys. Rev. B, 105 (2022) L081108.
[142] H. Li, U. Kumar, K. Sun, S.-Z. Lin, Spontaneous fractional Chern insulators in transition metal dichalcogenide moiré superlattices, Phys. Rev. Research, 3 (2021) L032070.
[143] V. Crépel, L. Fu, Anomalous Hall metal and fractional Chern insulator in twisted transition metal dichalcogenides, Phys. Rev. B, 107 (2023) L201109.
[144] T. Wang, M. Wang, W. Kim, S.G. Louie, L. Fu, M.P. Zaletel, Topology, magnetism and charge order in twisted MoTe2 at higher integer hole fillings, arXiv:2312.12531, DOI.
[145] C.-E. Ahn, W. Lee, K. Yananose, Y. Kim, G.Y. Cho, Non-Abelian fractional quantum anomalous Hall states and first Landau level physics of the second moiré band of twisted bilayer MoTe2, Phys. Rev. B, 110 (2024) L161109.
[146] Y.H. Zhang, D.N. Sheng, A. Vishwanath, SU(4) chiral spin liquid, exciton supersolid and electric detection in moire bilayers, Phys. Rev. Lett., 127 (2021).
[147] Y. Xu, K. Kang, K. Watanabe, T. Taniguchi, K.F. Mak, J. Shan, A tunable bilayer Hubbard model in twisted WSe2, Nat. Nanotechnol., 17 (2022) 934-939.
[148] C. Kuhlenkamp, W. Kadow, A. Imamoğlu, M. Knap, Chiral Pseudospin Liquids in Moir\'e Heterostructures, Phys. Rev. X, 14 (2024) 021013.
[149] L. Del Re, L. Classen, Field control of symmetry-broken and quantum disordered phases in frustrated moiré bilayers with population imbalance, Phys. Rev. Research, 6 (2024).
[150] L.F. Zhang, R.H. Ni, Y. Zhou, Controlling quantum phases of electrons and excitons in moire superlattices, J Appl Phys, 133 (2023).
[151] Z. Song, U.F.P. Seifert, Z.-X. Luo, L. Balents, Mott insulators in moiré transition metal dichalcogenides at fractional fillings: Slave-rotor mean-field theory, Phys. Rev. B, 108 (2023) 155109.
[152] J. Yu, J. Herzog-Arbeitman, M. Wang, O. Vafek, B.A. Bernevig, N. Regnault, Fractional Chern insulators versus nonmagnetic states in twisted bilayer MoTe2, Phys. Rev. B, 109 (2024) 045147.
[153] D. Zhai, C. Chen, C. Xiao, W. Yao, Time-reversal even charge hall effect from twisted interface coupling, Nat. Commun., 14 (2023) 1961.
[154] J. Zhu, D. Zhai, C. Xiao, W. Yao, Layer Hall counterflow as a model probe of magic-angle twisted bilayer graphene, Phys. Rev. B, 109 (2024) 155114.
[155] C. Chen, D. Zhai, C. Xiao, W. Yao, Crossed nonlinear dynamical Hall effect in twisted bilayers, Phys. Rev. Research, 6 (2024) L012059.
[156] J. Li, D. Zhai, C. Xiao, W. Yao, Dynamical chiral Nernst effect in twisted Van der Waals few layers, Quantum Front., 3 (2024) 11.
[157] H. Yu, X. Cui, X. Xu, W. Yao, Valley excitons in two-dimensional semiconductors,

Natl. Sci. Rev., 2 (2015) 57-70.
[158] G. Wang, A. Chernikov, M.M. Glazov, T.F. Heinz, X. Marie, T. Amand, B. Urbaszek, Colloquium: Excitons in atomically thin transition metal dichalcogenides, Rev. Mod. Phys., 90 (2018) 021001.
[159] A. Chernikov, T.C. Berkelbach, H.M. Hill, A. Rigosi, Y. Li, O.B. Aslan, D.R. Reichman, M.S. Hybertsen, T.F. Heinz, Exciton Binding Energy and Nonhydrogenic Rydberg Series in Monolayer WS2, Phys. Rev. Lett., 113 (2014) 076802.
[160] Z. Ye, T. Cao, K. O'Brien, H. Zhu, X. Yin, Y. Wang, S.G. Louie, X. Zhang, Probing excitonic dark states in single-layer tungsten disulphide, Nature, 513 (2014) 214-218.
[161] G. Wang, X. Marie, I. Gerber, T. Amand, D. Lagarde, L. Bouet, M. Vidal, A. Balocchi, B. Urbaszek, Giant Enhancement of the Optical Second-Harmonic Emission of WSe2 Monolayers by Laser Excitation at Exciton Resonances, Phys. Rev. Lett., 114 (2015) 097403.
[162] K. He, N. Kumar, L. Zhao, Z. Wang, K.F. Mak, H. Zhao, J. Shan, Tightly bound excitons in monolayer WSe2, Phys. Rev. Lett., 113 (2014) 026803.
[163] M.M. Ugeda, A.J. Bradley, S.-F. Shi, F.H.d. Jornada, Y. Zhang, D.Y. Qiu, W. Ruan, S.-K. Mo, Z. Hussain, Z.-X. Shen, F. Wang, S.G. Louie, M.F. Crommie, Giant bandgap renormalization and excitonic effects in a monolayer transition metal dichalcogenide semiconductor, Nat. Mater., 13 (2014) 1091-1095.
[164] C. Zhang, A. Johnson, C.-L. Hsu, L.-J. Li, C.-K. Shih, Direct Imaging of Band Profile in Single Layer MoS2 on Graphite: Quasiparticle Energy Gap, Metallic Edge States, and Edge Band Bending, Nano Lett., 14 (2014) 2443-2447.
[165] B. Zhu, X. Chen, X. Cui, Exciton Binding Energy of Monolayer WS2, Sci. Rep., 5 (2015) 9218.
[166] A.V. Stier, K.M. McCreary, B.T. Jonker, J. Kono, S.A. Crooker, Exciton diamagnetic shifts and valley Zeeman effects in monolayer WS2 and MoS2 to 65 Telsa, Nat. Commun., 7 (2016) 10643.
[167] A.V. Stier, N.P. Wilson, K.A. Velizhanin, J. Kono, X. Xu, S.A. Crooker, Magnetooptics of Exciton Rydberg States in a Monolayer Semiconductor, Phys. Rev. Lett., 120 (2018) 057405.
[168] S. Dong, M. Puppin, T. Pincelli, S. Beaulieu, D. Christiansen, H. Hübener, C.W. Nicholson, R.P. Xian, M. Dendzik, Y. Deng, Y.W. Windsor, M. Selig, E. Malic, A. Rubio, A. Knorr, M. Wolf, L. Rettig, R. Ernstorfer, Direct measurement of key exciton properties: Energy, dynamics, and spatial distribution of the wave function, Nat. Sci., 1 (2021) e10010.
[169] M.K.L. Man, J. Madéo1, C. Sahoo, K. Xie, M. Campbell, V. Pareek, A. Karmakar, E.L. Wong, A. Al-Mahboob, N.S. Chan, D.R. Bacon, X. Zhu, M.M.M. Abdelrasoul, X. Li, T.F. Heinz, F.H.d. Jornada, T. Cao, K.M. Dani, Experimental measurement of the intrinsic excitonic wave function, Sci. Adv., 7 (2021) eabg0192.
[170] S. Susarla, M.H. Naik, D.D. Blach, J. Zipfel, T. Taniguchi, K. Watanabe, L. Huang, R. Ramesh, F.H.d. Jornada, S.G. Louie, P. Ercius, A. Raja, Hyperspectral imaging of exciton confinement within a moiré unit cell with a subnanometer electron probe, Science, 378 (2022) 1235-1239.
[171] O. Karni, E. Barré, V. Pareek, J.D. Georgaras, M.K.L. Man, C. Sahoo, D.R. Bacon, X. Zhu, H.B. Ribeiro, A.L. O'Beirne, J. Hu, A. Al-Mahboob, M.M.M. Abdelrasoul, N.S. Chan, A. Karmakar, A.J. Winchester, B. Kim, K. Watanabe, T. Taniguchi, K. Barmak, J. Madéo, F.H.d. Jornada, T.F. Heinz, K.M. Dani, Structure of the moiré exciton captured by imaging its electron and hole, Nature, 603 (2022) 247-252.

[172] D.Y. Qiu, F.H. da Jornada, S.G. Louie, Optical Spectrum of ${\mathrm{MoS}}_{2}$: Many-Body Effects and Diversity of Exciton States, Phys. Rev. Lett., 111 (2013) 216805.
[173] S. Dufferwiel, S. Schwarz, F. Withers, A.A.P. Trichet, F. Li, M. Sich, O.D. Pozo-Zamudio, C. Clark, A. Nalitov, D.D. Solnyshkov, G. Malpuech, K.S. Novoselov, J.M. Smith, M.S. Skolnick, D.N. Krizhanovskii, A.I. Tartakovskii, Exciton-polaritons in van der Waals heterostructures embedded in tunable microcavities, Nat. Commun., 6 (2015) 8579.
[174] X. Liu, T. Galfsky, Z. Sun, F. Xia, E.-c. Lin, Y.-H. Lee, S. Kéna-Cohen, V.M. Menon, Strong light-matter coupling in two-dimensional atomic crystals, Nat. Photon., 9 (2015) 30-34.
[175] X. Liu, W. Bao, Q. Li, C. Ropp, Y. Wang, X. Zhang, Control of Coherently Coupled Exciton Polaritons in Monolayer Tungsten Disulphide, Phys. Rev. Lett., 119 (2017) 027403.
[176] S. Dhara, C. Chakraborty, K.M. Goodfellow, L. Qiu, T.A. O'Loughlin, G.W. Wicks, S. Bhattacharjee, A.N. Vamivakas, Anomalous dispersion of microcavity trion-polaritons, Nat. Phys., 14 (2017) 130-133.
[177] L.C. Flatten, Z. He, D.M. Coles, A.A.P. Trichet, A.W. Powell, R.A. Taylor, J.H. Warner, J.M. Smith, Room-temperature exciton-polaritons with two-dimensional WS2, Sci. Rep., 6 (2016) 33134.
[178] S. Wang, S. Li, T. Chervy, A. Shalabney, S. Azzini, E. Orgiu, J.A. Hutchison, C. Genet, P. Samorì, T.W. Ebbesen, Coherent Coupling of WS2 Monolayers with Metallic Photonic Nanostructures at Room Temperature, Nano Lett., 16 (2016) 4368-4374.
[179] Y.-J. Chen, J.D. Cain, T.K. Stanev, V.P. Dravid, N.P. Stern, Valley-polarized exciton-polaritons in a monolayer semiconductor, Nat. Photon., 11 (2017) 431-435.
[180] Z. Sun, J. Gu, A. Ghazaryan, Z. Shotan, C.R. Considine, M. Dollar, B. Chakraborty, X. Liu, P. Ghaemi, S. Kéna-Cohen, V.M. Menon, Optical control of room-temperature valley polaritons, Nat. Photon., 11 (2017) 491-496.
[181] S. Dufferwiel, T.P. Lyons, D.D. Solnyshkov, A.A.P. Trichet, F. Withers, S. Schwarz, G. Malpuech, J.M. Smith, K.S. Novoselov, M.S. Skolnick, D.N. Krizhanovskii, A.I. Tartakovskii, Valley-addressable polaritons in atomically thin semiconductors, Nat. Photon., 11 (2017) 497-501.
[182] N. Lundt, Ł. Dusanowski, E. Sedov, P. Stepanov, M.M. Glazov, S. Klembt, M. Klaas, J. Beierlein, Y. Qin, S. Tongay, M. Richard, A.V. Kavokin, S. Höfling, C. Schneider, Optical valley Hall effect for highly valley-coherent exciton-polaritons in an atomically thin semiconductor, Nat. Nanotechnol., 14 (2019) 770-775.
[183] S. Dufferwiel, T.P. Lyons, D.D. Solnyshkov, A.A.P. Trichet, A. Catanzaro, F. Withers, G. Malpuech, J.M. Smith, K.S. Novoselov, M.S. Skolnick, D.N. Krizhanovskii, A.I. Tartakovskii, Valley coherent exciton-polaritons in a monolayer semiconductor, Nat. Commun., 9 (2018) 4797.
[184] L. Qiu, C. Chakraborty, S. Dhara, A.N. Vamivakas, Room-temperature valley coherence in a polaritonic system, Nat. Commun., 10 (2019) 1513.
[185] M. Sidler, P. Back, O. Cotlet, A. Srivastava, T. Fink, M. Kroner, E. Demler, A. Imamoglu, Fermi polaron-polaritons in charge-tunable atomically thin semiconductors, Nat. Phys., 13 (2017) 255-261.
[186] P. Gong, H. Yu, Y. Wang, W. Yao, Optical selection rules for excitonic Rydberg series in the massive Dirac cones of hexagonal two-dimensional materials, Phys. Rev. B, 95 (2017) 125420.
[187] M.M. Glazov, L.E. Golub, G. Wang, X. Marie, T. Amand, B. Urbaszek, Intrinsic exciton-

state mixing and nonlinear optical properties in transition metal dichalcogenide monolayers, Phys. Rev. B, 95 (2017) 035311.
[188] K.F. Mak, K. He, J. Shan, T.F. Heinz, Control of valley polarization in monolayer MoS2 by optical helicity, Nat. Nanotechnol., 7 (2012) 494-498.
[189] H. Zeng, J. Dai, W. Yao, D. Xiao, X. Cui, Valley polarization in MoS2 monolayers by optical pumping, Nat. Nanotechnol., 7 (2012) 490-493.
[190] T. Cao, G. Wang, W. Han, H. Ye, C. Zhu, J. Shi, Q. Niu, P. Tan, E. Wang, B. Liu, J. Feng, Valley-selective circular dichroism of monolayer molybdenum disulphide, Nat. Commun., 3 (2012) 887.
[191] G. Sallen, L. Bouet, X. Marie, G. Wang, C.R. Zhu, W.P. Han, Y. Lu, P.H. Tan, T. Amand, B.L. Liu, B. Urbaszek, Robust optical emission polarization in MoS2 monolayers through selective valley excitation, Phys. Rev. B, 86 (2012) 081301(R).
[192] A.M. Jones, H. Yu, N.J. Ghimire, S. Wu, G. Aivazian, J.S. Ross, B. Zhao, J. Yan, D.G. Mandrus, D. Xiao, W. Yao, X. Xu, Optical generation of excitonic valley coherence in monolayer WSe2, Nat. Nanotechnol., 8 (2013) 634-638.
[193] D.Y. Qiu, T. Cao, S.G. Louie, Nonanalyticity, Valley Quantum Phases, and Lightlike Exciton Dispersion in Monolayer Transition Metal Dichalcogenides: Theory and First-Principles Calculations, Phys. Rev. Lett., 115 (2015) 176801.
[194] F. Wu, F. Qu, A.H. MacDonald, Exciton band structure of monolayer MoS2, Phys. Rev. B, 91 (2015) 075310.
[195] M.M. Glazov, T. Amand, X. Marie, D. Lagarde, L. Bouet, B. Urbaszek, Exciton fine structure and spin decoherence in monolayers of transition metal dichalcogenides, Phys. Rev. B, 89 (2014) 201302.
[196] T. Yu, M.W. Wu, Valley depolarization due to intervalley and intravalley electron-hole exchange interactions in monolayer MoS2, Phys. Rev. B, 89 (2014) 205303.
[197] M.Z. Maialle, E.A.d.A.e. Silva, L.J. Sham, Exciton spin dynamics in quantum wells, Phys. Rev. B, 47 (1993) 15776.
[198] H. Yu, G.-B. Liu, P. Gong, X. Xu, W. Yao, Dirac cones and Dirac saddle points of bright excitons in monolayer transition metal dichalcogenides, Nat. Commun., 5 (2014) 3876.
[199] M. Onga, Y. Zhang, T. Ideue, Y. Iwasa, Exciton Hall effect in monolayer MoS2, Nat. Mater., 16 (2017) 1193-1197.
[200] N. Ubrig, S. Jo, M. Philippi, D. Costanzo, H. Berger, A.B. Kuzmenko, A.F. Morpurgo, Microscopic Origin of the Valley Hall Effect in Transition Metal Dichalcogenides Revealed by Wavelength-Dependent Mapping, Nano Lett., 17 (2017) 5719-5725.
[201] X.-C. Yang, H. Yu, W. Yao, Chiral Excitonics in Monolayer Semiconductors on Patterned Dielectrics, Phys. Rev. Lett., 128 (2022) 217402.
[202] F. Wu, T. Lovorn, A.H. MacDonald, Topological Exciton Bands in Moiré Heterojunctions, Phys. Rev. Lett., 118 (2017) 147401.
[203] P. Rivera, H. Yu, K.L. Seyler, N.P. Wilson, W. Yao, X. Xu, Interlayer valley excitons in heterobilayers of transition metal dichalcogenides, Nat. Nanotechnol., 13 (2018) 1004-1015.
[204] H. Yu, Y. Wang, Q. Tong, X. Xu, W. Yao, Anomalous light cones and valley optical selection rules of interlayer excitons in twisted heterobilayers, Phys. Rev. Lett., 115 (2015) 187002.
[205] J.S. Ross, P. Rivera, J. Schaibley, E. Lee-Wong, H. Yu, T. Taniguchi, K. Watanabe, J.

Yan, D. Mandrus, D. Cobden, W. Yao, X. Xu, Interlayer Exciton Optoelectronics in a 2D Heterostructure p–n Junction, Nano Lett., 17 (2017) 638-643.
[206] C. Jin, E.C. Regan, D. Wang, M.I.B. Utama, C.-S. Yang, J. Cain, Y. Qin, Y. Shen, Z. Zheng, K. Watanabe, T. Taniguchi, S. Tongay, A. Zettl, F. Wang, Identification of spin, valley and moiré quasi-angular momentum of interlayer excitons, Nat. Phys., 15 (2019) 1140-1144.
[207] F. Wu, T. Lovorn, A.H. MacDonald, Theory of optical absorption by interlayer excitons in transition metal dichalcogenide heterobilayers, Phys. Rev. B, 97 (2018) 035306.
[208] P. Rivera, K.L. Seyler, H. Yu, J.R. Schaibley, J. Yan, D.G. Mandrus, W. Yao, X. Xu, Valley-polarized exciton dynamics in a 2D semiconductor heterostructure, Science, 351 (2016) 688-691.
[209] J. Horng, T. Stroucken, L. Zhang, E.Y. Paik, H. Deng, S.W. Koch, Observation of interlayer excitons in MoSe2 single crystals, Phys. Rev. B, 97 (2018) 241404.
[210] I.C. Gerber, E. Courtade, S. Shree, C. Robert, T. Taniguchi, K. Watanabe, A. Balocchi, P. Renucci, D. Lagarde, X. Marie, B. Urbaszek, Interlayer excitons in bilayer MoS2 with strong oscillator strength up to room temperature, Phys. Rev. B, 99 (2019) 035443.
[211] I. Niehues, A. Blob, T. Stiehm, S.M.d. Vasconcellos, R. Bratschitsch, Interlayer excitons in bilayer MoS2 under uniaxial tensile strain, Nanoscale, 11 (2019) 12788-12792.
[212] E.V. Calman, M.M. Fogler, L.V. Butov, S. Hu, A. Mishchenko, A.K. Geim, Indirect excitons in van der Waals heterostructures at room temperature, Nat. Commun., 9 (2018) 1895.
[213] N. Leisgang, S. Shree, I. Paradisanos, L. Sponfeldner, C. Robert, D. Lagarde, A. Balocchi, K. Watanabe, T. Taniguchi, X. Marie, R.J. Warburton, I.C. Gerber, B. Urbaszek, Giant Stark splitting of an exciton in bilayer MoS2, Nat. Nanotechnol., 15 (2020) 901-907.
[214] L. Zhang, Z. Zhang, F. Wu, D. Wang, R. Gogna, S. Hou, K. Watanabe, T. Taniguchi, K. Kulkarni, T. Kuo, S.R. Forrest, H. Deng, Twist-angle dependence of moiré excitons in WS2/MoSe2 heterobilayers, Nat. Commun., 11 (2020) 5888.
[215] J.R. Schaibley, P. Rivera, H. Yu, K.L. Seyler, J. Yan, D.G. Mandrus, T. Taniguchi, K. Watanabe, W. Yao, X. Xu, Directional interlayer spin-valley transfer in two-dimensional heterostructures, Nat. Commun., 7 (2016) 13747.
[216] E.M. Alexeev, D.A. Ruiz-Tijerina, M. Danovich, M.J. Hamer, D.J. Terry, P.K. Nayak, S. Ahn, S. Pak, J. Lee, J.I. Sohn, M.R. Molas, M. Koperski, K. Watanabe, T. Taniguchi, K.S. Novoselov, R.V. Gorbachev, H.S. Shin, V.I. Fal'ko, A.I. Tartakovskii, Resonantly hybridized excitons in moiré superlattices in van der Waals heterostructures, Nature, 567 (2019) 81-86.
[217] Y. Tang, J. Gu, S. Liu, K. Watanabe, T. Taniguchi, J. Hone, K.F. Mak, J. Shan, Tuning layer-hybridized moiré excitons by the quantum-confined Stark effect, Nat. Nanotech., 16 (2021) 52-57.
[218] W.-T. Hsu, B.-H. Lin, L.-S. Lu, M.-H. Lee, M.-W. Chu, L.-J. Li, W. Yao, W.-H. Chang, C.-K. Shih, Tailoring excitonic states of van der Waals bilayers through stacking configuration, band alignment, and valley spin, Sci. Adv., 5 (2019) eaax7407.
[219] E. Lorchat, M. Selig, F. Katsch, K. Yumigeta, S. Tongay, A. Knorr, C. Schneider, S. Höfling, Excitons in Bilayer MoS2 Displaying a Colossal Electric Field Splitting and Tunable Magnetic Response, Phys. Rev. Lett., 126 (2021) 037401.
[220] N. Peimyoo, T. Deilmann, F. Withers, J. Escolar, D. Nutting, T. Taniguchi, K. Watanabe, A. Taghizadeh, M.F. Craciun, K.S. Thygesen, S. Russo, Electrical tuning of

optically active interlayer excitons in bilayer MoS2, Nat. Nanotechnol., 16 (2021) 888-893.
[221] J. Sung, Y. Zhou, G. Scuri, V. Zólyomi, T.I. Andersen, H. Yoo, D.S. Wild, A.Y. Joe, R.J. Gelly, H. Heo, S.J. Magorrian, D. Bérubé, A.M.M. Valdivia, T. Taniguchi, K. Watanabe, M.D. Lukin, P. Kim, V.I. Fal'ko, H. Park, Broken mirror symmetry in excitonic response of reconstructed domains in twisted MoSe2/MoSe2 bilayers, Nat. Nanotechnol., 15 (2020) 750-754.
[222] A. Arora, M. Drüppel, R. Schmidt, T. Deilmann, R. Schneider, M.R. Molas, P. Marauhn, S.M.d. Vasconcellos, M. Potemski, M. Rohlfing, R. Bratschitsch, Interlayer excitons in a bulk van der Waals semiconductor, Nat. Commun., 8 (2017) 639.
[223] I. Paradisanos, S. Shree, A. George, N. Leisgang, C. Robert, K. Watanabe, T. Taniguchi, R.J. Warburton, A. Turchanin, X. Marie, I.C. Gerber, B. Urbaszek, Controlling interlayer excitons in MoS2 layers grown by chemical vapor deposition, Nat. Commun., 11 (2020) 2391.
[224] L. Zhang, F. Wu, S. Hou, Z. Zhang, Y.-H. Chou, K. Watanabe, T. Taniguchi, S.R. Forrest, H. Deng, Van der Waals heterostructure polaritons with moiré-induced nonlinearity, Nature, 591 (2021) 61-65.
[225] B. Datta, M. Khatoniar, P. Deshmukh, F. Thouin, R. Bushati, S.D. Liberato, S.K. Cohen, V.M. Menon, Highly nonlinear dipolar exciton-polaritons in bilayer MoS2, Nat. Commun., 13 (2022) 6341.
[226] C. Louca, A. Genco, S. Chiavazzo, T.P. Lyons, S. Randerson, C. Trovatello, P. Claronino, R. Jayaprakash, X. Hu, J. Howarth, K. Watanabe, T. Taniguchi, S.D. Conte, R. Gorbachev, D.G. Lidzey, G. Cerullo, O. Kyriienko, A.I. Tartakovskii, Interspecies exciton interactions lead to enhanced nonlinearity of dipolar excitons and polaritons in MoS2 homobilayers, Nat. Commun., 14 (2023) 3818.
[227] A. Ciarrocchi, D. Unuchek, A. Avsar, K. Watanabe, T. Taniguchi, A. Kis, Polarization switching and electrical control of interlayer excitons in two-dimensional van der Waals heterostructures, Nat. Photon., 13 (2019) 131-136.
[228] H. Yu, W. Yao, Electrically tunable topological transport of moiré polaritons, Sci. Bull., 65 (2020) 1555-1562.
[229] H. Zheng, D. Zhai, W. Yao, Anomalous Magneto-Optical Response and Chiral Interface of Dipolar Excitons at Twisted Valleys, Nano Lett., 22 (2022) 5466-5472.
[230] W.-T. Hsu, L.-S. Lu, P.-H. Wu, M.-H. Lee, P.-J. Chen, P.-Y. Wu, Y.-C. Chou, H.-T. Jeng, L.-J. Li, M.-W. Chu, W.-H. Chang, Negative circular polarization emissions from WSe2/MoSe2 commensurate heterobilayers, Nat. Commun., 9 (2018) 1356.
[231] K. Tran, G. Moody, F. Wu, X. Lu, J. Choi, K. Kim, A. Rai, D.A. Sanchez, J. Quan, A. Singh, J. Embley, A. Zepeda, M. Campbell, T. Autry, T. Taniguchi, K. Watanabe, N. Lu, S.K. Banerjee, K.L. Silverman, S. Kim, E. Tutuc, L. Yang, A.H. MacDonald, X. Li, Evidence for moiré excitons in van der Waals heterostructures, Nature, 567 (2019) 71-75.
[232] C. Jin, E.C. Regan, A. Yan, M.I.B. Utama, D. Wang, S. Zhao, Y. Qin, S. Yang, Z. Zheng, S. Shi, K. Watanabe, T. Taniguchi, S. Tongay, A. Zettl, F. Wang, Observation of moiré excitons in WSe2/WS2 heterostructure superlattices, Nature, 567 (2019) 76-80.
[233] K.L. Seyler, P. Rivera, H. Yu, N.P. Wilson, E.L. Ray, D.G. Mandrus, J. Yan, W. Yao, X. Xu, Signatures of moiré-trapped valley excitons in MoSe2/WSe2 heterobilayers, Nature, 567 (2019) 66-70.
[234] F. Lu, Q. Hu, Y. Xu, H. Yu, Coupled exciton internal and center-of-mass motions in two-dimensional semiconductors by a periodic electrostatic potential, Phys. Rev. B, 109

(2024) 165422.
[235] H.E. Hannachi, D. Elmaghraoui, S. Jaziri, Moiré interlayer exciton relative and center of mass motions coupling. Effect on $$1s-np$$ interlayer exciton THz transitions, Eur Phys J Plus, 138 (2023) 396.
[236] Y. Xu, C. Horn, J. Zhu, Y. Tang, L. Ma, L. Li, S. Liu, K. Watanabe, T. Taniguchi, J.C. Hone, J. Shan, K.F. Mak, Creation of moiré bands in a monolayer semiconductor by spatially periodic dielectric screening, Nat. Mater., 20 (2021) 645-649.
[237] P. Zhao, C. Xiao, W. Yao, Universal superlattice potential for 2D materials from twisted interface inside h-BN substrate, NPJ 2D Mater. Appl., 5 (2021) 38.
[238] D.S. Kim, R.C. Dominguez, R. Mayorga-Luna, D. Ye, J. Embley, T. Tan, Y. Ni, Z. Liu, M. Ford, F.Y. Gao, S. Arash, K. Watanabe, T. Taniguchi, S. Kim, C.-K. Shih, K. Lai, W. Yao, L. Yang, X. Li, Y. Miyahara, Electrostatic moiré potential from twisted hexagonal boron nitride layers, Nat. Mater., 23 (2024) 65-70.
[239] Q. Hu, Z. Zhan, H. Cui, Y. Zhang, F. Jin, X. Zhao, M. Zhang, Z. Wang, Q. Zhang, K. Watanabe, T. Taniguchi, X. Cao, W.-M. Liu, F. Wu, S. Yuan, Y. Xu, Observation of Rydberg moiré excitons, Science, 380 (2023) 1367-1372.
[240] Z. Zhang, J. Xie, W. Zhao, R. Qi, C. Sanborn, S. Wang, S. Kahn, K. Watanabe, T. Taniguchi, A. Zettl, M. Crommie, F. Wang, Engineering correlated insulators in bilayer graphene with a remote Coulomb superlattice, Nat. Mater., 23 (2024) 189-195.
[241] J. Gu, J. Zhu, P. Knuppel, K. Watanabe, T. Taniguchi, J. Shan, K.F. Mak, Remote imprinting of moiré lattices, Nat. Mater., 23 (2024) 219-223.
[242] M. He, J. Cai, H. Zheng, E. Seewald, T. Taniguchi, K. Watanabe, J. Yan, M. Yankowitz, A. Pasupathy, W. Yao, X. Xu, Dynamically tunable moiré exciton Rydberg states in a monolayer semiconductor on twisted bilayer graphene, Nat. Mater., 23 (2024) 224-229.
[243] S. Brem, C. Linderälv, P. Erhart, E. Malic, Tunable Phases of Moiré Excitons in van der Waals Heterostructures, Nano Lett., 20 (2020) 8534-8540.
[244] M. Brotons-Gisbert, H. Baek, A. Molina-Sánchez, A. Campbell, E. Scerri, D. White, K. Watanabe, T. Taniguchi, C. Bonato, B.D. Gerardot, Spin-layer locking of interlayer excitons trapped in moiré potentials, Nat. Mater., 19 (2020) 630-636.
[245] H. Baek, M. Brotons-Gisbert, Z.X. Koong, A. Campbell, M. Rambach, K. Watanabe, T. Taniguchi, B.D. Gerardot, Highly energy-tunable quantum light from moiré-trapped excitons, Sci. Adv., 6 (2020) eaba8526.
[246] H. Yu, W. Yao, Luminescence Anomaly of Dipolar Valley Excitons in Homobilayer Semiconductor Moiré Superlattices, Phys. Rev. X, 11 (2021) 021042.
[247] C. Ciuti, V. Savona, C. Piermarocchi, A. Quattropani, P. Schwendimann, Role of the exchange of carriers in elastic exciton-exciton scattering in quantum wells, Phys. Rev. B, 58 (1998) 7926.
[248] S.B.-T. de-Leon, B. Laikhtman, Exciton-exciton interactions in quantum wells: Optical properties and energy and spin relaxation, Phys. Rev. B, 63 (2001) 125306.
[249] T. Byrnes, P. Recher, Y. Yamamoto, Mott transitions of exciton polaritons and indirect excitons in a periodic potential, Phys. Rev. B, 81 (2010) 205312.
[250] V. Shahnazaryan, I. Iorsh, I.A. Shelykh, O. Kyriienko, Exciton-exciton interaction in transition-metal dichalcogenide monolayers, Phys. Rev. B, 96 (2017) 115409.
[251] V. Shahnazaryan, I.A. Shelykh, O. Kyriienko, Attractive Coulomb interaction of two-dimensional Rydberg excitons, Phys. Rev. B, 93 (2016) 245302.
[252] D. Erkensten, S. Brem, E. Malic, Exciton-exciton interaction in transition metal

dichalcogenide monolayers and van der Waals heterostructures, Phys. Rev. B, 103 (2021) 045426.
[253] V.A. Maslova, N.S. Voronova, Spatially-indirect and hybrid exciton-exciton interaction in MoS2 homobilayers, 2D Mater., 11 (2024) 025006.
[254] M. Combescot, R. Combescot, F. Dubin, Bose–Einstein condensation and indirect excitons: a review, Rep. Prog. Phys., 80 (2017) 066501.
[255] W. Li, X. Lu, S. Dubey, L. Devenica, A. Srivastava, Dipolar interactions between localized interlayer excitons in van der Waals heterostructures, Nat. Mater., 19 (2020) 624-629.
[256] M. Kremser, M. Brotons-Gisbert, J. Knörzer, J. Gückelhorn, M. Meyer, M. Barbone, A.V. Stier, B.D. Gerardot, K. Müller, J.J. Finley, Discrete interactions between a few interlayer excitons trapped at a MoSe2-WSe2 heterointerface, NPJ 2D Mater. Appl., 4 (2020) 8.
[257] L.A. Jauregui, A.Y. Joe, K. Pistunova, D.S. Wild, A.A. High, Y. Zhou, G. Scuri, K.D. Greve, A. Sushko, C.-H. Yu, T. Taniguchi, K. Watanabe, D.J. Needleman, M.D. Lukin, H. Park, P. Kim, Electrical control of interlayer exciton dynamics in atomically thin heterostructures, Science, 366 (2019) 870-875.
[258] L. Yuan, B. Zheng, J. Kunstmann, T. Brumme, A.B. Kuc, C. Ma, S. Deng, D. Blach, A. Pan, L. Huang, Twist-angle-dependent interlayer exciton diffusion in WS2-WSe2 heterobilayers, Nat. Mater., 19 (2020) 617-623.
[259] X. Sun, Y. Zhu, H. Qin, B. Liu, Y. Tang, T. Lü, S. Rahman, T. Yildirim, Y. Lu, Enhanced interactions of interlayer excitons in free-standing heterobilayers, Nature, 610 (2022) 478-484.
[260] Y. Hou, H. Yu, Dipolar interactions enhanced by two-dimensional dielectric screening in few-layer van der Waals structures, 2D Mater., 11 (2024) 025019.
[261] P. Cudazzo, I.V. Tokatly, A. Rubio, Dielectric screening in two-dimensional insulators: Implications for excitonic and impurity states in graphane, Phys. Rev. B, 84 (2011) 085406.
[262] T.C. Berkelbach, M.S. Hybertsen, D.R. Reichman, Theory of neutral and charged excitons in monolayer transition metal dichalcogenides, Phys. Rev. B, 88 (2013) 045318.
[263] I. Kylänpää, H.-P. Komsa, Binding energies of exciton complexes in transition metal dichalcogenide monolayers and effect of dielectric environment, Phys. Rev. B, 92 (2015) 205418.
[264] M.V.d. Donck, M. Zarenia, F.M. Peeters, Excitons and trions in monolayer transition metal dichalcogenides: A comparative study between the multiband model and the quadratic single-band model, Phys. Rev. B, 96 (2017) 035131.
[265] A.O. Slobodeniuk, M.R. Molas, Exciton spectrum in atomically thin monolayers: The role of hBN encapsulation, Phys. Rev. B, 108 (2023) 035427.
[266] M. Danovich, D.A. Ruiz-Tijerina, R.J. Hunt, M. Szyniszewski, N.D. Drummond, V.I. Fal'ko, Localized interlayer complexes in heterobilayer transition metal dichalcogenides, Phys. Rev. B, 97 (2018) 195452.
[267] S. Miao, T. Wang, X. Huang, D. Chen, Z. Lian, C. Wang, M. Blei, T. Taniguchi, K. Watanabe, S. Tongay, Z. Wang, D. Xiao, Y.-T. Cui, S.-F. Shi, Strong interaction between interlayer excitons and correlated electrons in WSe2/WS2 moiré superlattice, Nat. Commun., 12 (2021) 3608.
[268] T. Lahayef, C. Menotti, L. Santos, M. Lewenstein, T. Pfau, The physics of dipolar bosonic quantum gases, Rep. Prog. Phys., 72 (2009) 126401.

[269] H. Park, J. Zhu, X. Wang, Y. Wang, W. Holtzmann, T. Taniguchi, K. Watanabe, J. Yan, L. Fu, T. Cao, D. Xiao, D.R. Gamelin, H. Yu, W. Yao, X. Xu, Dipole ladders with large Hubbard interaction in a moiré exciton lattice, Nat. Phys., 19 (2023) 1286-1292.
[270] R. Xiong, J.H. Nie, S.L. Brantly, P. Hays, R. Sailus, K. Watanabe, T. Taniguchi, S. Tongay, C. Jin, Correlated insulator of excitons in WSe2/WS2 moiré superlattices, Science, 380 (2023) 860-864.
[271] Z. Lian, Y. Meng, L. Ma, I. Maity, L. Yan, Q. Wu, X. Huang, D. Chen, X. Chen, X. Chen, M. Blei, T. Taniguchi, K. Watanabe, S. Tongay, J. Lischner, Y.-T. Cui, S.-F. Shi, Valley-polarized exitonic Mott insulator in WS2/WSe2 moiré superlattice, Nat. Phys., 20 (2024) 34-39.
[272] B. Gao, D.G. Suárez-Forero, S. Sarkar, T.-S. Huang, D. Session, M.J. Mehrabad, R. Ni, M. Xie, P. Upadhyay, J. Vannucci, S. Mittal, K. Watanabe, T. Taniguchi, A. Imamoglu, Y. Zhou, M. Hafezi, Excitonic Mott insulator in a Bose-Fermi-Hubbard system of moiré WS2/WSe2 heterobilayer, Nat. Commun., 15 (2024) 2305.
[273] R. Xiong, S.L. Brantly, K. Su, J.H. Nie, Z. Zhang, R. Banerjee, H. Ruddick, K. Watanabe, T. Taniguchi, S.A. Tongay, C. Xu, C. Jin, Tunable exciton valley-pseudospin orders in moiré superlattices, Nat. Commun., 15 (2024) 4254.
[274] M. He, P. Rivera, D.V. Tuan, N.P. Wilson, M. Yang, T. Taniguchi, K. Watanabe, J. Yan, D.G. Mandrus, H. Yu, H. Dery, W. Yao, X. Xu, Valley phonons and exciton complexes in a monolayer semiconductor, Nat. Commun., 11 (2020) 618.
[275] Z. Li, T. Wang, C. Jin, Z. Lu, Z. Lian, Y. Meng, M. Blei, M. Gao, T. Taniguchi, K. Watanabe, T. Ren, T. Cao, S. Tongay, D. Smirnov, L. Zhang, S.-F. Shi, Momentum-Dark Intervalley Exciton in Monolayer Tungsten Diselenide Brightened via Chiral Phonon, ACS Nano, 13 (2019) 14107-14113.
[276] L. Wang, E.-M. Shih, A. Ghiotto, L. Xian, D.A. Rhodes, C. Tan, M. Claassen, D.M. Kennes, Y. Bai, B. Kim, K. Watanabe, T. Taniguchi, X. Zhu, J. Hone, A. Rubio, A.N. Pasupathy, C.R. Dean, Correlated electronic phases in twisted bilayer transition metal dichalcogenides, Nat. Mater., 19 (2020) 861-866.
[277] Y. Tang, L. Li, T. Li, Y. Xu, S. Liu, K. Barmak, K. Watanabe, T. Taniguchi, A.H. MacDonald, J. Shan, K.F. Mak, Simulation of Hubbard model physics in WSe2/WS2 moiré superlattices, Nature, 579 (2020) 353-358.
[278] T. Li, S. Jiang, L. Li, Y. Zhang, K. Kang, J. Zhu, K. Watanabe, T. Taniguchi, D. Chowdhury, L. Fu, J. Shan, K.F. Mak, Continuous Mott transition in semiconductor moiré superlattices, Nature, 597 (2021) 350-354.
[279] E.C. Regan, D. Wang, C. Jin, M.I.B. Utama, B. Gao, X. Wei, S. Zhao, W. Zhao, Z. Zhang, K. Yumigeta, M. Blei, J.D. Carlström, K. Watanabe, T. Taniguchi, S. Tongay, M. Crommie, A. Zettl, F. Wang, Mott and generalized Wigner crystal states in WSe2/WS2 moiré superlattices, Nature, 579 (2020) 359-363.
[280] X. Huang, T. Wang, S. Miao, C. Wang, Z. Li, Z. Lian, T. Taniguchi, K. Watanabe, S. Okamoto, D. Xiao, S.-F. Shi, Y.-T. Cui, Correlated insulating states at fractional fillings of the WS2/WSe2 moiré lattice, Nat. Phys., 17 (2021) 715-719.
[281] Y. Xu, S. Liu, D.A. Rhodes, K. Watanabe, T. Taniguchi, J. Hone, V. Elser, K.F. Mak, J. Shan, Correlated insulating states at fractional fillings of moiré superlattices, Nature, 587 (2020) 214-218.
[282] E. Liu, T. Taniguchi, K. Watanabe, N.M. Gabor, Y.-T. Cui, C.H. Lui, Excitonic and Valley-Polarization Signatures of Fractional Correlated Electronic Phases in a WSe2/WS2

Moiré Superlattice, Phys. Rev. Lett., 127 (2021) 037402.
[283] A.J. Campbell, M. Brotons-Gisbert, H. Baek, V. Vitale, T. Taniguchi, K. Watanabe, J. Lischner, B.D. Gerardot, Exciton-polarons in the presence of strongly correlated electronic states in a MoSe2/WSe2 moiré superlattice, NPJ 2D Mater. Appl., 6 (2022) 79.
[284] Z. Zhang, E.C. Regan, D. Wang, W. Zhao, S. Wang, M. Sayyad, K. Yumigeta, K. Watanabe, T. Taniguchi, S. Tongay, M. Crommie, A. Zettl, M.P. Zaletel, F. Wang, Correlated interlayer exciton insulator in heterostructures of monolayer WSe2 and moiré WS2/WSe2, Nat. Phys., 18 (2022) 1214-1220.
[285] J. Gu, L. Ma, S. Liu, K. Watanabe, T. Taniguchi, J.C. Hone, J. Shan, K.F. Mak, Dipolar excitonic insulator in a moiré lattice, Nat. Phys., 18 (2022) 395-400.
[286] D. Chen, Z. Lian, X. Huang, Y. Su, M. Rashetnia, L. Ma, L. Yan, M. Blei, L. Xiang, T. Taniguchi, K. Watanabe, S. Tongay, D. Smirnov, Z. Wang, C. Zhang, Y.-T. Cui, S.-F. Shi, Excitonic insulator in a heterojunction moiré superlattice, Nat. Phys., 18 (2022) 1171-1176.
[287] H. Li, S. Li, E.C. Regan, D. Wang, W. Zhao, S. Kahn, K. Yumigeta, M. Blei, T. Taniguchi, K. Watanabe, S. Tongay, A. Zettl, M.F. Crommie, F. Wang, Imaging two-dimensional generalized Wigner crystals, Nature, 597 (2021) 650-654.
[288] H. Li, S. Li, M.H. Naik, J. Xie, X. Li, E. Regan, D. Wang, W. Zhao, K. Yumigeta, M. Blei, T. Taniguchi, K. Watanabe, S. Tongay, A. Zettl, S.G. Louie, M.F. Crommie, F. Wang, Imaging local discharge cascades for correlated electrons in WS2/WSe2 moiré superlattices, Nat. Phys., 17 (2021) 1114-1119.
[289] C. Jin, Z. Tao, T. Li, Y. Xu, Y. Tang, J. Zhu, S. Liu, K. Watanabe, T. Taniguchi, J.C. Hone, L. Fu, J. Shan, K.F. Mak, Stripe phases in WSe2/WS2 moiré superlattices, Nat. Mater., 20 (2021) 940-944.
[290] H. Li, Z. Xiang, E. Regan, W. Zhao, R. Sailus, R. Banerjee, T. Taniguchi, K. Watanabe, S. Tongay, A. Zettl, M.F. Crommie, F. Wang, Mapping charge excitations in generalized Wigner crystals, Nat. Nanotechnol., 19 (2024) 618-623.
[291] T. Smoleński, P.E. Dolgirev, C. Kuhlenkamp, A. Popert, Y. Shimazaki, P. Back, X. Lu, M. Kroner, K. Watanabe, T. Taniguchi, I. Esterlis, E. Demler, A. Imamoğlu, Signatures of Wigner crystal of electrons in a monolayer semiconductor, Nature, 595 (2021) 53-57.
[292] Y. Shimazaki, I. Schwartz, K. Watanabe, T. Taniguchi, M. Kroner, A. Imamoğlu, Strongly correlated electrons and hybrid excitons in a moiré heterostructure, Nature, 580 (2020) 472-477.
[293] Y. Shimazaki, C. Kuhlenkamp, I. Schwartz, T. Smoleński, K. Watanabe, T. Taniguchi, M. Kroner, R. Schmidt, M. Knap, A. Imamoğlu, Optical Signatures of Periodic Charge Distribution in a Mott-like Correlated Insulator State, Phys. Rev. X, 11 (2021) 021027.
[294] Y. Zhou, J. Sung, E. Brutschea, I. Esterlis, Y. Wang, G. Scuri, R.J. Gelly, H. Heo, T. Taniguchi, K. Watanabe, G. Zaránd, M.D. Lukin, P. Kim, E. Demler, H. Park, Bilayer Wigner crystals in a transition metal dichalcogenide heterostructure, Nature, 595 (2021) 48-52.
[295] V.M. Bedanov, G.V. Gadiyak, Y.E. Lozovik, On a modified Lindemann-like criterion for 2D melting, Phys. Lett. A, 109 (1985) 289-291.
[296] G.M. Bruun, D.R. Nelson, Quantum hexatic order in two-dimensional dipolar and charged fluids, Phys. Rev. B, 89 (2014) 094112.
[297] J. Zhou, J. Tang, H. Yu, Melting of electronic/excitonic crystals in 2D semiconductor moiré patterns: A perspective from the Lindemann criterion, Chin. Phys. B, 32 (2023) 107308.
[298] N. Morales-Durán, A.H. MacDonald, P. Potasz, Metal-insulator transition in

transition metal dichalcogenide heterobilayer moiré superlattices, Phys. Rev. B, 103 (2021).

[299] Y. Zhou, D.N. Sheng, E.A. Kim, Quantum Phases of Transition Metal Dichalcogenide Moire Systems, Phys Rev Lett, 128 (2022) 157602.

[300] N. Morales-Duran, N.C. Hu, P. Potasz, A.H. MacDonald, Nonlocal Interactions in Moire Hubbard Systems, Phys Rev Lett, 128 (2022) 217202.

[301] N. Morales-Durán, P. Potasz, A.H. MacDonald, Magnetism and quantum melting in moiré-material Wigner crystals, Phys. Rev. B, 107 (2023).

[302] Y. Zhou, D.N. Sheng, E.A. Kim, Quantum Melting of Generalized Wigner Crystals in Transition Metal Dichalcogenide Moire Systems, Phys Rev Lett, 133 (2024) 156501.

[303] L. Bonsall, A.A. Maradudin, Some static and dynamical properties of a two-dimensional Wigner crystal, Phys. Rev. B, 15 (1977) 1959.

[304] R. Côté, A.H. MacDonald, Collective modes of the two-dimensional Wigner crystal in a strong magnetic field, Phys. Rev. B, 44 (1991) 8759.

[305] J.H. Jang, B.M. Hunt, L.N. Pfeiffer, K.W. West, R.C. Ashoori, Sharp tunnelling resonance from the vibrations of an electronicWigner crystal, Nat. Phys., 13 (2017) 340-344.

[306] H. Yu, J. Zhou, Phonons of electronic crystals in two-dimensional semiconductor moiré patterns, Nat. Sci., 3 (2023) e20220065.

[307] Z. Lian, D. Chen, Y. Meng, X. Chen, Y. Su, R. Banerjee, T. Taniguchi, K. Watanabe, S. Tongay, C. Zhang, Y.-T. Cui, S.-F. Shi, Exciton Superposition across Moiré States in a Semiconducting Moiré Superlattice, Nat. Commun., 14 (2023) 5042.

[308] N. Kumar, Q. Cui, F. Ceballos, D. He, Y. Wang, H. Zhao, Exciton-exciton annihilation in MoSe2 monolayers, Phys. Rev. B, 89 (2014) 125427.

[309] S. Mouri, Y. Miyauchi, M. Toh, W. Zhao, G. Eda, K. Matsuda, Nonlinear photoluminescence in atomically thin layered WSe2 arising from diffusion-assisted exciton-exciton annihilation, Phys. Rev. B, 90 (2014) 155449.

[310] D. Sun, Y. Rao, G.A. Reider, G. Chen, Y. You, L. Brézin, A.R. Harutyunyan, T.F. Heinz, Observation of Rapid Exciton-Exciton Annihilation in Monolayer Molybdenum Disulfide, Nano Lett., 14 (2014) 5625-5629.

[311] J. Wang, Q. Shi, E.-M. Shih, L. Zhou, W. Wu, Y. Bai, D. Rhodes, K. Barmak, J. Hone, C.R. Dean, X.-Y. Zhu, Diffusivity Reveals Three Distinct Phases of Interlayer Excitons in MoSe2/WSe2 Heterobilayers, Phys. Rev. Lett., 126 (2021) 106804.

[312] J. Wang, J. Ardelean, Y. Bai, A. Steinhoff, M. Florian, F. Jahnke, X. Xu, M. Kira, J. Hone, X.-Y. Zhu, Optical generation of high carrier densities in 2D semiconductor heterobilayers, Sci. Adv., 5 (2019) eaax0145.

[313] C. Lagoin, U. Bhattacharya, T. Grass, R.W. Chhajlany, T. Salamon, K. Baldwin, L. Pfeiffer, M. Lewenstein, M. Holzmann, F. Dubin, Extended Bose-Hubbard model with dipolar excitons, Nature, 609 (2022) 485-489.

[314] Y. Bai, Y. Li, S. Liu, Y. Guo, J. Pack, J. Wang, C.R. Dean, J. Hone, X. Zhu, Evidence for Exciton Crystals in a 2D Semiconductor Heterotrilayer, Nano Lett., 23 (2023) 11621-11629.

[315] Y. Zeng, Z. Xia, R. Dery, K. Watanabe, T. Taniguchi, J. Shan, K.F. Mak, Exciton density waves in Coulomb-coupled dual moire lattices, Nat Mater, 22 (2023) 175-179.

[316] Y. Aharonov, A. Stern, Origin of the geometric forces accompanying Berry's geometric potentials, Phys. Rev. Lett., 69 (1992) 3593.

[317] A. Graf, F. Piéchon, Berry curvature and quantum metric in Phys. Rev. B, 104 (2021) 085114.
[318] N. Morales-Durán, N. Wei, J. Shi, A.H. MacDonald, Magic Angles and Fractional Chern Insulators in Twisted Homobilayer Transition Metal Dichalcogenides, Phys. Rev. Lett., 132 (2024) 096602.
[319] J. Shi, N. Morales-Durán, E. Khalaf, A.H. MacDonald, Adiabatic approximation and Aharonov-Casher bands in twisted homobilayer transition metal dichalcogenides, Phys. Rev. B, 110 (2024) 035130.
[320] E. Thompson, K.T. Chu, F. Mesple, X.-W. Zhang, C. Hu, Y. Zhao, H. Park, J. Cai, E. Anderson, K. Watanabe, T. Taniguchi, J. Yang, J.-H. Chu, X. Xu, T. Cao, D. Xiao, M. Yankowitz, Microscopic signatures of topology in twisted MoTe2, Nat. Phys., DOI 10.1038/s41567-025-02877-x(2025).
[321] F. Zhang, N. Morales-Durán, Y. Li, W. Yao, J.-J. Su, Y.-C. Lin, C. Dong, X. Liu, F.-X.R. Chen, H. Kim, K. Watanabe, T. Taniguchi, X. Li, J.A. Robinson, A.H. Macdonald, C.-K. Shih, Experimental signature of layer skyrmions and implications for band topology in twisted WSe2 bilayers, Nat. Phys., DOI 10.1038/s41567-025-02876-y(2025).
[322] F.D.M. Haldane, Model for a Quantum Hall Effect without Landau Levels: Condensed-Matter Realization of the "Parity Anomaly", Phys. Rev. Lett., 61 (1988) 2015.
[323] W.-X. Qiu, B. Li, X.-J. Luo, F. Wu, Interaction-Driven Topological Phase Diagram of Twisted Bilayer MoTe2, Phys. Rev. X, 13 (2023) 041026.
[324] C.L. Kane, E.J. Mele, Quantum Spin Hall Effect in Graphene, Phys. Rev. Lett., 95 (2005) 226801.
[325] K. Kang, B. Shen, Y. Qiu, Y. Zeng, Z. Xia, K. Watanabe, T. Taniguchi, J. Shan, K.F. Mak, Evidence of the fractional quantum spin Hall effect in moiré MoTe2, Nature, 628 (2024) 522-526.
[326] K. Kang, Y. Qiu, K. Watanabe, T. Taniguchi, J. Shan, K.F. Mak, Double quantum spin Hall phase in moir\'e WSe2, Nano Lett., 24 (2024) 14901–14907.
[327] D. Zhai, W. Yao, Ultrafast control of moiré pseudo-electromagnetic field in homobilayer semiconductors, Nat. Sci., 2 (2022) e20210101.
[328] E. Anderson, F.-R. Fan, J. Cai, W. Holtzmann, T. Taniguchi, K. Watanabe, D. Xiao, W. Yao, X. Xu, Programming correlated magnetic states with gate-controlled moiré geometry, Science, 381 (2023) 325-330.
[329] H. Zheng, D. Zhai, C. Xiao, W. Yao, Layer Coherence Origin of Planar Hall Effect: From Charge to Multipole and Valley, Nano Lett, 25 (2025) 10096-10101.
[330] J. Dalibard, F. Gerbier, G. Juzeliūnas, P. Öhberg, Colloquium: Artificial gauge potentials for neutral atoms, Rev. Mod. Phys., 83 (2011) 1523.
[331] N. Goldman, G. Juzeliūnas, P. Öhberg, I.B. Spielman, Light-induced gauge fields for ultracold atoms, Rep. Prog. Phys., 77 (2014) 126401.
[332] D. Zhai, W. Yao, Layer Pseudospin Dynamics and Genuine Non-Abelian Berry Phase in Inhomogeneously Strained Moiré Pattern, Phys. Rev. Lett., 125 (2020) 266404.
[333] L. Onsager, Reciprocal Relations in Irreversible Processes. I, Phys. Rev., 37 (1931) 405-426.
[334] T. Stauber, T. Low, G. Gómez-Santos, Chiral Response of Twisted Bilayer Graphene, Phys. Rev. Lett., 120 (2018) 046801.
[335] T. Stauber, T. Low, G. Gómez-Santos, Linear response of twisted bilayer graphene: Continuum versus tight-binding models, Phys. Rev. B, 98 (2018) 195414.

[336] T. Stauber, J. González, G. Gómez-Santos, Change of chirality at magic angles of twisted bilayer graphene, Phys. Rev. B, 102 (2020) 081404.
[337] D. Xiao, M.-C. Chang, Q. Niu, Berry phase effects on electronic properties, Rev. Mod. Phys., 82 (2010) 1959.
[338] C.-J. Kim, A. Sánchez-Castillo, Z. Ziegler, Y. Ogawa, C. Noguez, J. Park, Chiral atomically thin films, Nat. Nanotech., 11 (2016) 520-524.
[339] E.S. Morell, L. Chico, L. Brey, Twisting dirac fermions: circular dichroism in bilayer graphene, 2D Mater., 4 (2017) 035015.
[340] F.-R. Fan, C. Xiao, W. Yao, Intrinsic dipole Hall effect in twisted MoTe2: magnetoelectricity and contact-free signatures of topological transitions, Nat. Commun., 15 (2024) 7997.
[341] J.P. Provost, G. Vallee, Riemannian structure on manifolds of quantum states, Commun.Math. Phys., 76 (1980) 289-301.
[342] A. Gao, Y.-F. Liu, C. Hu, J.-X. Qiu, C. Tzschaschel, B. Ghosh, S.-C. Ho, D. Bérubé, R. Chen, H. Sun, Z. Zhang, X.-Y. Zhang, Y.-X. Wang, N. Wang, Z. Huang, C. Felser, A. Agarwal, T. Ding, H.-J. Tien, A. Akey, J. Gardener, B. Singh, K. Watanabe, T. Taniguchi, K.S. Burch, D.C. Bell, B.B. Zhou, W. Gao, H.-Z. Lu, A. Bansil, H. Lin, T.-R. Chang, L. Fu, Q. Ma, N. Ni, S.-Y. Xu, Layer Hall effect in a 2D topological axion antiferromagnet, Nature, 595 (2021) 521-525.
[343] R. Chen, H.-P. Sun, M. Gu, C.-B. Hua, Q. Liu, H.-Z. Lu, X.C. Xie, Layer Hall effect induced by hidden Berry curvature in antiferromagnetic insulators, Natl. Sci. Rev., 11 (2024) nwac140.
[344] S. Li, M. Gong, S. Cheng, H. Jiang, X.C. Xie, Dissipationless layertronics in axion insulator MnBi2Te4, Natl. Sci. Rev., 11 (2024) nwad262.
[345] N. Wang, D. Kaplan, Z. Zhang, T. Holder, N. Cao, A. Wang, X. Zhou, F. Zhou, Z. Jiang, C. Zhang, S. Ru, H. Cai, K. Watanabe, T. Taniguchi, B. Yan, W. Gao, Quantum-metric-induced nonlinear transport in a topological antiferromagnet, Nature, 621 (2023) 487-492.
[346] A. Gao, Y.-F. Liu, J.-X. Qiu, B. Ghosh, T.V. Trevisan, Y. Onishi, C. Hu, T. Qian, H.-J. Tien, S.-W. Chen, M. Huang, D. Bérubé, H. Li, C. Tzschaschel, T. Dinh, Z. Sun, S.-C. Ho, S.-W. Lien, B. Singh, K. Watanabe, T. Taniguchi, D.C. Bell, H. Lin, T.-R. Chang, C.R. Du, A. Bansil, L. Fu, N. Ni, P.P. Orth, Q. Ma, S.-Y. Xu, Quantum metric nonlinear Hall effect in a topological antiferromagnetic heterostructure, Science, 381 (2023) 181-186.
[347] C. Wang, Y. Gao, D. Xiao, Intrinsic Nonlinear Hall Effect in Antiferromagnetic Tetragonal CuMnAs, Phys. Rev. Lett., 127 (2021) 277201.
[348] H. Liu, J. Zhao, Y.-X. Huang, W. Wu, X.-L. Sheng, C. Xiao, S.A. Yang, Intrinsic Second-Order Anomalous Hall Effect and Its Application in Compensated Antiferromagnets, Phys. Rev. Lett., 127 (2021) 277202.
[349] Y. Gao, S.A. Yang, Q. Niu, Field Induced Positional Shift of Bloch Electrons and Its Dynamical Implications, Phys. Rev. Lett., 112 (2014) 166601.
[350] J.-X. Hu, C. Zeng, Y. Yao, Colossal layer Nernst effect in twisted moir\'e layers, Phys. Rev. B, 109 (2024) L201403.
[351] H. Zheng, D. Zhai, C. Xiao, W. Yao, Interlayer Electric Multipoles Induced by In-Plane Field from Quantum Geometric Origins, Nano Lett., 24 (2024) 8017-8023.
[352] M. Huang, Z. Wu, J. Hu, X. Cai, E. Li, L. An, X. Feng, Z. Ye, N. Lin, K.T. Law, N. Wang, Giant nonlinear Hall effect in twisted bilayer WSe2, Natl. Sci. Rev., 10 (2022) nwac232.

[353] J.-X. Hu, C.-P. Zhang, Y.-M. Xie, K.T. Law, Nonlinear Hall effects in strained twisted bilayer WSe2, Commun. Phys., 5 (2022) 255.
[354] I. Sodemann, L. Fu, Quantum Nonlinear Hall Effect Induced by Berry Curvature Dipole in Time-Reversal Invariant Materials, Phys. Rev. Lett., 115 (2015) 216806.
[355] Z.Z. Du, H.-Z. Lu, X.C. Xie, Nonlinear Hall effects, Nat. Rev. Phys., 3 (2021) 744-752.
[356] P.C. Adak, S. Sinha, A. Agarwal, M.M. Deshmukh, Tunable moiré materials for probing Berry physics and topology, Nat. Rev. Mater., 9 (2024) 481-498.
[357] L. Du, Z. Huang, J. Zhang, F. Ye, Q. Dai, H. Deng, G. Zhang, Z. Sun, Nonlinear physics of moiré superlattices, Nat. Mater., 23 (2024) 1179-1192.
[358] T. Devakul, L. Fu, Quantum Anomalous Hall Effect from Inverted Charge Transfer Gap, Phys. Rev. X, 12 (2022) 021031.
[359] L. Rademaker, Spin-orbit coupling in transition metal dichalcogenide heterobilayer flat bands, Phys. Rev. B, 105 (2022) 195428.
[360] Y. Su, H. Li, C. Zhang, K. Sun, S.-Z. Lin, Massive Dirac fermions in moiré superlattices: A route towards topological flat minibands and correlated topological insulators, Phys. Rev. Research, 4 (2022) L032024.
[361] Y.-W. Chang, Y.-C. Chang, Quantum anomalous Hall effect and electric-field-induced topological phase transition in AB-stacked MoTe2/WSe2 moiré heterobilayers, Phys. Rev. B, 106 (2022) 245412.
[362] H. Pan, M. Xie, F. Wu, S.D. Sarma, Topological Phases in AB-Stacked MoTe2/WSe2: z2 Topological Insulators, Chern Insulators, and Topological Charge Density Waves, Phys. Rev. Lett., 129 056804.
[363] Z. Dong, Y.-H. Zhang, Excitonic Chern insulator and kinetic ferromagnetism in a MoTe2/WSe2 moiré bilayer, Phys. Rev. B, 107 (2023) L081101.
[364] M. Xie, H. Pan, F. Wu, S.D. Sarma, Nematic Excitonic Insulator in Transition Metal Dichalcogenide Moiré Heterobilayers, Phys. Rev. Lett., 131 (2023) 046402.
[365] Y.-M. Xie, C.-P. Zhang, K.T. Law, Topological px+ipy intervalley coherent state in moiré MoTe2/WSe2 heterobilayers, Phys. Rev. B, 110 (2024) 045115.
[366] J.-X. Hu, Y.-M. Xie, K.T. Law, Berry curvature, spin Hall effect, and nonlinear optical response in moiré transition metal dichalcogenide heterobilayers, Phys. Rev. B, 107 (2023) 075424.
[367] Z. Tao, B. Shen, S. Jiang, T. Li, L. Li, L. Ma, W. Zhao, J. Hu, K. Pistunova, K. Watanabe, T. Taniguchi, T.F. Heinz, K.F. Mak, J. Shan, Valley-Coherent Quantum Anomalous Hall State in AB-Stacked MoTe2/WSe2 Bilayers, Phys. Rev. X, 14 (2024) 011004.
[368] Z. Tao, B. Shen, W. Zhao, N.C. Hu, T. Li, S. Jiang, L. Li, K. Watanabe, T. Taniguchi, A.H. MacDonald, J. Shan, K.F. Mak, Giant Spin Hall Effect in AB-Stacked MoTe2/WSe2 Bilayers, Nat. Nanotechnol., 19 (2024) 28-33.
[369] C.L. Tschirhart, E. Redekop, L. Li, T. Li, S. Jiang, T. Arp, O. Sheekey, T. Taniguchi, K. Watanabe, M.E. Huber, K.F. Mak, J. Shan, A.F. Young, Intrinsic spin Hall torque in a moiré Chern magnet, Nat. Phys., 19 (2023) 807-813.
[370] W. Zhao, K. Kang, Y. Zhang, P. Knüppel, Z. Tao, L. Li, C.L. Tschirhart, E. Redekop, K. Watanabe, T. Taniguchi, A.F. Young, J. Shan, K.F. Mak, Realization of the Haldane Chern insulator in a moiré lattice, Nat. Phys., 20 (2024) 275-280.
[371] W. Zhao, B. Shen, Z. Tao, Z. Han, K. Kang, K. Watanabe, T. Taniguchi, K.F. Mak, J. Shan, Gate-tunable heavy fermions in a moiré Kondo lattice, Nature, 616 (2023) 61-65.
[372] W. Zhao, B. Shen, Z. Tao, S. Kim, P. Knüppel, Z. Han, Y. Zhang, K. Watanabe, T.

Taniguchi, D. Chowdhury, J. Shan, K.F. Mak, Emergence of ferromagnetism at the onset of moiré Kondo breakdown, Nat. Phys., DOI 10.1038/s41567-024-02636-4(2024).
[373] D. Guerci, J. Wang, J. Zang, J. Cano, J.H. Pixley, A. Millis, Chiral Kondo lattice in doped MoTe2/WSe2 bilayers, Sci. Adv., 9 eade7701.
[374] H. Jia, B. Ma, R.L. Luo, G. Chen, Double exchange, itinerant ferromagnetism, and topological Hall effect in moir\'e heterobilayer, Phys. Rev. Research, 5 (2023) L042033.
[375] F. Xie, L. Chen, Q. Si, Kondo effect and its destruction in heterobilayer transition metal dichalcogenides, Phys. Rev. Research, 6 (2024) 013219.
[376] J.F. Mendez-Valderrama, S. Kim, D. Chowdhury, Correlated topological mixed-valence insulators in moiré heterobilayers, Phys. Rev. B, 110 (2024).
[377] D. Guerci, K.P. Lucht, V. Crépel, J. Cano, J.H. Pixley, A. Millis, Topological Kondo semimetal and insulator in AB-stacked heterobilayer transition metal dichalcogenides, Phys. Rev. B, 110 (2024).
[378] P. Streda, Theory of quantised Hall conductivity in two dimensions, J. Phys. C: Solid State Phys., 15 (1982) L717.
[379] A. Abouelkomsan, A.P. Reddy, L. Fu, E.J. Bergholtz, Band mixing in the quantum anomalous Hall regime of twisted semiconductor bilayers, Phys. Rev. B, 109 (2024) L121107.
[380] J. Dong, J. Wang, P.J. Ledwith, A. Vishwanath, D.E. Parker, Composite Fermi Liquid at Zero Magnetic Field in Twisted MoTe2, Phys. Rev. Lett., 131 (2023) 136502.
[381] H. Goldman, A.P. Reddy, N. Paul, L. Fu, Zero-Field Composite Fermi Liquid in Twisted Semiconductor Bilayers, Phys. Rev. Lett., 131 (2023) 136501.
[382] E. Anderson, J. Cai, A.P. Reddy, H. Park, W. Holtzmann, K. Davis, T. Taniguchi, K. Watanabe, T. Smolenski, A. Imamoğlu, T. Cao, D. Xiao, L. Fu, W. Yao, X. Xu, Trion sensing of a zero-field composite Fermi liquid, Nature, 635 (2024) 590-595.
[383] Z. Ji, H. Park, M.E. Barber, C. Hu, K. Watanabe, T. Taniguchi, J.-H. Chu, X. Xu, Z.-X. Shen, Local probe of bulk and edge states in a fractional Chern insulator, Nature, 635 (2024) 578-583.
[384] E. Redekop, C. Zhang, H. Park, J. Cai, E. Anderson, O. Sheekey, T. Arp, G. Babikyan, S. Salters, K. Watanabe, T. Taniguchi, M.E. Huber, X. Xu, A.F. Young, Direct magnetic imaging of fractional Chern insulators in twisted MoTe2, Nature, 635 (2024) 584-589.
[385] Y. Wang, J. Choe, E. Anderson, W. Li, J. Ingham, E.A. Arsenault, Y. Li, X. Hu, T. Taniguchi, K. Watanabe, X. Roy, D. Basov, D. Xiao, R. Queiroz, J.C. Hone, X. Xu, X.Y. Zhu, Hidden states and dynamics of fractional fillings in twisted MoTe(2) bilayers, Nature, 641 (2025) 1149-1155.
[386] A.P. Reddy, L. Fu, Toward a global phase diagram of the fractional quantum anomalous Hall effect, Phys. Rev. B, 108 (2023) 245159.
[387] N. Regnault, B.A. Bernevig, Fractional Chern Insulator, Phys. Rev. X, 1 (2011) 021014.
[388] B.A. Bernevig, N. Regnault, Emergent many-body translational symmetries of Abelian and non-Abelian fractionally filled topological insulators, Phys. Rev. B, 85 (2012) 075128.
[389] E.J. Bergholtz, Z. Liu, Topological Flat Band Models and Fractional Chern Insulators, Int. J. Mod. Phys. B, 27 (2013) 1330017.
[390] S.A. Parameswaran, R. Roy, S.L. Sondhi, Fractional quantum Hall physics in topological flat bands, C. R. Phys, 14 (2013) 816-839.
[391] L. Zheng, L. Feng, W. Yong-Shi, Exotic electronic states in the world of flat bands:

From theory to material, Chin. Phys. B, 23 (2014) 077308.
[392] F.D.M. Haldane, "Fractional statistics" in arbitrary dimensions: A generalization of the Pauli principle, Phys. Rev. Lett., 67 (1991) 937.
[393] H. Li, F.D.M. Haldane, Entanglement Spectrum as a Generalization of Entanglement Entropy: Identification of Topological Order in Non-Abelian Fractional Quantum Hall Effect States, Phys. Rev. Lett., 101 (2008) 010504.
[394] J. Wang, J. Cano, A.J. Millis, Z. Liu, B. Yang, Exact Landau Level Description of Geometry and Interaction in a Flatband, Phys. Rev. Lett., 127 (2021) 246403.
[395] P.J. Ledwith, A. Vishwanath, D.E. Parker, Vortexability: A unifying criterion for ideal fractional Chern insulators, Phys. Rev. B, 108 (2023) 205144.
[396] Z. Liu, B. Mera, M. Fujimoto, T. Ozawa, J. Wang, Theory of Generalized Landau Levels and Implication for non-Abelian States, arXiv:2405.14479, DOI (2024).
[397] M. Fujimoto, D.E. Parker, J. Dong, E. Khalaf, A. Vishwanath, P. Ledwith, Higher Vortexability: Zero-Field Realization of Higher Landau Levels, Phys Rev Lett, 134 (2025) 106502.
[398] J. Wang, S. Klevtsov, Z. Liu, Origin of model fractional Chern insulators in all topological ideal flatbands: Explicit color-entangled wave function and exact density algebra, Phys. Rev. Research, 5 (2023).
[399] H. Li, Y. Su, Y.B. Kim, H.-Y. Kee, K. Sun, S.-Z. Lin, Contrasting twisted bilayer graphene and transition metal dichalcogenides for fractional Chern insulators: An emergent gauge picture, Phys. Rev. B, 109 (2024) 245131.
[400] K. Kolář, K. Yang, F. von Oppen, C. Mora, Hofstadter spectrum of Chern bands in twisted transition metal dichalcogenides, Phys. Rev. B, 110 (2024) 115114.
[401] B. Li, F. Wu, Variational mapping of Chern bands to Landau levels: Application to fractional Chern insulators in twisted MoTe2, Phys. Rev. B, 111 (2025).
[402] J. Shi, J. Cano, N. Morales-Durán, Effects of Berry curvature on ideal band magnetorotons, arXiv:2503.15900, DOI (2025).
[403] W. Yang, D. Zhai, T. Tan, F.R. Fan, Z. Lin, W. Yao, Fractional Quantum Anomalous Hall Effect in a Singular Flat Band, Phys Rev Lett, 134 (2025) 196501.
[404] Z. Lin, W. Yang, H. Lu, D. Zhai, W. Yao, Fractional Chern insulator states in an isolated flat band of zero Chern number, arXiv:2505.09009, DOI (2025).
[405] D. Bauer, S. Talkington, F. Harper, B. Andrews, R. Roy, Fractional Chern insulators with a non-Landau level continuum limit, Phys. Rev. B, 105 (2022).
[406] B. Andrews, M. Raja, N. Mishra, M.P. Zaletel, R. Roy, Stability of fractional Chern insulators with a non-Landau level continuum limit, Phys. Rev. B, 109 (2024).
[407] G. Shavit, Y. Oreg, Quantum Geometry and Stabilization of Fractional Chern Insulators Far from the Ideal Limit, Phys Rev Lett, 133 (2024) 156504.
[408] B.A. Foutty, C.R. Kometter, T. Devakul, A.P. Reddy, K. Watanabe, T. Taniguchi, L. Fu, B.E. Feldman, Mapping twist-tuned multiband topology in bilayer WSe2, Science, 384 (2024) 343-347.
[409] H. Park, J.Q. Cai, E. Anderson, X.W. Zhang, X.Y. Liu, W. Holtzmann, W.J. Li, C. Wang, C.W. Hu, Y.Z. Zhao, T. Taniguchi, K. Watanabe, J.H. Yang, D. Cobden, J.H. Chu, N. Regnault, B.A. Bernevig, L. Fu, T. Cao, D. Xiao, X.D. Xu, Ferromagnetism and topology of the higher flat band in a fractional Chern insulator, Nat. Phys., 21 (2025) 549–555.
[410] F. Xu, X. Chang, J. Xiao, Y. Zhang, F. Liu, Z. Sun, N. Mao, N. Peshcherenko, J. Li, K. Watanabe, T. Taniguchi, B. Tong, L. Lu, J. Jia, D. Qian, Z. Shi, Y. Zhang, X. Liu, S. Jiang, T.

Li, Interplay between topology and correlations in the second moiré band of twisted bilayer MoTe2, Nat. Phys., 21 (2025) 542-548.
[411] L. An, H. Pan, W.X. Qiu, N. Wang, S. Ru, Q. Tan, X. Dai, X. Cai, Q. Shang, X. Lu, H. Jiang, X. Lyu, S. Yang, K. Watanabe, T. Taniguchi, F. Wu, W.B. Gao, Observation of ferromagnetic phase in the second moire band of twisted MoTe(2), Nat Commun, 16 (2025) 5131.
[412] P. Knuppel, J. Zhu, Y. Xia, Z. Xia, Z. Han, Y. Zeng, K. Watanabe, T. Taniguchi, J. Shan, K.F. Mak, Correlated states controlled by a tunable van Hove singularity in moire WSe(2) bilayers, Nat Commun, 16 (2025) 1959.
[413] A. Ghiotto, E.-M. Shih, G.S.S.G. Pereira, D.A. Rhodes, B. Kim, J. Zang, A.J. Millis, K. Watanabe, T. Taniguchi, J.C. Hone, L. Wang, C.R. Dean, A.N. Pasupathy, Quantum criticality in twisted transition metal dichalcogenides, Nature, 597 (2021) 345-349.
[414] Y. Xia, Z. Han, K. Watanabe, T. Taniguchi, J. Shan, K.F. Mak, Superconductivity in twisted bilayer WSe(2), Nature, 637 (2025) 833-838.
[415] Y. Guo, J. Pack, J. Swann, L. Holtzman, M. Cothrine, K. Watanabe, T. Taniguchi, D.G. Mandrus, K. Barmak, J. Hone, A.J. Millis, A. Pasupathy, C.R. Dean, Superconductivity in 5.0 degrees twisted bilayer WSe(2), Nature, 637 (2025) 839-845.
[416] Y.T. Hsu, F.C. Wu, S. Das Sarma, Spin-valley locked instabilities in moire transition metal dichalcogenides with conventional and higher-order Van Hove singularities, Phys. Rev. B, 104 (2021).
[417] M. Bélanger, J. Fournier, D. Sénéchal, Superconductivity in the twisted bilayer transition metal dichalcogenide WSe2: A quantum cluster study, Phys. Rev. B, 106 (2022).
[418] A. Wietek, J. Wang, J.W. Zang, J. Cano, A. Georges, A. Millis, Tunable stripe order and weak superconductivity in the Moire Hubbard model, Phys. Rev. Research, 4 (2022).
[419] F. Chen, D.N. Sheng, Singlet, triplet, and pair density wave superconductivity in the doped triangular-lattice moire system, Phys. Rev. B, 108 (2023).
[420] L. Klebl, A. Fischer, L. Classen, M.M. Scherer, D.M. Kennes, Competition of density waves and superconductivity in twisted tungsten diselenide, Phys. Rev. Research, 5 (2023).
[421] Y.M. Wu, Z. Wu, H. Yao, Pair-Density-Wave and Chiral Superconductivity in Twisted Bilayer Transition Metal Dichalcogenides, Phys Rev Lett, 130 (2023) 126001.
[422] M. Zegrodnik, A. Biborski, Mixed singlet-triplet superconducting state within the moire model applied to twisted bilayer WSe2, Phys. Rev. B, 108 (2023).
[423] B. Zhou, Y.H. Zhang, Chiral and nodal superconductors in the t-J model with valley contrasting flux on a triangular moire lattice, Phys. Rev. B, 108 (2023).
[424] W. Akbar, A. Biborski, L. Rademaker, M. Zegrodnik, Topological superconductivity with mixed singlet-triplet pairing in moiré transition metal dichalcogenide bilayers, Phys. Rev. B, 110 (2024).
[425] M. Christos, P.M. Bonetti, M.S. Scheurer, Approximate symmetries, insulators, and superconductivity in continuum-model description of twisted WSe $_2$, arXiv:2407.02393, DOI (2024).
[426] A. Fischer, L. Klebl, V. Crépel, S. Ryee, A. Rubio, L. Xian, T.O. Wehling, A. Georges, D.M. Kennes, A.J. Millis, Theory of intervalley-coherent AFM order and topological superconductivity in tWSe $_2$, arXiv:2412.14296, DOI (2024).
[427] D. Guerci, D. Kaplan, J. Ingham, J. Pixley, A.J. Millis, Topological superconductivity from repulsive interactions in twisted WSe $_2$, arXiv:2408.16075, DOI (2024).
[428] N. Myerson-Jain, C. Xu, Superconductor-Insulator Transition in the TMD moir\'{e}

systems and the Deconfined Quantum Critical Point, arXiv:2406.12971, DOI (2024).
[429] W. Qin, W.-X. Qiu, F. Wu, Kohn-Luttinger Mechanism of Superconductivity in Twisted Bilayer WSe$_2$: Gate-Tunable Unconventional Pairing Symmetry, arXiv:2409.16114, DOI (2024).
[430] C. Schrade, L. Fu, Nematic, chiral, and topological superconductivity in twisted transition metal dichalcogenides, Phys. Rev. B, 110 (2024).
[431] C. Tuo, M.-R. Li, Z. Wu, W. Sun, H. Yao, Theory of Topological Superconductivity and Antiferromagnetic Correlated Insulators in Twisted Bilayer WSe $_2$, arXiv:2409.06779, DOI (2024).
[432] A.V. Chubukov, C.M. Varma, Quantum criticality and superconductivity in twisted transition metal dichalcogenides, Phys. Rev. B, 111 (2025).
[433] S. Kim, J.F. Mendez-Valderrama, X. Wang, D. Chowdhury, Theory of correlated insulators and superconductor at nu = 1 in twisted WSe(2), Nat Commun, 16 (2025) 1701.
[434] F. Xie, L. Chen, S. Sur, Y. Fang, J. Cano, Q. Si, Superconductivity in Twisted WSe_2 from Topology-Induced Quantum Fluctuations, Phys Rev Lett, 134 (2025) 136503.
[435] J. Zhu, Y.-Z. Chou, Y. Huang, S.D. Sarma, In-plane magnetic field-induced orbital FFLO superconductivity in twisted WSe $_2$ homobilayers, arXiv:2503.18946, DOI (2025).
[436] J. Zhu, Y.Z. Chou, M. Xie, S. Das Sarma, Superconductivity in twisted transition metal dichalcogenide homobilayers, Phys. Rev. B, 111 (2025).
[437] F. Xu, Z. Sun, J. Li, C. Zheng, C. Xu, J. Gao, T. Jia, K. Watanabe, T. Taniguchi, B. Tong, L. Lu, J. Jia, Z. Shi, S. Jiang, Y. Zhang, Y. Zhang, S. Lei, X. Liu, T. Li, Signatures of unconventional superconductivity near reentrant and fractional quantum anomalous Hall insulators, arXiv:2504.06972, DOI (2025).
[438] G. Moore, N. Read, Nonabelions in the fractional quantum hall effect, Nucl. Phys. B, 360 (1991) 362-396.
[439] N. Read, D. Green, Paired states of fermions in two dimensions with breaking of parity and time-reversal symmetries and the fractional quantum Hall effect, Phys. Rev. B, 61 (2000) 10267-10297.
[440] M. Levin, B.I. Halperin, B. Rosenow, Particle-Hole Symmetry and the Pfaffian State, Phys. Rev. Lett., 99 (2007) 236806.
[441] S.-S. Lee, S. Ryu, C. Nayak, M.P.A. Fisher, Particle-Hole Symmetry and the $\ensuremath{\nu}=\frac{5}{2}$ Quantum Hall State, Phys. Rev. Lett., 99 (2007) 236807.
[442] N. Read, E. Rezayi, Beyond paired quantum Hall states: Parafermions and incompressible states in the first excited Landau level, Phys. Rev. B, 59 (1999) 8084-8092.
[443] C. Wang, X.W. Zhang, X. Liu, J. Wang, T. Cao, D. Xiao, Higher Landau-Level Analogs and Signatures of Non-Abelian States in Twisted Bilayer MoTe_2, Phys Rev Lett, 134 (2025) 076503.
[444] F. Chen, W.W. Luo, W. Zhu, D.N. Sheng, Robust non-Abelian even-denominator fractional Chern insulator in twisted bilayer MoTe(2), Nat Commun, 16 (2025) 2115.
[445] A.P. Reddy, N. Paul, A. Abouelkomsan, L. Fu, Non-Abelian Fractionalization in Topological Minibands, Phys. Rev. Lett., 133 (2024) 166503.
[446] M. Levin, A. Stern, Fractional Topological Insulators, Phys. Rev. Lett., 103 (2009) 196803.
[447] Y.-H. Zhang, Vortex Spin Liquid with Fractional Quantum Spin Hall Effect in Moir\'e Chern Bands, Phys. Rev. Lett., 133 (2024) 106502.
[448] V. Crépel, A. Millis, Spinon Pairing Induced by Chiral In-Plane Exchange and the

Stabilization of Odd-Spin Chern Number Spin Liquid in Twisted ${\mathrm{MoTe}}_{2}$, Phys. Rev. Lett., 133 (2024) 146503.
[449] Y.-H. Zhang, Non-Abelian and Abelian descendants of a vortex spin liquid: Fractional quantum spin Hall effect in twisted ${\mathrm{MoTe}}_{2}$, Phys. Rev. B, 110 (2024) 155102.
[450] Y.-Z. Chou, S. Das Sarma, Composite helical edges from Abelian fractional topological insulators, Phys. Rev. B, 110 (2024) 155117.
[451] I. Sodemann Villadiego, Halperin states of particles and holes in ideal time reversal invariant pairs of Chern bands and the fractional quantum spin Hall effect in moir\'e ${\mathrm{MoTe}}_{2}$, Phys. Rev. B, 110 (2024) 045114.
[452] P. Mai, J. Zhao, P.W. Phillips, Incipient quantum spin Hall insulator under strong correlations, arXiv:2409.07557, DOI (2024).
[453] C.M. Jian, M. Cheng, C.K. Xu, Minimal Fractional Topological Insulator in Half-Filled Conjugate Moire<acute accent> Chern Bands, Phys. Rev. X, 15 (2025).
[454] A. Abouelkomsan, L. Fu, Non-Abelian spin Hall insulator, Phys. Rev. Research, 7 (2025).
[455] J. May-Mann, A. Stern, T. Devakul, Theory of half-integer fractional quantum spin Hall edges, Phys. Rev. B, 111 (2025).
[456] Y.H. Kwan, G. Wagner, J. Yu, A.K. Dagnino, Y. Jiang, X. Xu, B.A. Bernevig, T. Neupert, N. Regnault, When Could Abelian Fractional Topological Insulators Exist in Twisted MoTe $ _2 $(and Other Systems), arXiv:2407.02560, DOI (2024).
[457] D.N. Sheng, A.P. Reddy, A. Abouelkomsan, E.J. Bergholtz, L. Fu, Quantum Anomalous Hall Crystal at Fractional Filling of Moir\'e Superlattices, Phys. Rev. Lett., 133 (2024) 066601.
[458] X.-Y. Song, C.-M. Jian, L. Fu, C. Xu, Intertwined fractional quantum anomalous Hall states and charge density waves, Phys. Rev. B, 109 (2024) 115116.
[459] F.-R. Fan, C. Xiao, W. Yao, Orbital Chern insulator at $\nu=-2$ in twisted MoTe2, Phys. Rev. B, 109 (2024) L041403.
[460] B. Li, W.-X. Qiu, F. Wu, Electrically tuned topology and magnetism in twisted bilayer MoTe2 at $\nu_h=1$, Phys. Rev. B, 109 (2024) L041106.
[461] X. Liu, Y. He, C. Wang, X.-W. Zhang, T. Cao, D. Xiao, Gate-Tunable Antiferromagnetic Chern Insulator in Twisted Bilayer Transition Metal Dichalcogenides, Phys. Rev. Lett., 132 (2024) 146401.
[462] A. Abouelkomsan, E.J. Bergholtz, S. Chatterjee, Multiferroicity and Topology in Twisted Transition Metal Dichalcogenides, Phys. Rev. Lett., 133 (2024) 026801.
[463] Z. Lu, T. Han, Y. Yao, A.P. Reddy, J. Yang, J. Seo, K. Watanabe, T. Taniguchi, L. Fu, L. Ju, Fractional quantum anomalous Hall effect in multilayer graphene, Nature, 626 (2024) 759-764.
[464] J. Xie, Z. Huo, X. Lu, Z. Feng, Z. Zhang, W. Wang, Q. Yang, K. Watanabe, T. Taniguchi, K. Liu, Z. Song, X.C. Xie, J. Liu, X. Lu, Tunable fractional Chern insulators in rhombohedral graphene superlattices, Nat Mater, DOI 10.1038/s41563-025-02225-7(2025).
[465] Z. Lu, T. Han, Y. Yao, Z. Hadjri, J. Yang, J. Seo, L. Shi, S. Ye, K. Watanabe, T. Taniguchi, L. Ju, Extended quantum anomalous Hall states in graphene/hBN moire superlattices, Nature, 637 (2025) 1090-1095.
[466] Q. Gao, J.K. Dong, P. Ledwith, D. Parker, E. Khalaf, Untwisting Moire Physics: Almost Ideal Bands and Fractional Chern Insulators in Periodically Strained Monolayer Graphene,

Phys. Rev. Lett., 131 (2023).
[467] Y. Liu, Z. Zhu, Engineering Fractional Chern Insulators through Periodic Strain in Monolayer Graphene and Transition Metal Dichalcogenides, arXiv:2407.12411, DOI (2024).
[468] J. Mao, S.P. Milovanović, M. Anđelković, X. Lai, Y. Cao, K. Watanabe, T. Taniguchi, L. Covaci, F.M. Peeters, A.K. Geim, Y. Jiang, E.Y. Andrei, Evidence of flat bands and correlated states in buckled graphene superlattices, Nature, 584 (2020) 215-220.
[469] C. Nayak, S.H. Simon, A. Stern, M. Freedman, S. Das Sarma, Non-Abelian anyons and topological quantum computation, Rev. Mod. Phys., 80 (2008) 1083-1159.
[470] J. Alicea, P. Fendley, Topological Phases with Parafermions: Theory and Blueprints, Annu. Rev. Condens. Matter Phys., 7 (2016) 119-139.
[471] N.H. Lindner, E. Berg, G. Refael, A. Stern, Fractionalizing Majorana Fermions: Non-Abelian Statistics on the Edges of Abelian Quantum Hall States, Phys. Rev. X, 2 (2012) 041002.
[472] M. Cheng, Superconducting proximity effect on the edge of fractional topological insulators, Phys. Rev. B, 86 (2012) 195126.
[473] D.J. Clarke, J. Alicea, K. Shtengel, Exotic non-Abelian anyons from conventional fractional quantum Hall states, Nat. Commun., 4 (2013) 1348.
[474] Y. Jia, F. Yuan, G. Cheng, Y. Tang, G. Yu, T. Song, P. Wang, R. Singha, A.J. Uzan-Narovlansky, M. Onyszczak, K. Watanabe, T. Taniguchi, N. Yao, L.M. Schoop, S. Wu, Surface-confined two-dimensional mass transport and crystal growth on monolayer materials, Nat. Synth., 3 (2024) 386-393.
[475] Y. Jia, G. Yu, T. Song, F. Yuan, A.J. Uzan, Y. Tang, P. Wang, R. Singha, M. Onyszczak, Z.J. Zheng, K. Watanabe, T. Taniguchi, L.M. Schoop, S. Wu, Superconductivity from On-Chip Metallization on 2D Topological Chalcogenides, Phys. Rev. X, 14 (2024) 021051.
[476] Y. Jia, T. Song, Z.J. Zheng, G. Cheng, A.J. Uzan, G. Yu, Y. Tang, C.J. Pollak, F. Yuan, M. Onyszczak, K. Watanabe, T. Taniguchi, S. Lei, N. Yao, L.M. Schoop, N.P. Ong, S. Wu, Anomalous superconductivity in twisted MoTe(2) nanojunctions, Sci Adv, 11 (2025) eadq5712.
[477] J.M. Park, Y. Cao, K. Watanabe, T. Taniguchi, P. Jarillo-Herrero, Tunable strongly coupled superconductivity in magic-angle twisted trilayer graphene, Nature, 590 (2021) 249-255.
[478] Z. Hao, A.M. Zimmerman, P. Ledwith, E. Khalaf, D.H. Najafabadi, K. Watanabe, T. Taniguchi, A. Vishwanath, P. Kim, Electric field–tunable superconductivity in alternating-twist magic-angle trilayer graphene, Science, 371 (2021) 1133-1138.
[479] X. Zhang, K.-T. Tsai, Z. Zhu, W. Ren, Y. Luo, S. Carr, M. Luskin, E. Kaxiras, K. Wang, Correlated Insulating States and Transport Signature of Superconductivity in Twisted Trilayer Graphene Superlattices, Phys. Rev. Lett., 127 (2021) 166802.
[480] A. Uri, S.C. de la Barrera, M.T. Randeria, D. Rodan-Legrain, T. Devakul, P.J.D. Crowley, N. Paul, K. Watanabe, T. Taniguchi, R. Lifshitz, L. Fu, R.C. Ashoori, P. Jarillo-Herrero, Superconductivity and strong interactions in a tunable moiré quasicrystal, Nature, 620 (2023) 762-767.
[481] C.-Y. Hao, Z. Zhan, P.A. Pantaleón, J.-Q. He, Y.-X. Zhao, K. Watanabe, T. Taniguchi, F. Guinea, L. He, Robust flat bands in twisted trilayer graphene moiré quasicrystals, Nat. Commun., 15 (2024) 8437.
[482] Z. Lian, D. Chen, L. Ma, Y. Meng, Y. Su, L. Yan, X. Huang, Q. Wu, X. Chen, M. Blei, T.

Taniguchi, K. Watanabe, S. Tongay, C. Zhang, Y.-T. Cui, S.-F. Shi, Quadrupolar excitons and hybridized interlayer Mott insulator in a trilayer moiré superlattice, Nat. Commun., 14 (2023) 4604.
[483] W. Li, Z. Hadjri, L.M. Devenica, J. Zhang, S. Liu, J. Hone, K. Watanabe, T. Taniguchi, A. Rubio, A. Srivastava, Quadrupolar–dipolar excitonic transition in a tunnel-coupled van der Waals heterotrilayer, Nat. Mater., 22 (2023) 1478-1484.
[484] L. Yu, K. Pistunova, J. Hu, K. Watanabe, T. Taniguchi, T.F. Heinz, Observation of quadrupolar and dipolar excitons in a semiconductor heterotrilayer, Nat. Mater., 22 (2023) 1485-1491.
[485] Y. Xie, Y. Gao, F. Chen, Y. Wang, J. Mao, Q. Liu, S. Chu, H. Yang, Y. Ye, Q. Gong, J. Feng, Y. Gao, Bright and Dark Quadrupolar Excitons in the ${\mathrm{WSe}}_{2}/{\mathrm{MoSe}}_{2}/{\mathrm{WSe}}_{2}$ Heterotrilayer, Phys. Rev. Lett., 131 (2023) 186901.